\documentclass[12pt]{iopart}

\expandafter\let\csname eqalign\endcsname\relax
\expandafter\let\csname equation*\endcsname\relax
\expandafter\let\csname endequation*\endcsname\relax

\usepackage{algorithm2e}
\usepackage{algorithmic}
\usepackage{bm}
\usepackage{graphicx}
 \usepackage{mathtools, nccmath}
 \usepackage[makeroom]{cancel}
 \usepackage{xcolor}
 \usepackage{amsfonts}
\usepackage{placeins}
\usepackage{comment}
\usepackage{grffile}

\DeclarePairedDelimiter{\nint}\lfloor\rceil
\begin{document}

\title[Excitation of TAE modes by an electromagnetic antenna using ORB5]{Linear and nonlinear excitation of TAE modes by external electromagnetic perturbations using ORB5}

\author{Mohsen Sadr$^{1}$, Alexey Mishchenko$^{2}$, Thomas Hayward-Schneider$^{3}$, Axel Koenies$^{2}$, Alberto Bottino$^{3}$, Alessandro Biancalani$^{4,3}$, Peter Donnel$^{5}$, Emmanuel Lanti$^{1}$,
Laurent Villard$^{1}$}

\address{$^{1}$Swiss Plasma Center, EPFL, CH-1015 Lausanne, Switzerland}
\address{$^{2}$Max-Planck Institute f\"{u}r Plasmaphysik, Greifswald, Germany}
\address{$^{3}$Max-Planck Institute f\"{u}r Plasmaphysik, Garching, Germany}
\address{$^{4}$
L{\'e}onard de Vinci P{\^o}le Universitaire, Research Center, Paris la D{\'e}fense, France}
\address{$^{5}$CEA, IRFM, Saint-Paul-lez-Durance, F-13108, France}

\ead{msadr@mit.edu}
\vspace{10pt}
\begin{indented}
\item[]\today
\end{indented}

\begin{abstract}
The excitation of toroidicity induced Alfv{\'e}n eigenmodes (TAEs) using prescribed external electromagnetic perturbations (hereafter ``antenna") acting on a confined toroidal plasma  as well as its nonlinear couplings to other modes in the system is studied. The antenna is described by an electrostatic potential resembling the target TAE mode structure along with its corresponding parallel electromagnetic potential computed from Ohm's law. Numerically stable long-time linear simulations are achieved by integrating the antenna within the framework of a mixed representation and pullback scheme [A. Mishchenko, et al., Comput. Phys. Commun. \textbf{238} (2019) 194]. By decomposing the plasma electromagnetic potential into symplectic and Hamiltonian parts and using Ohm's law, the destabilizing contribution of the potential gradient parallel to the magnetic field is canceled in the equations of motion. Besides evaluating the frequencies as well as growth/damping rates of excited modes compared to referenced TAEs, we study the interaction of antenna-driven modes with fast particles and indicate their margins of instability. Furthermore, we show  first nonlinear simulations in the presence of a TAE-like antenna exciting other TAE modes, as well as Global Alfv\'en Eigenmodes (GAE)
having different toroidal wave numbers from that of the antenna.
\end{abstract}

%
%
%
%
%

\section{Introduction}
Since the toroidicity induced Alfv{\'e}n eigenmodes (TAEs) \cite{cheng1986low} can become destabilized when fast particles are included in the confined plasma \cite{fu1989excitation,cheng1991alpha,biglari1992resonant,villard1995global}, analysis of the mode's structure for a given system of equations is of great interest. Ideally, from studying the mode behavior, one may predict and avoid the region of TAE instabilities that may be triggered in experiments. Although the growth rate and the frequency of the mode can be measured in simulations/experiments when excited in the presence of fast particles, 
the associated damping rate is out of reach unless the modes are excited by an external source (antenna). Hence, one may deploy an antenna to 
resonantly excite
TAE modes and measure the damping rates as the antenna is switched off, e.g. see the experiments performed in the Joint European Torus (JET) tokamak \cite{fasoli2000fast}.
\\ \ \\
 The TAE modes have been studied extensively in the literature through simulations of hybrid fluid-kinetic or gyrokinetic codes \cite{briguglio1995hybrid,nishimura2009excitation,bass2010gyrokinetic,konies2018benchmark,zhang2012global}. In particular, the Linear Gyrokinetic Shear Alfvén Physics (LIGKA) code \cite{lauber2003linear,lauber2005kinetic,lauber2007ligka} was developed in order to study the growth and damping of eigenmodes in the presence of  energetic particles and a numerical antenna.
\\ \ \\
\noindent Motivated by previous works \cite{zhang2012global,nabais2018tae}, in this study we further develop an electrostatic and electromagnetic antenna \cite{ohana2020using} in the ORB5 code \cite{lanti2020orb5}, i.e., a global electromagnetic gyrokinetic code using the PIC approach in toroidal geometry, in order to excite the localized TAEs at a specific gap position
in the vicinity of the rational $q=(|m|+1/2)/|n|$ surface. Here, $m$ and $n$ are the poloidal and toroidal mode numbers, respectively, and $q$ denotes the safety factor. The antenna is devised as an electrostatic potential and its electromagnetic counterpart is computed by solving Ohm's law. In order to allow simulations with large time step sizes, the self-consistent antenna field has been integrated into the equations of motion in the mixed-variable formulation as well as the celebrated pullback scheme of $\delta f$ method \cite{MISHCHENKO2019194}.
\\ \ \\
\noindent The content of this manuscript is distributed as follows. In section~\ref{sec:Rev_ORB5_gyrokinetic}, a short review of the mixed-variable formulation and the governing equations of ORB5, as well as the deployed solution algorithm, is provided. Next, the description of the antenna and its integration in the equations of motion is explained in section~\ref{sec:antenna}. Then in section~\ref{sec:results},  excitation of the TAE mode using the antenna is shown with several linear and nonlinear simulation results. Here, the frequency and damping rate of the mode in the linear setting is measured and compared to the results of fast particle simulations. Furthermore, for a plasma that is TAE-excited with an antenna,  the margins of instability as a function of fast particle density have been studied. In nonlinear simulations using a $n=6$ antenna, the excitation of a $n=2$ TAE, as well as axisymmetric ($n=0$) and non-axisymmetric ($n=1$) Global Alfv\'en Modes (GAEs) is demonstrated. Concluding remarks and outlook are made in section~\ref{sec:conclusion}.
\section{Review: Gyrokinetic model solved by ORB5}
\label{sec:Rev_ORB5_gyrokinetic}
\subsection{Mixed-variable collisionless kinetic equation solved by ORB5}
In this section we review the governing equations that are solved with the global  gyrokinetic particle-in-cell code ORB5 \cite{lanti2020orb5} in the framework of the pullback scheme \cite{MISHCHENKO2019194}. First, the velocity distribution function is reduced to $f(\bm R, v_{||}, \mu;t)$ where $\bm R$ denotes the coordinates of gyrocenter, $v_{||}$ is the component of velocity parallel to magnetic field, and $\mu$ is the magnetic moment.  By decomposing the reduced velocity distribution function associated with each species $f_s$ into background control variate $F_{0s}$ and the remaining $\delta f_s$, i.e., $f_s=F_{0s}+\delta f_s$, ORB5 solves the evolution of $\delta f_s$ for each species following the gyrokinetic Vlasov-Maxwell system of equations, i.e.,
\begin{equation}
\frac{\partial \delta f_s}{\partial t} + \dot{\bm R} \cdot \frac{\partial \delta f_s}{\partial \bm R}\Big|_{v_{||}} +\dot{v}_{||} \frac{\partial \delta f_s}{\partial v_{||}} = -\dot{\bm R}^{(1)}  \cdot \frac{\partial F_{0s}}{\partial \bm R}\Big|_{\epsilon} - \dot{\epsilon}^{(1)} \frac{\partial F_{0s}}{\partial \epsilon}~,
\label{eq:kinetic_eq}
\end{equation}
using the method of characteristics. Here, $[\dot{\bm R},\dot{v}_{||}]$ and $\epsilon = v_{||}^2/2+\mu B$ are the gyrocenter trajectories and the energy of particle, respectively. The zeroth-order gyrocenter characteristics follow the common formulation
\begin{flalign}
\dot{\bm R}^{(0)} &= v_{||}\bm b^* + \frac{1}{q B_{||}^*} \bm b \times \mu \nabla B
\label{eq:R0},\\
\text{and}\ \ \ \dot{v}_{||}^{(0)} &= -\frac{\mu}{m} \bm b^* \cdot \nabla B.
\label{eq:v||0}
\end{flalign}
\noindent Here, the mixed-variable formulation is considered  for the  perturbation  of trajectories which are computed from perturbed electrostatic potential $\phi$, parallel electromagnetic potentials $A_{||}$ and magnetic field $\bm B$ in the linear field approximation (ignoring multiplication of perturbed fields), i.e.,
\begin{flalign}
\dot{\bm R}^{(1),\ \mathrm{lin. \ fields}}_\mathrm{plasma} &= \frac{\bm b}{B_{||}^*} \times \nabla \langle \phi-v_{||} A_{||}^{(s)}-v_{||} A_{||}^{(h)}  \rangle - \frac{q_s}{m_s} \langle A_{||}^{(h)} \rangle  \bm b^*,
\label{eq:position_update_lin}\\
 \dot{v}_{||,\ \mathrm{plasma}}^{(1),\ \mathrm{lin. \ fields}} &= - \frac{q_s}{m_s}\left[ 
\bm b^*\cdot \nabla \langle \phi-v_{||} A_{||}^{(h)}  \rangle
+ \frac{\partial \langle A_{||}^{(s)} \rangle }{\partial t}\right]
- \mu \frac{\bm b\times \nabla B}{B_{||}^*} \cdot \nabla \langle A_{||}^{(s)} \rangle, \label{eq:vpardot_update_lin} \\
\mathrm{and}\ \ \
\dot{\epsilon}^{(1),\ \mathrm{lin. \ fields}}_{\mathrm{plasma}} =&  v_{||} \dot{v}_{||}^{(1),\ \mathrm{lin. \ fields}} + \mu \dot{\bm R}^{(1),\ \mathrm{lin. \ fields}} \cdot \nabla B \nonumber \\
=&
-\frac{q_s}{m_s} \left[ m_s \mu \frac{\bm b \times \nabla B}{q_s B_{||}^*} + \frac{m_s v_{||}^2}{q_s B_{||}^*} \nabla \times \bm b \right] \cdot \nabla \langle \phi \rangle \nonumber
\\
&+\frac{q_s}{m_s} v_{||}\left[ v_{||}\bm b +  m_s \mu \frac{\bm b\times \nabla B}{q_s B_{||}^*} + \frac{m_s v_{||}^2}{q_s B_{||}^*} \nabla \times \bm b \right] \cdot \nabla \langle A_{||}^{(h)} \rangle  \nonumber
\\
&+\frac{q_s}{m_s} \mu B \left[
\nabla \cdot \bm b- \frac{m_s v_{||}}{qB_{||}^*} \frac{\nabla \times \bm B}{B^2} \cdot     \nabla B\right]\langle A_{||}^{(h)}\rangle,
\label{eq:epsdot_update_lin}
\end{flalign}
where $\mu=v_\perp^2/(2B)$ is the magnetic moment, $m_s$ is the mass of particle, $q_s$ indicates the charge of the particle,
\begin{flalign}
B_{||}^*&=\bm b \cdot \nabla \bm A^*,\\
\bm b^*&=\nabla \times \bm A^*/B_{||}^*,\\
\mathrm{and}\ \ \ \bm A^*&=\bm A+\frac{m_s v_{||}}{q_s} \bm b~.
\end{flalign}
In this formulation, $\bm A$ denotes the magnetic potential
corresponding to the equilibrium magnetic field, i.e., $\bm B = \nabla  \times \bm A$,  and $\bm b$ indicates the unit vector  $\bm b = \bm B/B$ where $B=|\bm B|$ is the magnitude of magnetic field at a given point in space. Furthermore,  $A_{||}^{(h)}$  and $A_{||}^{(s)}$ denote the Hamiltonian and the symplectic parts of the perturbed magnetic potential, respectively, following the decomposition $A_{||}=A_{||}^{(s)}+A_{||}^{(h)}$. Moreover, the notation $\langle (.) \rangle$ indicates the quantity of interest is gyro-averaged, i.e., $\langle (.) \rangle=\int (.)(\bm R+\bm \rho) d\alpha /(2\pi) $ where $\bm \rho$ is  the gyroradius of the particle and $\alpha$ the gyro-phase.
\\ \ \\
\noindent In the fully nonlinear setting, the dominant nonlinear contribution originates from
\begin{flalign}
\bm B^* &= \bm B + \frac{mv_{||}}{q} \nabla \times \bm b + \nabla \langle  A_{||}^{(s)} \rangle  \times \bm b\\
\text{where}\ \ \ \bm b^* &= \frac{\bm B^*}{B_{||}^*} = \bm b_0^* + \frac{\nabla \langle A_{||}^{(s)} \rangle \times \bm b  }{B_{||}^*}\\
\bm b_0^* &\approx \bm b + \frac{mv_{||}}{q B_{||}^*} \nabla \times \bm b~.
\end{flalign}
Hence, the outcome equations of motion in the mixed-variable formulation become nonlinear in terms of potentials, i.e.,
\begin{flalign}
\dot{\bm R}^{(1),\ \mathrm{nonlin.\ fields}} &= \frac{\bm b}{B_{||}^*} \times \nabla \langle \phi-v_{||} A_{||}^{(s)}-v_{||} A_{||}^{(h)}  \rangle - \frac{q_s}{m_s} \langle A_{||}^{(h)} \rangle \left(  \bm b^*_0 + \frac{\nabla \langle A_{||}^{(s)} \rangle \times \bm b }{B_{||}^*} \right),
\label{eq:position_update_nonlin}\\
 \dot{v}_{||}^{(1),\ \mathrm{nonlin.\ fields}} =& - \frac{q_s}{m_s}\left[ 
\left(  \bm b^*_0 + \frac{\nabla \langle A_{||}^{(s)} \rangle \times \bm b }{B_{||}^*} \right) \cdot \nabla \langle \phi-v_{||} A_{||}^{(h)}  \rangle
+ \frac{\partial \langle A_{||}^{(s)} \rangle }{\partial t}\right] \nonumber \\
&- \mu \frac{\bm b\times \nabla B}{B_{||}^*} \cdot \nabla \langle A_{||}^{(s)} \rangle, \label{eq:vpardot_update_nonlin} \\
\mathrm{and}\ \ \
\dot{\epsilon}^{(1),\ \mathrm{nonlin.\ fields}} =&  v_{||} \dot{v}_{||}^{(1),\ \mathrm{nonlin.\ fields}} + \mu \dot{\bm R}^{(1),\ \mathrm{nonlin.\ fields}} \cdot \nabla B \nonumber \\
=&
-\frac{q_s}{m_s} \left[ m_s \mu \frac{\bm b \times \nabla B}{q_s B_{||}^*} + \frac{m_s v_{||}^2}{q_s B_{||}^*} \nabla \times \bm b + v_{||} \frac{\nabla \langle A_{||}^{(s)} \rangle \times \bm b}{B_{||}^*} \right] \cdot \nabla \langle \phi \rangle \nonumber
\\
&+\frac{q_s}{m_s} v_{||}\left[ v_{||}\bm b +  m_s \mu \frac{\bm b\times \nabla B}{q_s B_{||}^*} + \frac{m_s v_{||}^2}{q_s B_{||}^*} \nabla \times \bm b  +v_{||} \frac{\nabla \langle A_{||}^{(s)} \rangle \times \bm b}{B_{||}^*} \right] \cdot \nabla \langle A_{||}^{(h)} \rangle  \nonumber
\\
&+\frac{q_s}{m_s} \mu B \left[
\nabla \cdot \bm b- \frac{m_s v_{||}}{qB_{||}^*} \frac{\nabla \times \bm B}{B^2} \cdot     \nabla B  
-\frac{\nabla \langle A_{||}^{(s)} \rangle }{B_{||}^*} \cdot \frac{\bm b \times \nabla B}{B} \right]\langle A_{||}^{(h)}\rangle.
\label{eq:epsdot_update_nonlin}
\end{flalign}
\\
\noindent In order to have a self-contained system of equations, we need to introduce closures for the field quantities. The perturbed electrostatic potential $\phi$ is computed from the quasineutrality equation, here with the polarization density expressed in the long wavelength approximation 
\begin{equation}
    - \nabla \cdot \left[ \left(\sum_{s=i,f} \frac{q_s^2 n_s}{T_s} \rho_s^2\right) \nabla_\perp \phi  \right] = \sum_{s=i,e,f} q_s n_{1,s},
    \label{eq:quas_neut}
\end{equation}
where $n_{1,s} = \int  \delta f_s  \delta (\bm R + \bm \rho - \bm x) d^6 Z$ indicates the perturbed gyrocenter density, $\rho_s=\sqrt{m_s T}/(q_s B)$ the thermal gyroradius, and $d^6Z=B_{||}^* d\bm R\  dv_{||}\  d\mu\  d\alpha$ the phase-space infinitesimal volume. Once the electrostatic potential $\phi$ is computed, the symplectic part of parallel electromagnetic potential is obtained by solving Ohm's law
\begin{equation}
\frac{\partial }{\partial t} A_{||}^{(s)} + \bm b \cdot \nabla \phi =0,
\label{eq:Ohm_law}
\end{equation}
and the Hamiltonian part of electromagnetic potential $A_{||}^{(h)}$ from the mixed-variable parallel Ampere's law 
\begin{flalign}
    &\left( \sum_{s=i,e,f} \frac{\beta_s}{\rho_s^2}-\nabla_\perp^2 \right) A_{||}^{(h)} = \mu_0 \sum_{s=i,e,f} j_{||,1s} + \nabla_\perp^2 A_{||}^{(s)}
    \label{eq:Amp_law},\\
    \mathrm{and}\ \ \ &j_{||,1s} = \int  v_{||} \delta f_s \delta(\bm R+\bm \rho -\bm x) d^6 Z
\end{flalign}
where $j_{||,1s}$ is the perturbed parallel gyrocenter current. 
The arbitrary splitting of the magnetic potential into parts and fixing the Hamiltonian part at the end of each time step allows containing the contribution of the skin term  $\beta_e A_{||}^{(h)}/\rho_e^2$ which is associated with the cancellation problem, see \cite{mishchenko2014pullback} for details.

\subsection{Solution algorithm}
\label{sec:pullback}
The high dimensionality of mixed distribution function is dealt with using particle method. We discretize $\delta f$ in the phase space with markers
\begin{flalign}
\delta f_s^{(m)} (\bm R, v_{||}, \mu; t) = \sum_{i=1}^{N_p} \frac{1}{2\pi B_{\parallel,i}^\ast} w_{s,i}(t) \delta(\bm R- \bm R_i) \delta (v_{||}-v_{||,i}) \delta (\mu - \mu_i)
\end{flalign}
where $\delta(.)$ indicates the Dirac delta function, $N_p$ indicates the number of markers, and $w_{s,i}$ is the weight associated with $i$th marker of species $s$. Having discretized the solution domain in the configuration space $\Omega \subset \mathbb{R}^3$ with an appropriately refined mesh, the densities in each computational cell can be estimated from the markers. Given the density field from particles,  other required field quantities can be evaluated using standard Finite Element method equipped with B-splines as the basis function to solve quasi-neutrality equation, Ohm's law and Ampere's law, respectively. In particular
\begin{flalign}
\phi(s,\theta, \varphi, t) &\approx \sum_{ljk} \hat{\phi}_{ljk} (t) \Lambda_{ljk}(s,\theta, \varphi)  ,\\
A_{||}^{(s)}(s,\theta, \varphi, t) &\approx \sum_{ljk}
\hat{A}_{||,ljk}^{(s)}(t) \Lambda_{ljk}(s,\theta, \varphi),\\
\mathrm{and}\ \ \ A_{||}^{(h)}(s,\theta, \varphi, t) &\approx \sum_{ljk}
\hat{A}_{||,ljk}^{(h)}(t) \Lambda_{ljk}(s,\theta, \varphi),
\end{flalign}
where $\bm \Lambda$ is the tensor product of usual B-splines $\lambda$ in each direction, i.e., $\Lambda_{ljk}(s,\theta, \varphi)=\lambda_{l}(s)\lambda_{k}(\theta)\lambda_{k}(\varphi)$, and $\hat{(.)}$ indicates the projected coefficient. For detailed numerics of field computations, the reader is refered to \cite{lanti2020orb5}.
\\ \ \\
\noindent Once electrostatic and electromagnetic potentials are computed, the particle representation of distribution function is evolved by solving the zeroth-order evolution equations for the linear simulations, i.e., Eqs.~\eqref{eq:R0} and \eqref{eq:v||0}. In the nonlinear simulations, besides zeroth-order equations equations of motion, the first-order evolution equations of motion   Eqs.~\eqref{eq:position_update_lin}-\eqref{eq:epsdot_update_lin} or Eqs.~\eqref{eq:position_update_nonlin}-\eqref{eq:epsdot_update_nonlin} are solved for linear field approximation and nonlinear fields, respectively. Note that in this study, the nonlinear simulations are performed including the nonlinear field contributions. The outcome system of ordinary differential equations is solved using the standard fourth-order Runge–Kutta method. 
\\ \ \\
The pullback scheme enters the solution algorithm  by updating  symplectic and Hamiltonian parts of $A_{||}$ as well as updating the symplectic part of distribution function at the end of each iteration. This separation combined with the control variate nature of $\delta f$ method allows for long and stable simulations. Short explanations of the pullback scheme for both linear and nonlinear settings are provided in Algorithms~\ref{alg:pullback_lin}-\ref{alg:pullback_nonlin}, respectively \cite{MISHCHENKO2019194}.
\\ \ \\
\begin{algorithm}[H]
\SetAlgoLined
 Initialize markers in phase space\;
 \While{$t<t_\mathrm{final}$}{
  \For{$k=1,...,4$ \text{step of Runge-Kutta scheme}}{
  - Compute $\phi$, $A_{||}^{(s)}$ and $A_{||}^{(h)}$ by solving Eqs.~\eqref{eq:quas_neut}-\eqref{eq:Amp_law}\;
  - Push particles according to zeroth-order equations of motion Eqs.~\eqref{eq:R0}- \eqref{eq:v||0}\;
  - Apply boundary conditions\;
   }
   - Update mixed-variable $\delta f$ with  $\delta f_s^{(m)} \leftarrow \delta f_s^{(m)}+\dfrac{q_s\langle A_{||}^{(h)}\rangle}{m_s} \dfrac{\partial F_{0s}}{\partial v_{||}}$\;
   - Update $A_{||}$ decomposition, i.e,  $A_{||}^{(s)}\leftarrow A_{||}^{(s)}+A_{||}^{(h)}$ and
   $A_{||}^{(h)}\leftarrow0$\;
  - $t=t+\Delta t$\;
 }
 \caption{The $\delta f$ solution algorithm used in ORB5 within the linear pullback scheme.}
 \label{alg:pullback_lin}
\end{algorithm}
\ \\ \ \\
\begin{algorithm}[H]
\SetAlgoLined
 Initialize markers in the phase space\;
 \While{$t<t_\mathrm{final}$}{
  \For{$k=1,...,4$ \text{step of Runge-Kutta scheme}}{
  - Compute $\phi$, $A_{||}^{(s)}$ and $A_{||}^{(h)}$ by solving Eqs.~\eqref{eq:quas_neut}-\eqref{eq:Amp_law}\;
  - Push particles according to the zeroth and first-order equations of motion Eqs.~\eqref{eq:R0}-\eqref{eq:v||0} and Eqs.~\eqref{eq:position_update_nonlin}-\eqref{eq:epsdot_update_nonlin}\;
  - Apply boundary conditions\;
   }
   - Transform the phase-space coordinates keeping the particle weights, i.e., $v_{||}^{(m),\ \mathrm{new}} \leftarrow v_{||}^{(m),\ \mathrm{old}} - \dfrac{q_s}{m_s} \langle A_{||}^{(h)} \rangle$ and $f_{1s}^{(m)}(v_{||}^{(m),\mathrm{new}})\leftarrow f_{1s}^{(m)}(v_{||}^{(m),\mathrm{old}})$ \;
   - Update $A_{||}$ decomposition, i.e,  $A_{||}^{(s)}\leftarrow  A_{||}^{(s)}+A_{||}^{(h)}$ and
   $A_{||}^{(h)}\leftarrow0$\;
  - $t=t+\Delta t$\;
 }
 \caption{The $\delta f$ solution algorithm used in ORB5 within the nonlinear pullback scheme.}
 \label{alg:pullback_nonlin}
\end{algorithm}
\ \\ 
\section{Antenna}
\label{sec:antenna}
The idea is to excite the eigenmode of interest using an external targeted force/drift in the equations of motion \cite{brunner1998global}. This can be achieved by describing the external force as gradient of a potential which resembles the target mode structure and frequency. Having described such an external potential, one can extend the plasma's equations of motion including the antenna field by substitution $\phi$ with $\phi_\mathrm{ant}+\phi_\mathrm{plasma}$ as well as $A_{||}$ with $A_{||,\mathrm{ant}}+A_{||,\mathrm{plasma}}$.

\subsection{Devising an electrostatic potential for antenna}
\label{sec:electrostatic_antenna}
Consider antenna as an electrostatic potential described by
\begin{flalign}
\phi_\mathrm{ant}(s,\theta, \varphi; t) =&  \mathrm{Re}\left[ \sum_{I\in \mathcal{T}} h_{I}(s)  A_{I} e^{\hat{i}( I_1 \theta + I_2 \varphi +\Phi_{I}  )}(c_1 + c_2 e^{\hat{i}w_\mathrm{ant}t} ) \right]
\label{eq:ant_pot_modes}\\
=&(c_1 + c_2 \cos(\omega_\mathrm{ant}t)) \underbrace{\left( 
 \sum_{I\in \mathcal{T}}  h_{I}(s) A_{I} \cos{( I_1 \theta + I_2 \varphi +\Phi_{I} )}
\right)}_{S_{\mathrm{ant},1}(s,\theta, \varphi)} \nonumber
\\
&+c_2 \sin(\omega_\mathrm{ant}t) \underbrace{\left( 
 \sum_{I\in \mathcal{T}}  h_{I}(s) A_{I} \cos{( I_1 \theta + I_2 \varphi +\Phi_{I}  +\frac{\pi}{2} )}
\right)}_{S_{\mathrm{ant},2}(s,\theta, \varphi)},
\label{eq:ant_pot_modes_decomposed}
\end{flalign}
where the set $\mathcal{T}=\{ (m_1,n_1),... \}$ contains all the targeted mode number pairs, $A_{I}$ denotes the mode coefficient for $I$th mode, $\Phi_{I}$ is its  phase offsets, $c_1,c_2\in \mathbb{R}$ are the coefficients of the static and oscillating components, respectively, $\hat{i}$ is the imaginary number, and $\omega_\mathrm{ant}$ is the antenna's frequency. The radial profile $h(s)$ is chosen as an input. Since field properties in ORB5 are stored in the B-spline basis function, we need to project $\phi_\mathrm{ant}$ to $\bm \Lambda$. The presented decomposition of potential into time dependent and independent parts in  Eq.~\eqref{eq:ant_pot_modes_decomposed} saves computational time since the projections of the time independent parts of $\phi_\mathrm{ant}$, i.e., $S_1$ and $S_2$,  can be computed at the beginning of simulation.
\\ \ \\
Following ORB5's approach of evaluation of field properties at the particle positions, the antenna's potential described in Eq.~\eqref{eq:ant_pot_modes} needs to be projected into B-Spline basis function
\begin{flalign}
S_\mathrm{ant} = \sum_{ijk}  \hat{S}_\mathrm{ant,ijk} \Lambda_{ijk}(s,\theta, \varphi)
\end{flalign}
using the weak form
\begin{flalign}
\int S_\mathrm{ant} \psi dV = \int \hat{S}_\mathrm{ant} \psi dV\ \ \ \ \forall \psi;
\end{flalign}
where B-spline basis functions are taken for the test function $\psi$. The projected coefficients are obtained by solving the linear system of equations
\begin{flalign}
&\sum_{i^\prime j^\prime k^\prime} M_{ijk\ i^\prime j^\prime k^\prime} \hat{S}_{\mathrm{ant},i^\prime j^\prime k^\prime} dV = \underbrace{\int S_{\mathrm{ant}} \Lambda_{ijk} dV}_{b_{\mathrm{ant},imn}},\\
\mathrm{where}\ \ \ 
&M_{ijk\  i^\prime j^\prime k^\prime}  = \int J(s,\theta)  \Lambda_{ijk} (s,\theta, \varphi) \Lambda_{i^\prime j^\prime k^\prime } (s,\theta, \varphi) ds d\theta d \varphi,
\end{flalign}
and $dV=d^3 \bm x$ is the infinitesimal volume in configuration space. This system of equations is solved in discrete Fourier space 
\begin{flalign}
\sum_{i^\prime=1}^{N_s+p} \sum_{n^\prime=n_\mathrm{min}}^{n_\mathrm{max}} \sum_{m^\prime=-n^\prime q_{i^\prime}-\Delta m}^{-n^\prime q_{i^\prime}+\Delta m} 
\tilde{M}_{ijk\  i^\prime j^\prime k^\prime} \tilde{\hat{S}}_{\mathrm{ant},i^\prime m^\prime n^\prime} = \tilde{b}_{\mathrm{ant},imn} 
\end{flalign}
where the integrals are computed numerically using Gaussian quadrature rule \cite{ohana2020using}. 

\subsection{Electromagnetic antenna}
Motivated by the pullback scheme and mitigation cancellation, instead of considering an arbitrary electromagnetic potential for antenna, we impose $A_{\mathrm{ant},||}=A_{\mathrm{ant},||}^{(s)}+A_{\mathrm{ant},||}^{(h)}$ where the symplectic part follows Ohm's law
\begin{flalign}
\frac{\partial}{\partial t} A_{\mathrm{ant},||}^{(s)} + \bm b \cdot \nabla \phi_\mathrm{ant} = 0,
\label{eq:Ohm_ant}
\end{flalign}
and the Hamiltonian part $A_{\mathrm{ant},||}^{(h)}=0$. Here, the $\phi_\mathrm{ant}$ is an arbitrary electrostatic potential described according to the target mode of interest, as described in section~\ref{sec:electrostatic_antenna}.

\subsection{Integrating electrostatic antenna in the equations of motion}
\label{sec:antenna_equations_of_motion}
The equations of motion for an electrostatic antenna described can be naturally developed by substituting $\phi+\phi_\mathrm{ant}$ instead of $\phi$ in the plasma's equation of motion Eqs.~\eqref{eq:position_update_lin}-\eqref{eq:epsdot_update_lin} leading to

\begin{flalign}
\dot{\bm R}^{(1),\ \mathrm{lin. \ fields}} &=\dot{\bm R}^{(1),\ \mathrm{lin. \ fields}}_\mathrm{plasma}
  {+\frac{\bm b}{B_{||}^*} \times \nabla \langle \phi_\text{ant} \rangle} \label{eq:R1dot_lin_pullback_phi_ant},
 \\
 \dot{v}_{||}^{(1),\ \mathrm{lin. \ fields}} &= \dot{v}_{||,\mathrm{plasma}}^{(1),\ \mathrm{lin. \ fields}} 
{- \frac{q_s}{m_s} 
\bm b^*\cdot \nabla \langle \phi_\text{ant} \rangle },
\label{eq:vpar1dot_lin_pullback_phi_ant}
\\
\dot{\epsilon}^{(1),\ \mathrm{lin. \ fields}} &=  \dot{\epsilon}^{(1),\ \mathrm{lin. \ fields}}_\mathrm{plasma}
{-\frac{q_s}{m_s} \left[\frac{v_{||} \bm B}{B_{||}^*}+ m_s \mu \frac{\bm b \times \nabla B}{q_s B_{||}^*} + \frac{m_s v_{||}^2}{q_s B_{||}^*} \nabla \times \bm b \right] \cdot \nabla \langle \phi_\text{ant} \rangle}~.
\label{eq:eps1dot_lin_pullback_phi_ant}
\end{flalign}
for the pullback scheme linearized with respect to the field perturbations and 
\begin{flalign}
\dot{\bm R}^{(1),\ \mathrm{nonlin.\ fields}} =&\dot{\bm R}^{(1),\  \mathrm{nonlin.\ fields}}_\mathrm{plasma}
  {+\frac{\bm b}{B_{||}^*} \times \nabla \langle \phi_\text{ant} \rangle},
 \\
 \dot{v}_{||}^{(1),\ \mathrm{nonlin.\ fields}} =& \dot{v}_{||,\ \mathrm{plasma}}^{(1),\  \mathrm{nonlin.\ fields}} 
{- \frac{q_s}{m_s} 
\bm b^*\cdot \nabla \langle \phi_\text{ant} \rangle },\\
\dot{\epsilon}^{(1),\ \mathrm{nonlin.\ fields}} =&  \dot{\epsilon}^{(1),\ \mathrm{nonlin.\ fields}}_\mathrm{plasma}
\nonumber \\
&-\frac{q_s}{m_s} \left[\frac{v_{||} \bm B}{B_{||}^*}+ m_s \mu \frac{\bm b \times \nabla B}{q_s B_{||}^*} + \frac{m_s v_{||}^2}{q_s B_{||}^*} \nabla \times \bm b + v_{||} \frac{\nabla \langle A_{||}^{(s)}  \rangle \times \bm b  }{B_{||}^*} \right] \cdot \nabla \langle \phi_\text{ant} \rangle~.
\end{flalign}
for the nonlinear pullback scheme. Here, the subscript $(.)_\mathrm{plasma}$ indicates the plasma contribution to the equations of motion and energy, i.e., Eqs.~\eqref{eq:position_update_lin}-\eqref{eq:epsdot_update_lin} for linear field approximation and Eqs.~\eqref{eq:position_update_nonlin}-\eqref{eq:epsdot_update_nonlin} for nonlinear formulation. 
\\ \ \\
It can be of interest to perform numerical simulations in the presence of the antenna where the plasma response is linearized. For consistency, one needs to use the linearized pullback scheme,  Eqs~\eqref{eq:R1dot_lin_pullback_phi_ant}-\eqref{eq:eps1dot_lin_pullback_phi_ant}, where the terms corresponding to the plasma response in the equations of motion, i.e. $\dot{\bm R}^{(1)\ \mathrm{lin. fields}}_{\mathrm{plasma}}$ and $\dot{v}^{(1)\ \mathrm{lin. fields}}_{\mathrm{plasma}}$ in the left hand side of kinetic equation \eqref{eq:kinetic_eq}, are set to zero.
\\ \ \\
\noindent Note that appearance of $v_{||} \bm B \cdot \nabla \langle \phi_\mathrm{ant} \rangle $ causes a numerical challenge as it pushes particles further in the parallel direction. Motivated by the treatment of such terms in the pullback scheme, next we consider equations of motion for the electromagnetic antenna.

\subsection{Integrating electromagnetic antenna in the equations of motion}
In order to obtain long and stable simulations of TAE excitation with the antenna, we use Ohm's law to introduce the electromagnetic antenna in the equations of motion and mimic the treatment of the cancellation problem in the mixed-variable formulation, see section~\ref{sec:pullback}. We derive equations of motion corresponding to antenna's field by natural extension of plasma's equations of motion where we  substitute $\phi+\phi_\mathrm{ant}$ instead of plasma's electrostatic potential $\phi$ and $A_{||}^{(s)}+A_{||,\mathrm{ant}}^{(s)}$ instead of $A_{||}^{(s)}$. Hence, one obtains for the linear field approximation

\begin{flalign}
\dot{\bm R}^{(1),\ \mathrm{lin. \ fields}} =& \dot{\bm R}^{(1),\ \mathrm{lin. \ fields}}_\mathrm{plasma}
{+\frac{\bm b}{B_{||}^*} \times \nabla \langle \phi_\text{ant} \rangle}
 {-\frac{\bm b}{B_{||}^*} \times \nabla \langle v_{||} A_{||,\text{ant}}^{(s)} \rangle},
 \label{eq:R1dot_lin_pullback_electromagnetic_antenna}
 \\
 \dot{v}_{||}^{(1),\ \mathrm{lin. \ fields}} =&  \dot{v}_{||,\mathrm{plasma}}^{(1),\ \mathrm{lin. \ fields}}
{- \frac{q_s}{m_s} 
(\bm b^*-\bm b)\cdot \nabla \langle \phi_\text{ant} \rangle  
}
- \frac{\mu}{B_{||}^*} \bm b\times \nabla B \cdot \nabla \langle A_{||,\text{ant}}^{(s)} \rangle,
\label{eq:vapr1dot_lin_pullback_electromagnetic_antenna}
\\
\mathrm{and}\ \ \ \dot{\epsilon}^{(1),\ \mathrm{lin. \ fields}} =& \dot{\epsilon}^{(1),\ \mathrm{lin. \ fields}}_\mathrm{plasma}
 -\frac{q_s}{m_s} \left[ m_s \mu \frac{\bm b \times \nabla B}{q_s B_{||}^*} + \frac{m_s v_{||}^2}{q_s B_{||}^*} \nabla \times \bm b \right] \cdot \nabla \langle \phi_\text{ant} \rangle. 
 \label{eq:eps1dot_lin_pullback_electromagnetic_antenna}
\end{flalign}
Here, Ohm's law on the antenna Eq.~\eqref{eq:Ohm_ant} was deployed to cancel out destabilizing terms, i.e. cancellation of $\partial A_{||,\mathrm{ant}}^{(s)}/\partial t$ terms. More precisely, we found that this procedure allowed us to preserve the benefits provided by the pullback scheme, resulting in the possibility to perform numerically stable simulations with large time steps. In the case of nonlinear mixed variable formulation, one obtains
\begin{flalign}
\dot{\bm R}^{(1),\ \mathrm{nonlin.\ fields}} =&\dot{\bm R}^{(1),\  \mathrm{nonlin.\ fields}}_\mathrm{plasma}
  {+\frac{\bm b}{B_{||}^*} \times \nabla \langle \phi_\text{ant} - v_{||} A_{||,\mathrm{ant}}^{(s)}  \rangle}  - \frac{q}{m} \langle A_{||}^{(h)}  \rangle \frac{ \nabla \langle A_{||,\mathrm{ant}}^{(s)} \rangle \times \bm b }{B_{||}^*}   ,
  \label{eq:Rdot_antenna_nonlin}
 \\
 \dot{v}_{||}^{(1),\ \mathrm{nonlin.\ fields}} =& \dot{v}_{||,\ \mathrm{plasma}}^{(1),\  \mathrm{nonlin.\ fields}} 
{- \left( \frac{v_{||}}{B_{||}^*} \nabla \times \bm b+\frac{q}{m} \frac{\nabla \langle A_{||}^{(s)} \rangle \times \bm b }{B_{||}^*} \right)\cdot \nabla \langle \phi_\text{ant} \rangle }\nonumber \\
&{-\frac{q}{m} \frac{\nabla \langle A_{||,\mathrm{ant}}^{(s)} \rangle \times \bm b }{B_{||}^*} \cdot \nabla \langle \phi - v_{||} A_{||}^{(h)} \rangle }
{-\frac{q}{m} \frac{\nabla \langle A_{||,\mathrm{ant}}^{(s)} \rangle \times \bm b }{B_{||}^*} \cdot \nabla \langle \phi_\mathrm{ant} \rangle } \nonumber \\
&- \mu \frac{\bm b \times \nabla B}{B_{||}^*} \cdot \nabla \langle A_{||,\mathrm{ant}}^{(s)} \rangle 
,
\label{eq:vdot_antenna_nonlin}
\\
\text{and}\ \dot{\epsilon}^{(1),\ \mathrm{nonlin.\ fields}} =&  \dot{\epsilon}^{(1),\ \mathrm{nonlin.\ fields}}_\mathrm{plasma}
\nonumber \\
&-\frac{q_s}{m_s} \left[\frac{v_{||} \bm B}{B_{||}^*}+ m_s \mu \frac{\bm b \times \nabla B}{q_s B_{||}^*} + \frac{m_s v_{||}^2}{q_s B_{||}^*} \nabla \times \bm b + v_{||} \frac{\nabla \langle A_{||}^{(s)}  \rangle \times \bm b  }{B_{||}^*} \right] \cdot \nabla \langle \phi_\text{ant} \rangle \nonumber \\
&-\frac{q_s}{m_s} \left[ v_{||} \frac{\nabla \langle A_{||,\mathrm{ant}}^{(s)}  \rangle \times \bm b  }{B_{||}^*} \right] \cdot \nabla \langle \phi \rangle
-\frac{q_s}{m_s} \left[ v_{||} \frac{\nabla \langle A_{||,\mathrm{ant}}^{(s)}  \rangle \times \bm b  }{B_{||}^*} \right] \cdot \nabla \langle \phi_\mathrm{ant} \rangle\nonumber \\
&+\frac{q_s}{m_s} v_{||} \left[ v_{||} \frac{\nabla \langle A_{||,\mathrm{ant}}^{(s)}  \rangle \times \bm b}{B_{||}^*} \right]\cdot \nabla \langle A_{||}^{(h)} \rangle
- \frac{q_s}{m_s} \mu B \left[ \frac{\nabla \langle A_{||,\mathrm{ant}}^{(s)} \rangle}{B_{||}^*} \cdot \frac{\bm b \times \nabla B}{B} \right] \langle A_{||}^{(h)} \rangle
~.
\label{eq:epsilon_antenna_nonlin}
\end{flalign}
Often, one might be interested in exciting a mode using an electromagnetic antenna with linearized plasma response. In that case, it is necessary to use the linearized pullback scheme,  Eqs~\eqref{eq:R1dot_lin_pullback_electromagnetic_antenna}-\eqref{eq:eps1dot_lin_pullback_electromagnetic_antenna}, in which the plasma response terms in the equations of motion, i.e. $\dot{\bm R}^{(1)\ \mathrm{lin. fields}}_{\mathrm{plasma}}$ and $\dot{v}^{(1)\ \mathrm{lin. fields}}_{\mathrm{plasma}}$ in the left hand side of kinetic equation \eqref{eq:kinetic_eq}, are ignored.
\section{Results}
\label{sec:results}
In this section, we deploy the described antenna in ORB5 in order to excite TAE modes in the linear setting, i.e. section~\ref{sec:lin_sims}, where we measure the frequency and the damping rate, i.e. in  section~\ref{sec:damp_TAE}, as well as study the margins of instability in the presence of fast particles, i.e. in section~\ref{sec:ant_fast}. Then, nonlinear simulation results of  exciting various TAE and GAE modes using the antenna are presented  in section~\ref{sec:nonlin}. 
\subsection{Simulation setting}
\begin{figure}
  	\centering
   \includegraphics[scale=0.8]{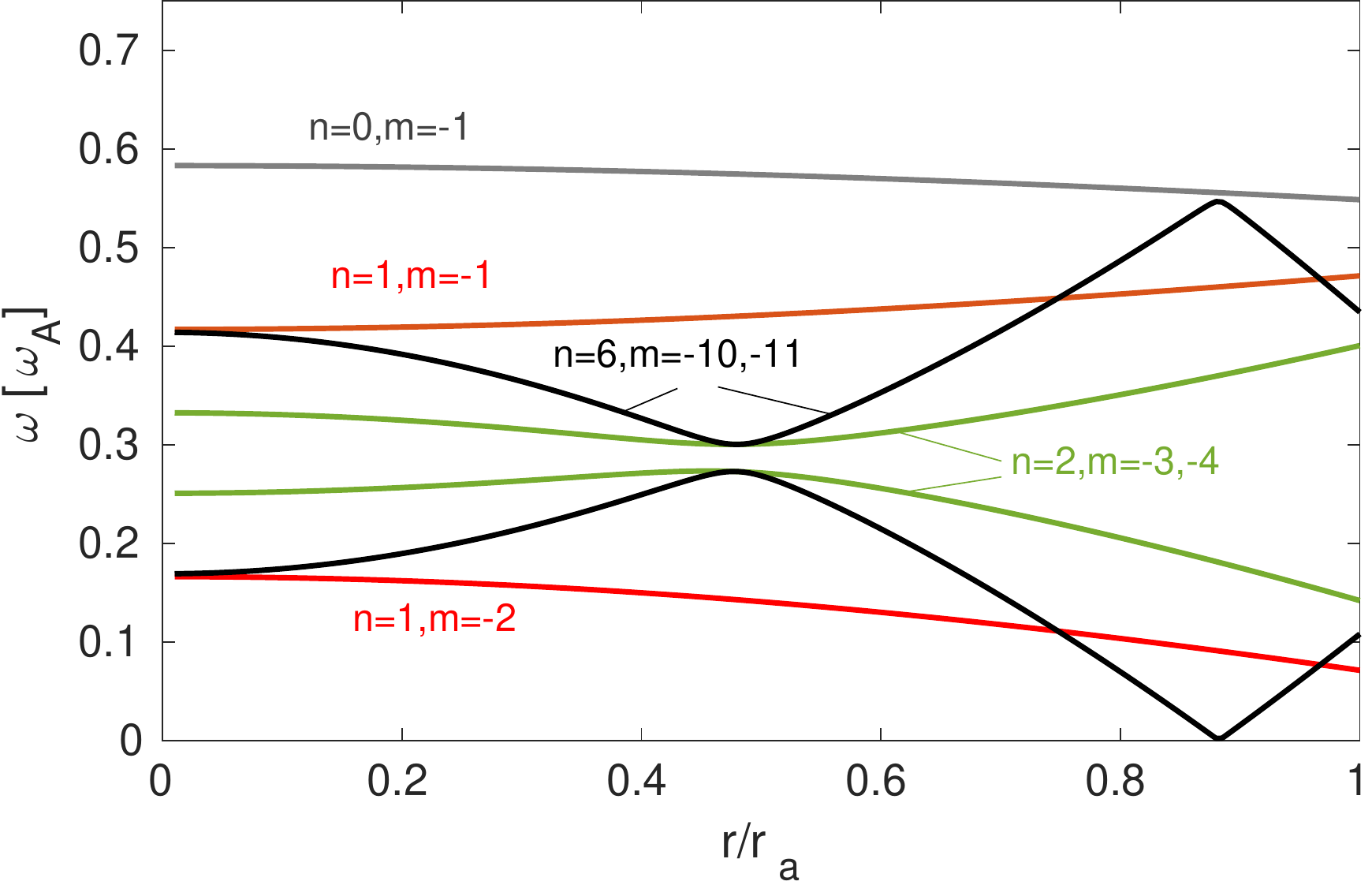}
  \caption{The Alfv{\'e}n continua of ITPA test case for $n=0,1,2$ and $6$ toroidal mode numbers.}
  \label{fig:alfven_continua}
\end{figure}

\begin{figure}
  	\centering
  	\begin{tabular}{cc}
  \includegraphics[scale=1.]{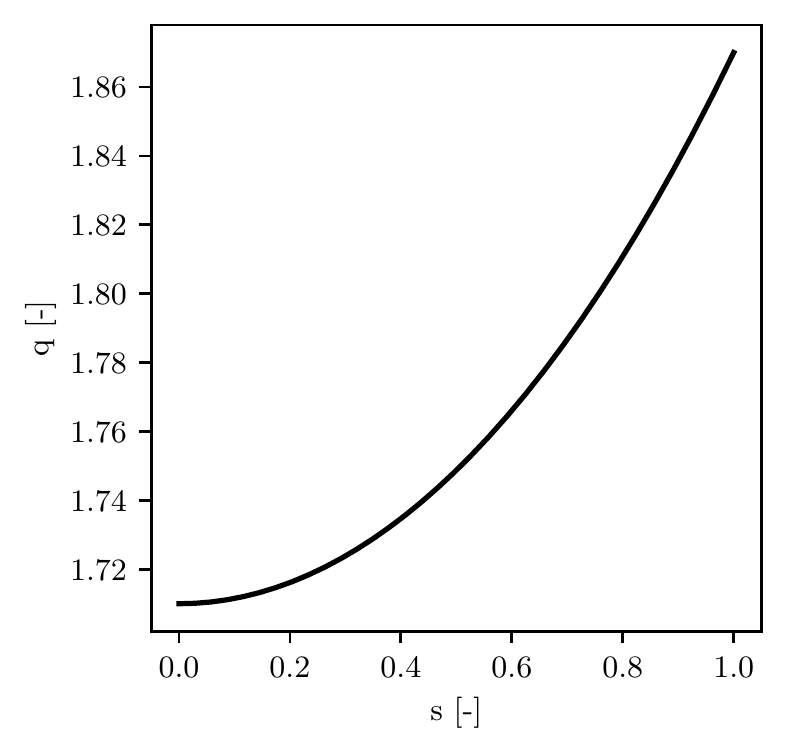} &
   \includegraphics[scale=1.]{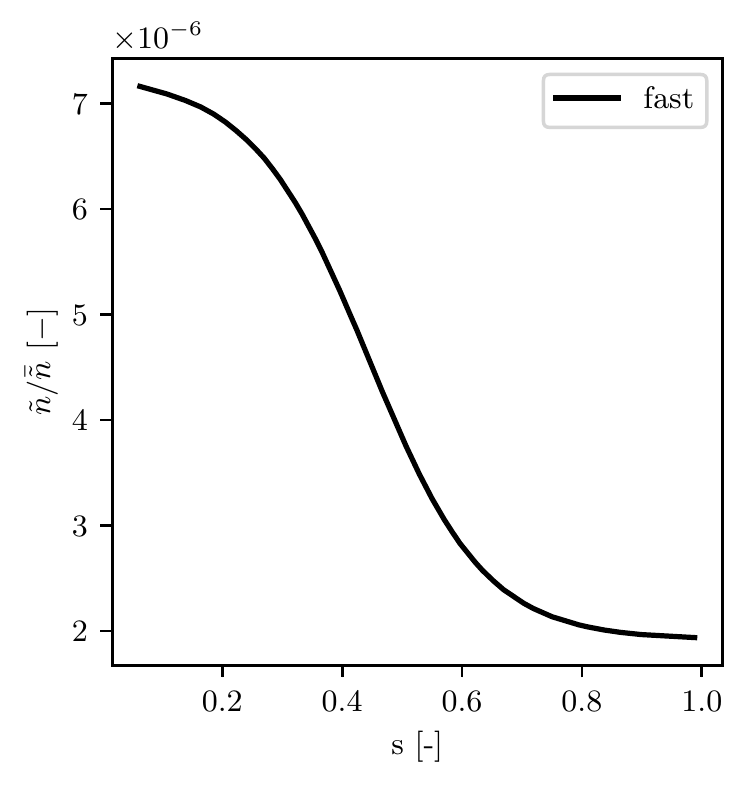}
   \\
   (a) & (b)
   \end{tabular}
  \caption{The radial profile of (a) safety factor $q$ and (b) the normalized number density of fast particles in the ITPA test case.}
  \label{fig:qprofile_nfastprofile}
\end{figure}

\noindent Here, we consider the international cross-code reference test case ``ITPA-TAE" \cite{konies2018benchmark}, where the Toroidal Alfv{\'e}n Eigenmode with the toroidal mode number $n =  6$ and the dominant poloidal mode numbers $m = -10$ and $m = -11$ is considered in the linear regime. Further investigations indicated that there is another TAE gap mode in this test case with $n = 2$ and $m=-4,-3$, see Fig.~\ref{fig:alfven_continua} for the Alfv{\'e}n continua of ITPA test case.
These modes are studied in tokamak-size geometry with the minor radius $r_a = 1 \ \mathrm{m}$,  major radius $R_0 = 10 \ \mathrm{m}$,  magnetic field on  axis $B_0 = 3\  \mathrm{T}$, and the safety factor profile $q(r) = 1.71 + 0.16(r/r_a)^2$, where $r$ is the geometric radial coordinate.  We consider flat background plasma profiles with the ion number density $n_\mathrm{i}=2 \times 10^{19}\  \mathrm{m}^{-3}$, a mass ratio of $m_\mathrm{i}/m_\mathrm{e}=200$, ion and electron temperatures of $T_\mathrm{i} = T_\mathrm{e} = 1\ \mathrm{keV}$ which correspond to $\beta_\mathrm{bulk} = 2\mu_0(n_\mathrm{i} T_\mathrm{i} + n_\mathrm{e} T_\mathrm{e})/B_0^2 \approx 0.18 \%$. Here, $\mu_0$ is the permeability of the vacuum and $n_s$ is the number density of species $s$ which is related to the mass density $\rho_s$ via $\rho_s = n_s m_s$. In some cases we deploy fast particles with Maxwellian distribution in velocity space and flat radial profile of temperature $T_\mathrm{f}=400\ \mathrm{keV}$ and number density with the profile
\begin{flalign}
\tilde{n} = n_{0\mathrm f} \exp\left(-w \kappa_{n} \tanh( \frac{s-a}{w})\right)
\ \ \ \text{and} \ \ \ 
n_f = \tilde{n}/\bar{\tilde{n}}
\end{flalign}
where $\bar{(.)}$ indicates the annular averaged value, and the values for the parameters $w=0.2,\ \kappa_{n}=3.333,\ a=0.5,\ \ \mathrm{and}\ n_{0\mathrm f}=0.0031$ are taken from the reference \cite{konies2018benchmark}. See Fig.~\ref{fig:qprofile_nfastprofile} for the profiles of initial density of fast particles and safety factor in the radial direction $s=\sqrt{\psi/\psi_\mathrm{edge}}$. Here, $\psi$ is the poloidal magnetic flux and $\psi_\mathrm{edge}$ denotes its value at the radial edge.
\\ \ \\
\noindent Following a convergence study for the linear simulations, the simulation results reported here are obtained using $N_\mathrm{e}=2\times 10^7$ markers for electrons,  $N_\mathrm{i}=10^7$ markers for ions, a grid of size $N_s\times N_\theta \times N_\phi=320 \times 256 \times 128$, and time step size of $\Delta t =50\  \omega_{ci}^{-1}$. 
ORB5 uses the inverse of the ion-cyclotron frequency $\omega_{ci}^{-1}=q_i B_0/(m_ic)$ as the unit for time where $c$ is the speed of light. However, here we show the simulation results in the inverse of Alfv{\'e}n frequency $\omega_A^{-1}$ as the relevant unit for time where $\omega_A={v_{A0}}/{R_0}$ with Alfv{\'e}n velocity on the axis $v_{A0}=B_0/\sqrt{\mu_0 \rho_0}$, background plasma mass density $\rho_0$ and magnetic field strength $B_0$ evaluated on the axis $s=0$. Note the ratio $\omega_{ci}/\omega_A \approx 196.5$ for the ITPA test case.
In case of having fast  particles, $N_\mathrm{f}=2\times 10^7$ markers are considered. Here, we have used $10$ times more particles in the nonlinear simulations compared to the linear ones. Furthermore, we impose a Fourier filter that only includes the poloidal modes within $|m+\nint{nq}|<5$ where  $\nint{.}$ indicates the nearest integer, and  Dirichlet boundary condition at the axis and in the edge for the potentials. The kinetic equation is solved within an annular of $s\in [0.01,0.99]$.

\subsection{Linear simulations}
\label{sec:lin_sims}
Here, we consider the electromagnetic antenna in the linear setting and excite TAE modes in order to study their damping rate and frequency. In particular, we consider the zeroth-order equations of motion Eqs~\eqref{eq:R0}-\eqref{eq:v||0} and the first-order electromagnetic antenna contribution within linear pullback scheme Eqs.~\eqref{eq:R1dot_lin_pullback_electromagnetic_antenna}-\eqref{eq:eps1dot_lin_pullback_electromagnetic_antenna}, while setting the plasma response to zero in the equations of motion, i.e., $\dot{\bm R}^{(1)\ \mathrm{lin. fields}}_{\mathrm{plasma}}$ and $\dot{v}^{(1)\ \mathrm{lin. fields}}_{\mathrm{plasma}}$ in the left hand side of kinetic equation \eqref{eq:kinetic_eq}. 

\subsubsection{Excitation of TAE modes}
\label{sec:excite_TAE_linear}
\begin{figure}
  	\centering
	\begin{tabular}{cc}
  \includegraphics[scale=0.8]{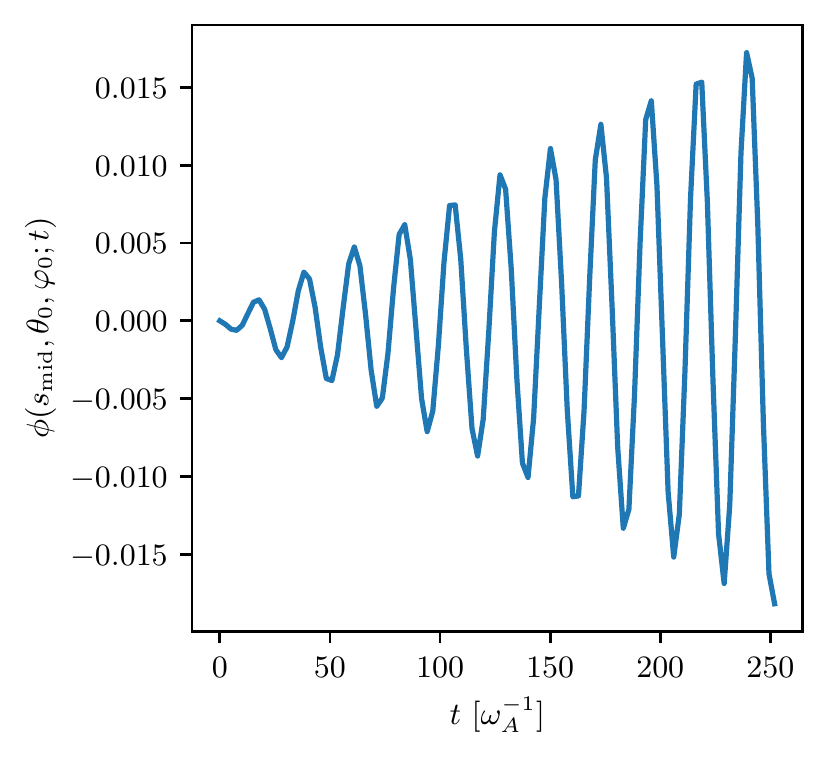}
  & \includegraphics[scale=0.8]{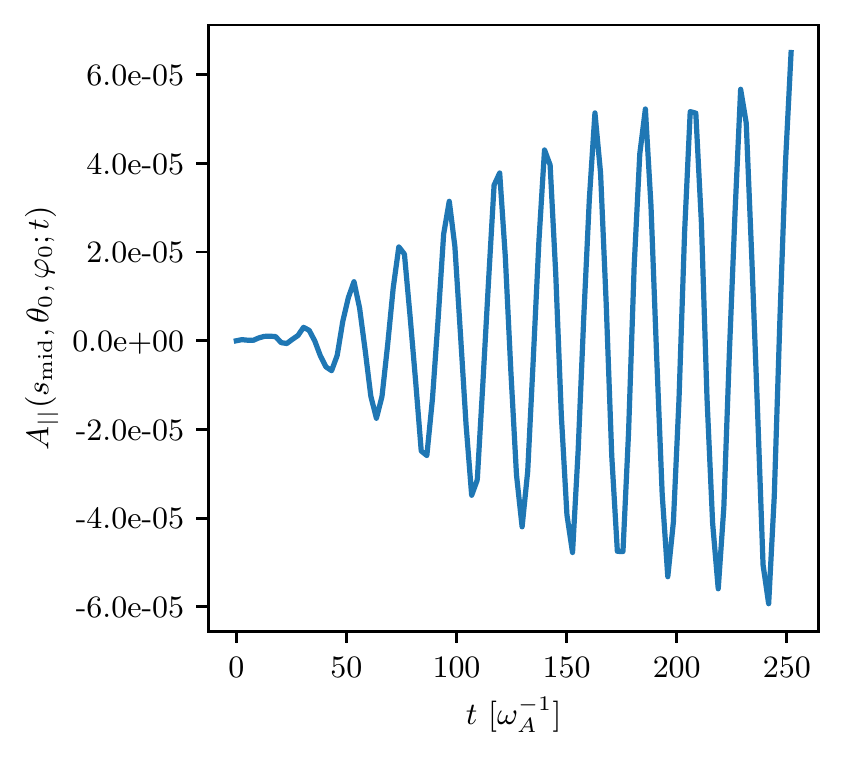}\\
  (a)  & (b)\\
  \includegraphics[scale=0.5]{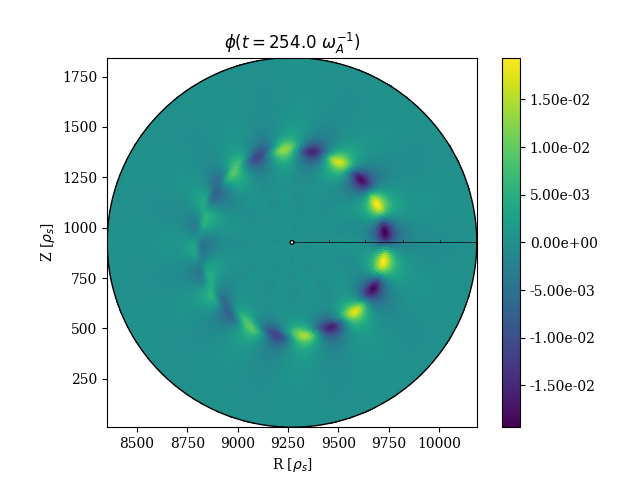}
  & \includegraphics[scale=0.5]{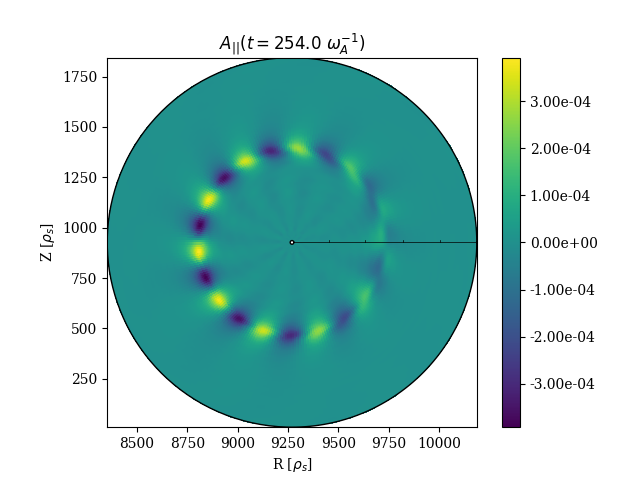}
  \\
  (c)  & (d)
  \end{tabular}
  \caption{Linear growth of (a) electrostatic and  (b) electromagnetic potentials evaluated at the mid point $s=1/2$ on the axis $\theta_0=\varphi_0=0$ for the $n=6$ TAE mode excitation with an electromagnetic antenna. The potentials for the given toroidal angle is depicted in (c) and (d).}
  \label{fig:linear_growth_rate}
\end{figure}

\begin{figure}
  	\centering
	\begin{tabular}{cc}
  \includegraphics[scale=0.8]{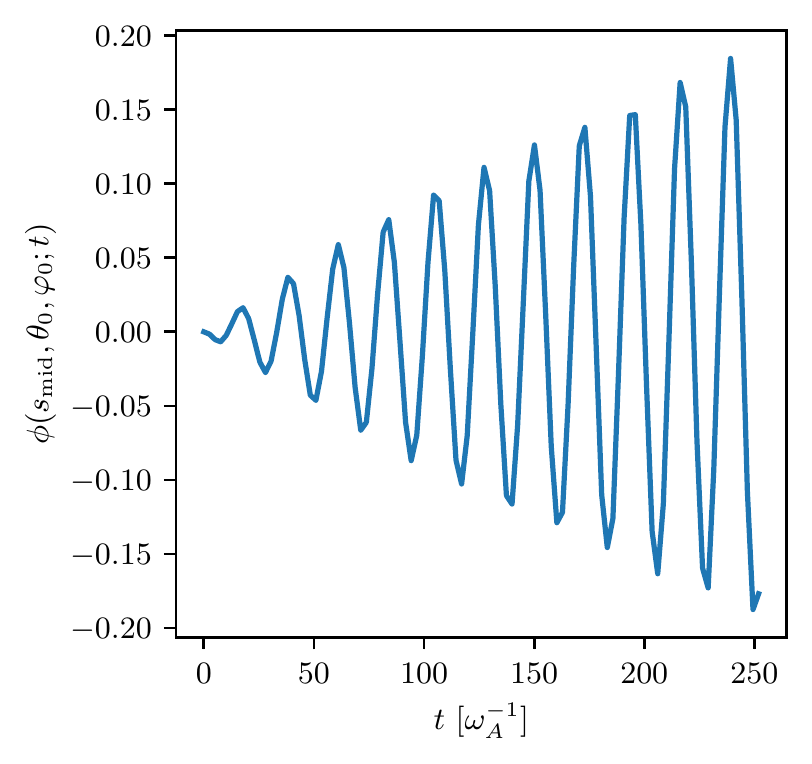}
  & \includegraphics[scale=0.8]{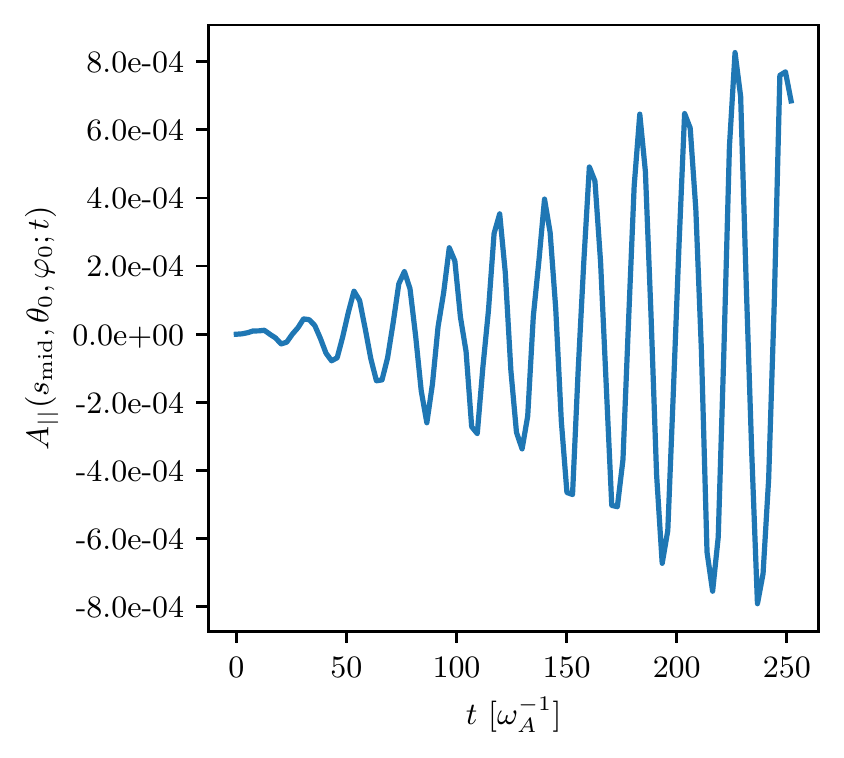}\\
  (a)  & (b)\\
  \includegraphics[scale=0.5]{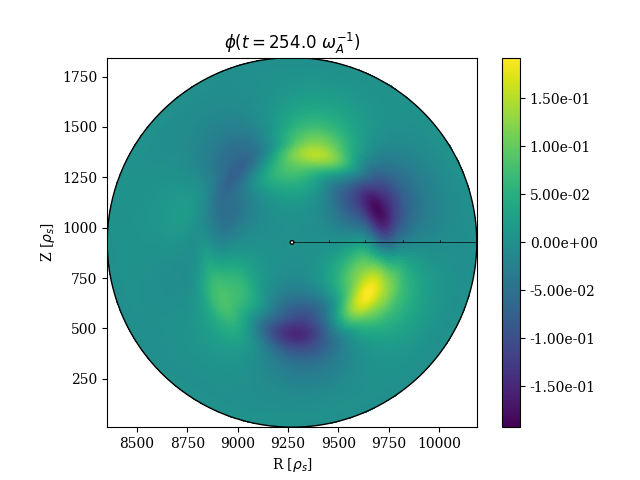}
  & \includegraphics[scale=0.5]{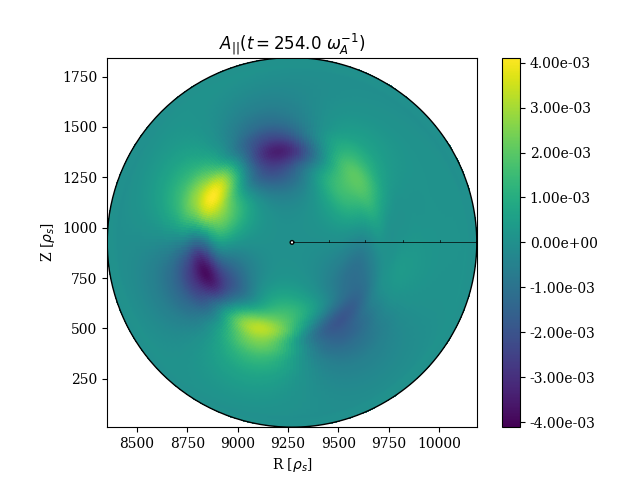}
  \\
  (c)  & (d)
  \end{tabular}
  \caption{Linear growth of (a) electrostatic and  (b) electromagnetic potentials evaluated at the mid point $s=1/2$ on the axis $\theta_0=\varphi_0=0$ for the $n=2$ TAE mode excitation with an electromagnetic antenna. The potentials for a given toroidal angle is depicted in (c) and (d).}
  \label{fig:linear_growth_rate_n2}
\end{figure}

\begin{figure}
  	\centering
	\begin{tabular}{cc}
  \includegraphics[scale=0.8]{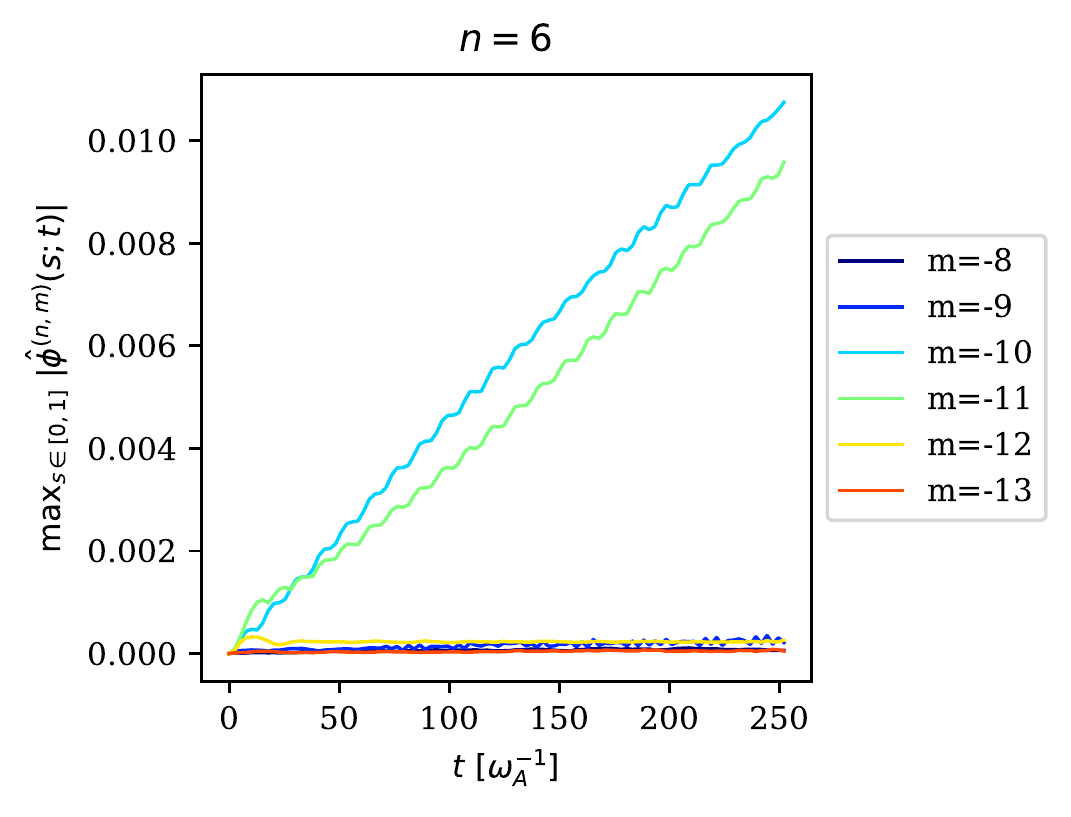}
  & \includegraphics[scale=0.8]{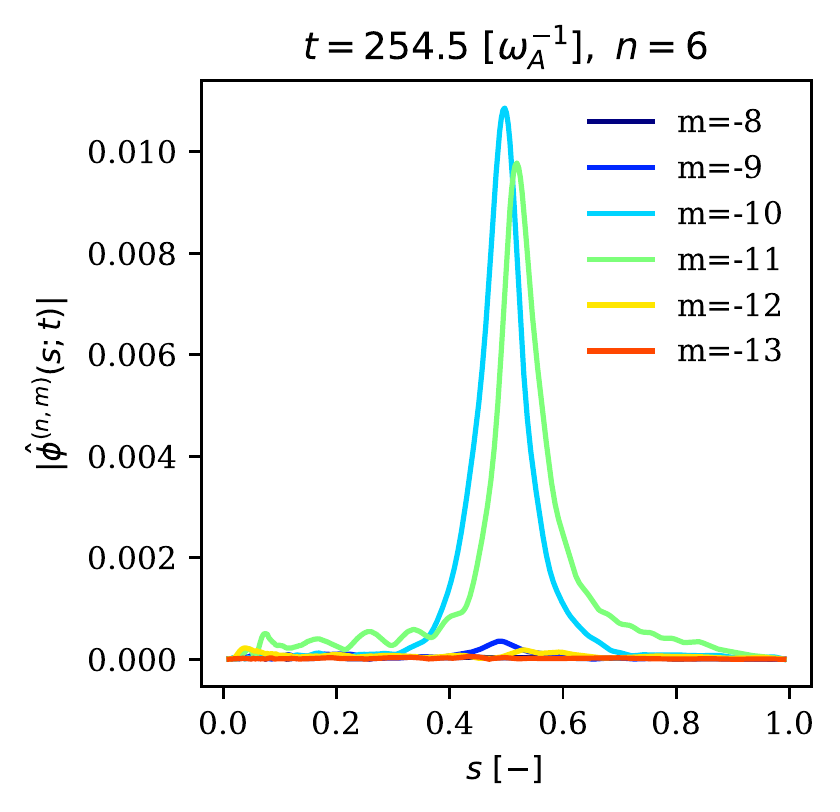}\\
  (a)  & (b)\\
    \includegraphics[scale=0.8]{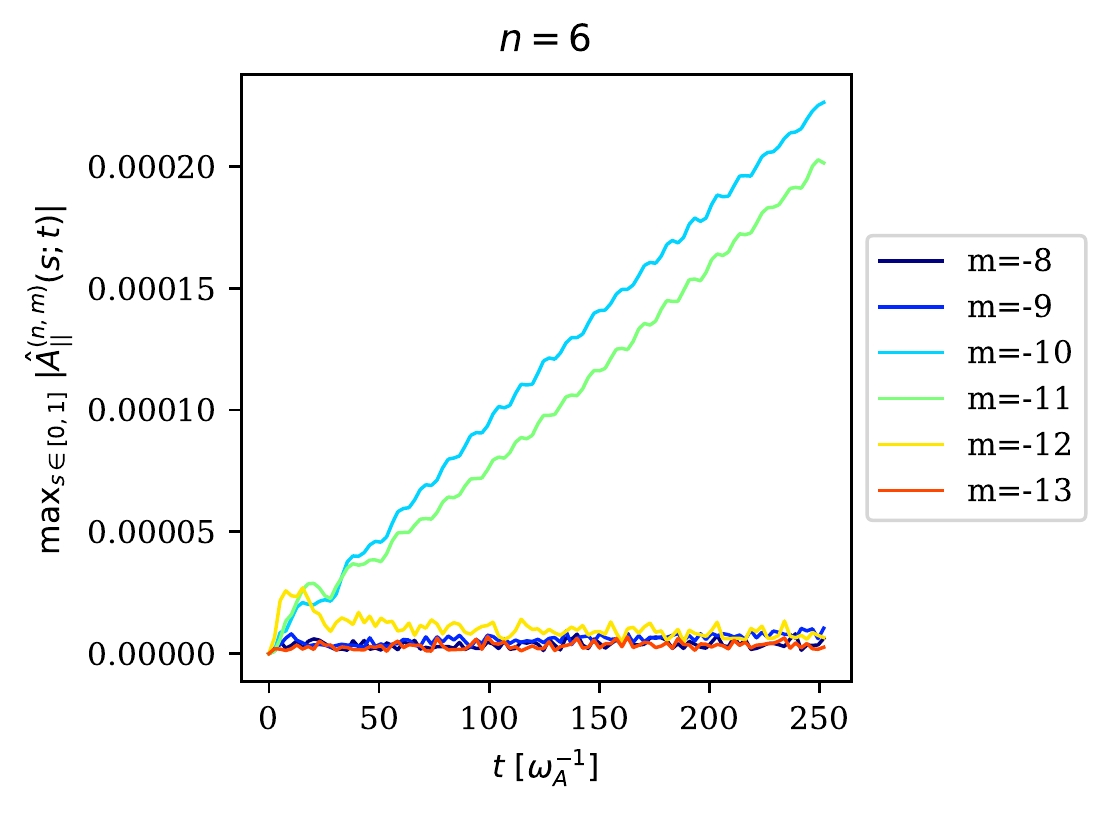}
  & \includegraphics[scale=0.8]{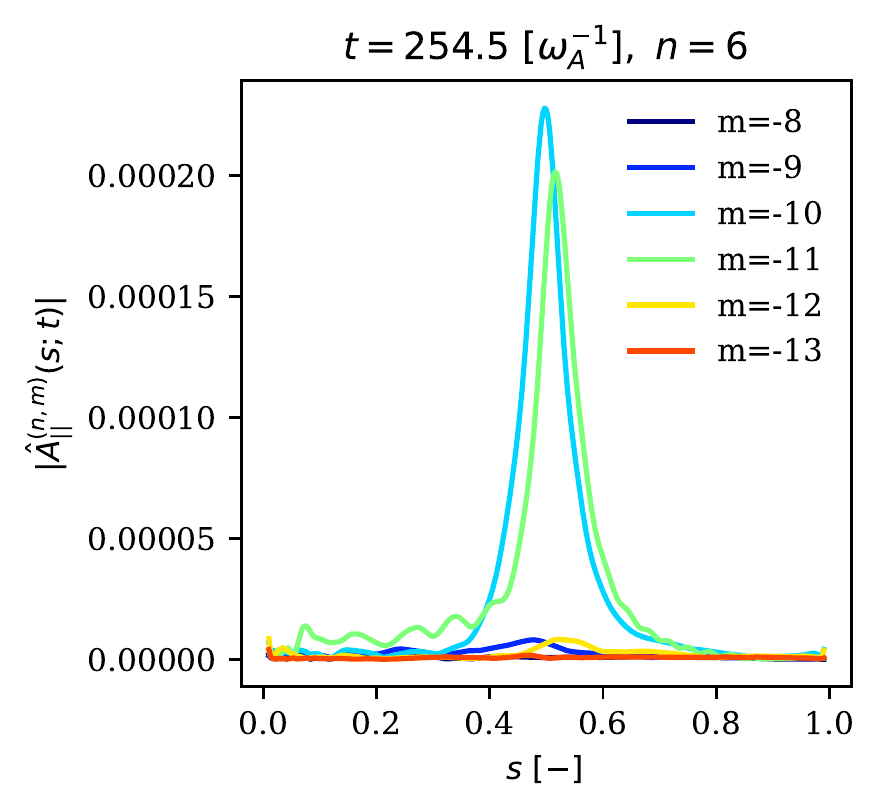}
  \\
  (c)  & (d)
  \end{tabular}
  \caption{Evolution of discrete Fourier transform of (a) electrostatic and  (c) electromagnetic potentials in maximum norm that is taken in the radial direction for the $n=6$ TAE mode excitation with an electromagnetic antenna.  The outcome radial profile of (b) electrostatic and  (d) electromagnetic potentials are shown at final time $t=50000\ \omega_{ci}^{-1}\approx 254.5\ \omega_A^{-1}$.}
  \label{fig:profile_linear_growth_mode_coeff}
\end{figure}

\begin{figure}
  	\centering
	\begin{tabular}{cc}
  \includegraphics[scale=0.8]{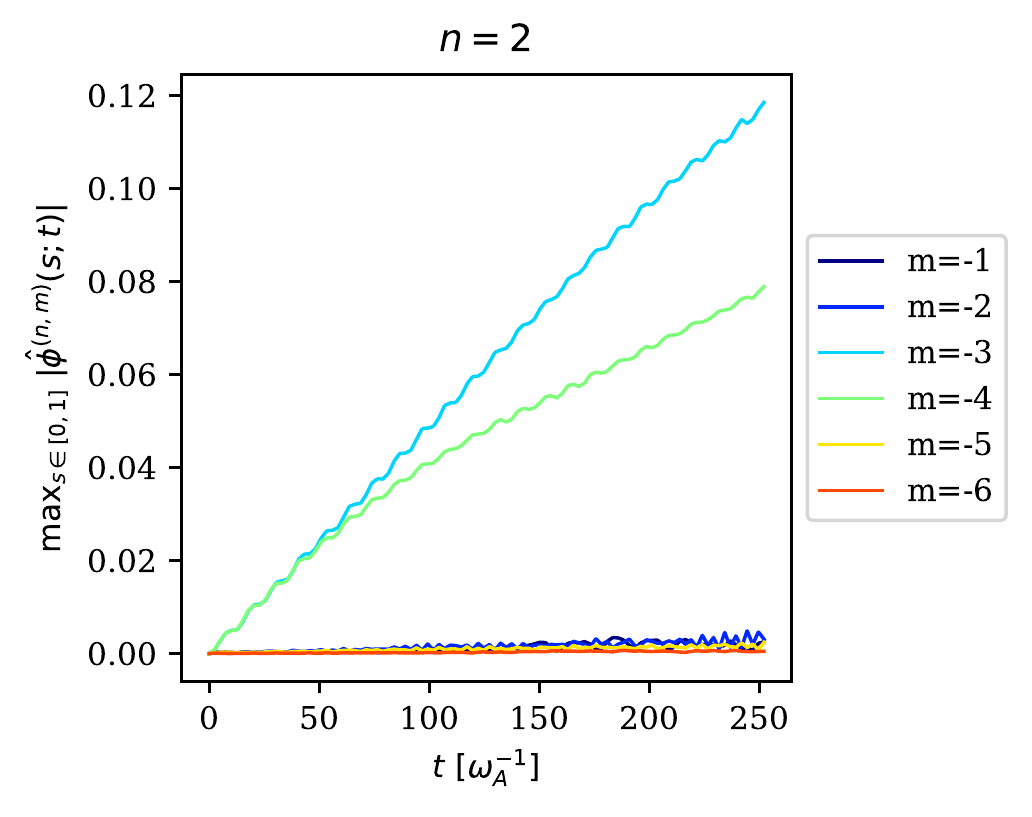}
  & \includegraphics[scale=0.8]{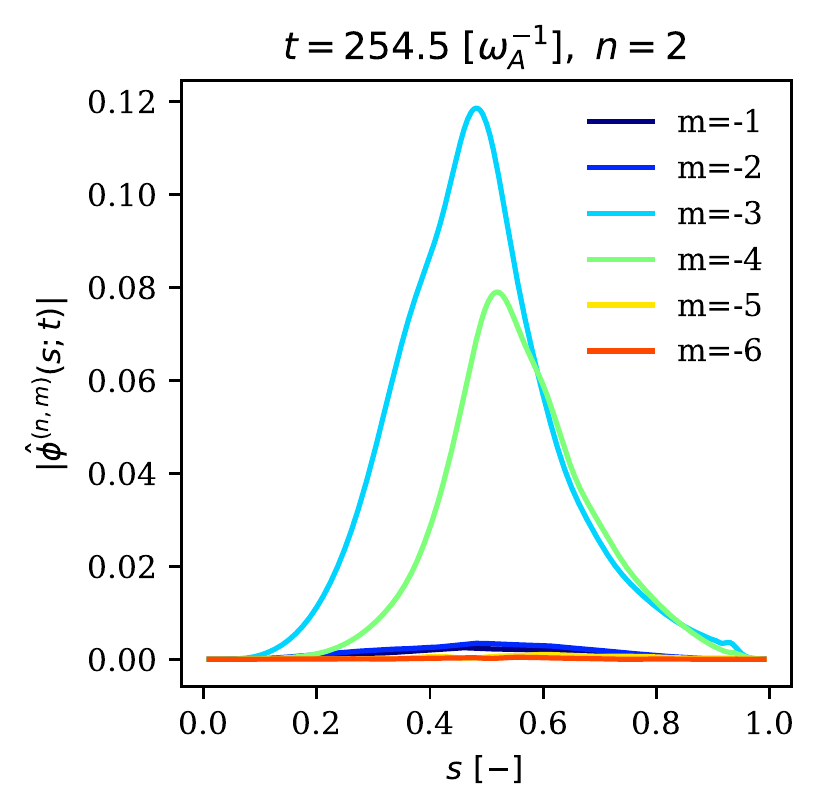}\\
  (a)  & (b)\\
  \includegraphics[scale=0.8]{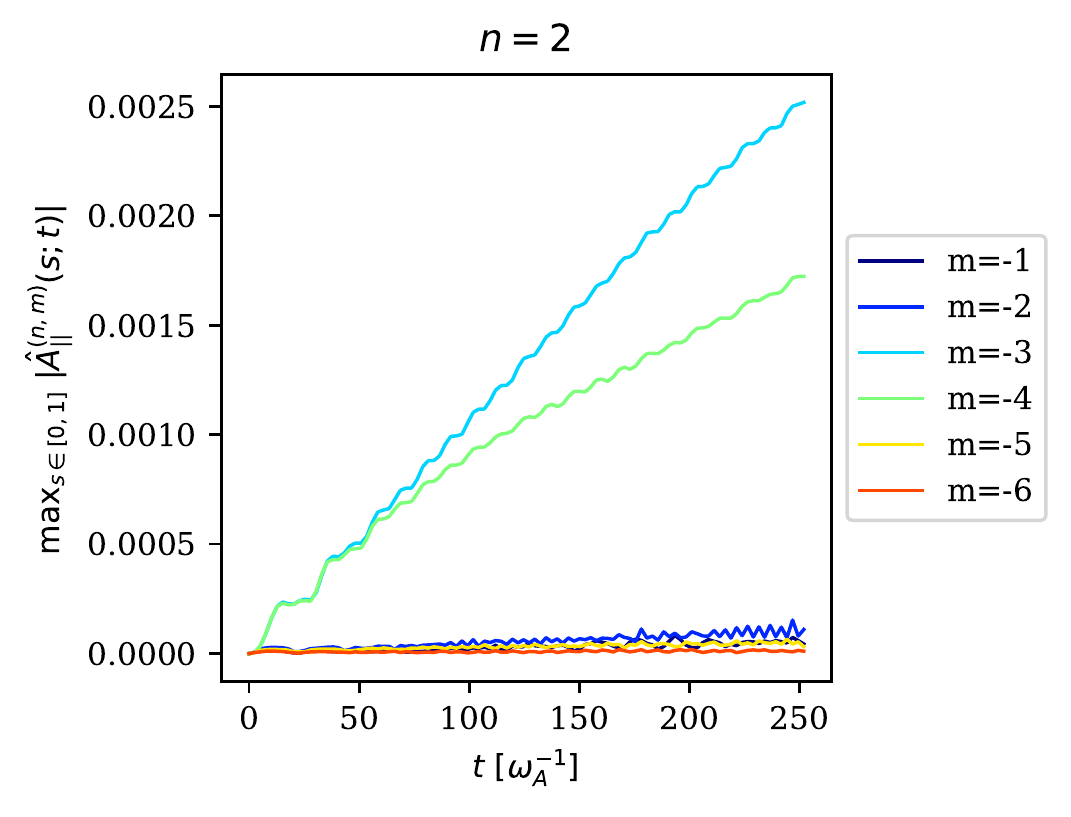}
  & \includegraphics[scale=0.8]{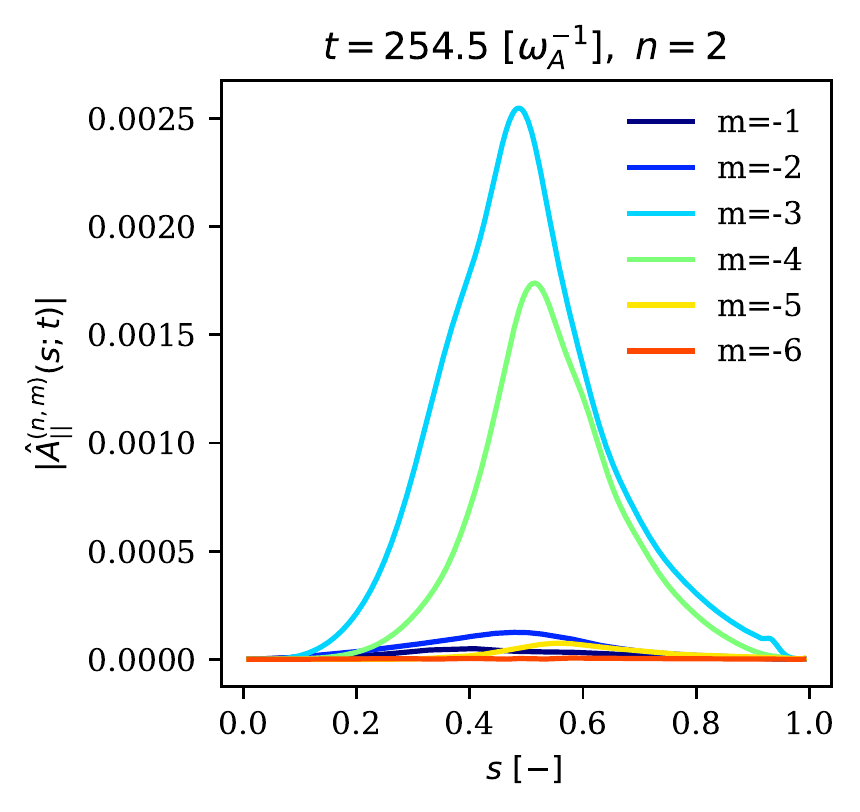}
  \\
  (c)  & (d)
  \end{tabular}
  \caption{Evolution of discrete Fourier transform of (a) electrostatic and  (c) electromagnetic potentials in maximum norm that is taken in the radial direction for the $n=2$ TAE mode excitation with an electromagnetic antenna.  The outcome radial profile of (b) electrostatic and  (d) electromagnetic potentials are shown at final time $t=50000\ \omega_{ci}^{-1}\approx 254.5\ \omega_A^{-1}$.}
  \label{fig:profile_linear_growth_mode_coeff_n2}
\end{figure}

\ \\ \ \\
\noindent In order to excite the $n=6$ TAE mode, let us consider the antenna with
$\mathcal{T}=\{ (n_1,m_1), ...\}=\{(6,-11), (6,-10)\}$, the frequency $\omega_\mathrm{ant}=-0.28 \omega_{A}\approx -0.00143\ \omega_{ci}$, and Gaussian radial profiles 
\begin{flalign}
h_i(s) &= a_i \exp\left({-(s-s_{0,i})^2/\delta_i^2}\right)\ \ \ \text{for}\ i=1,2\\
\text{where}\ \ \ 
a_1 &= 0.01,\ \ s_{0,1}=0.48, \ \ \  \delta_1=0.087~,\\
\text{and}\ \ \ 
a_2 &= 0.01,\ \ s_{0,2}=0.52,\ \ \  \delta_2=0.082~.
\end{flalign}
Similarly, for the $n=2$ TAE mode, we consider the antenna with
$\mathcal{T}=\{ (n_1,m_1), ...\}=\{(2,-4), (2,-3)\}$, the frequency $\omega_\mathrm{ant}\approx -0.28 \omega_{A}\approx -0.00143\ \omega_{ci}$ and radial profile
\begin{flalign}
h_i(s) &= a_i \exp\left({-(s-s_{0,i})^2/\delta_i^2}\right)\ \ \ \text{for}\ i=1,2\\
\text{where}\ \ \ 
a_1 &= 0.1,\ \ s_{0,1}=0.5, \ \ \  \delta_1=0.1~,\\
\text{and}\ \ \ 
a_2 &= 0.1,\ \ s_{0,2}=0.5,\ \ \  \delta_2=0.1~.
\end{flalign}
The parameters of the Gaussian shape functions were found by fitting the  function in the radial direction to the TAE excited simulations obtained with fast ion particles. Choosing such excitation parameters leads to the excitation of target modes in the plasma as shown in Figs.~\ref{fig:linear_growth_rate}-\ref{fig:linear_growth_rate_n2}. The time traces and the radial profiles of the discrete Fourier transform of the electrostatic and magnetic potentials, shown in Figs.~\ref{fig:profile_linear_growth_mode_coeff}-\ref{fig:profile_linear_growth_mode_coeff_n2}, indicate that the antenna has excited the target modes. 

\begin{figure}
  	\centering
	\begin{tabular}{cc}
\includegraphics[scale=0.9]{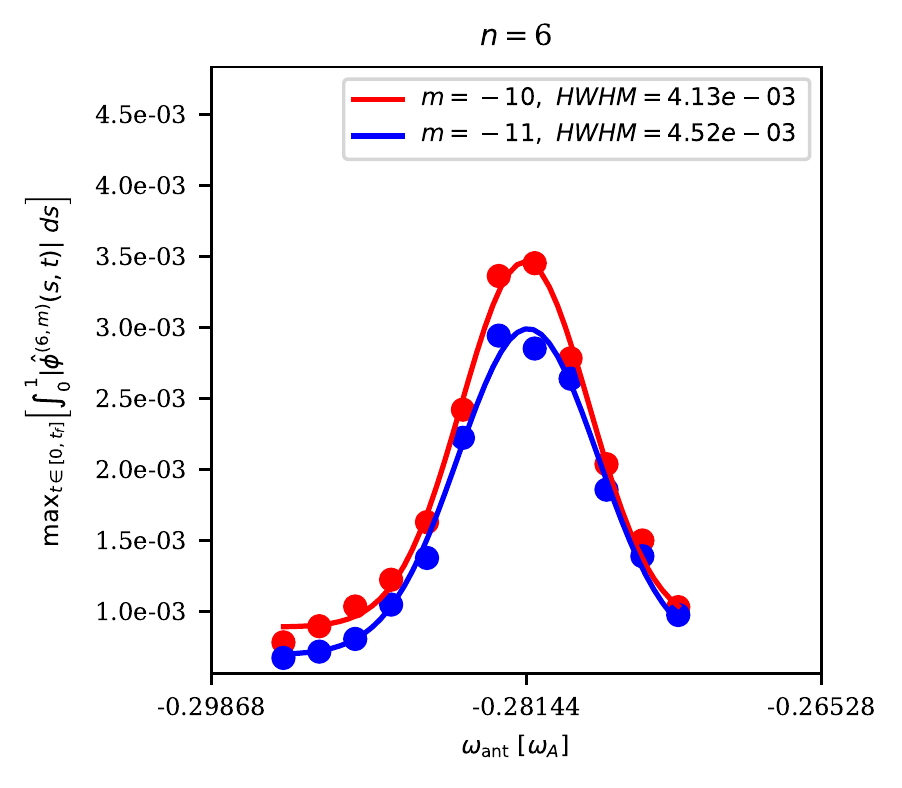}
  &
  \includegraphics[scale=0.9]{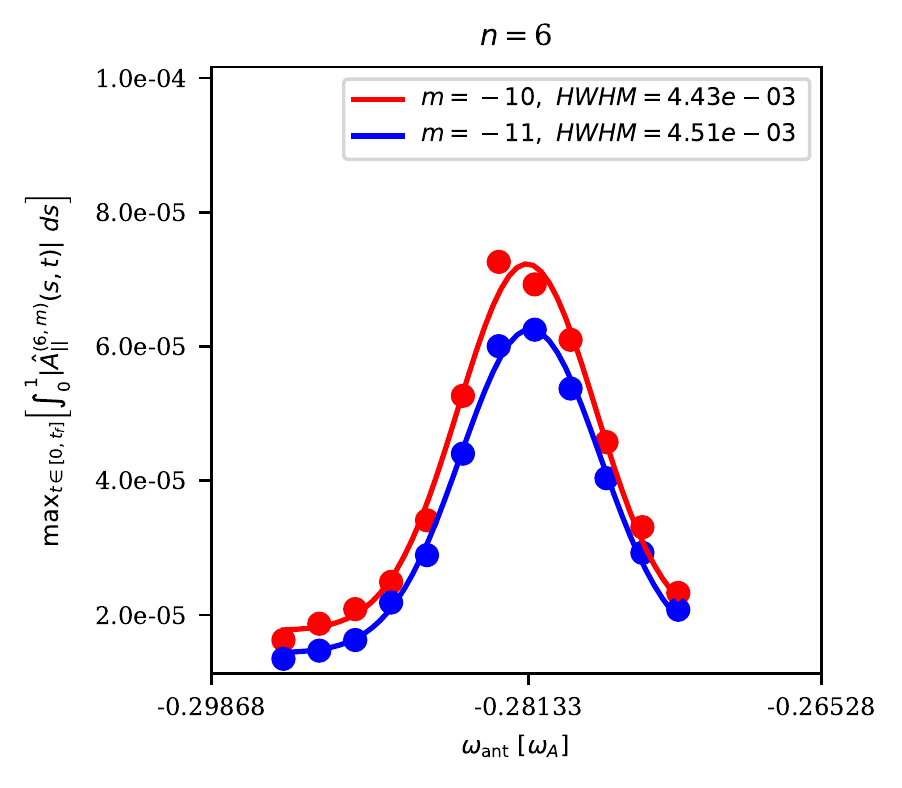}
  \\
  (a)  & (b)
  \\
  \includegraphics[scale=0.9]{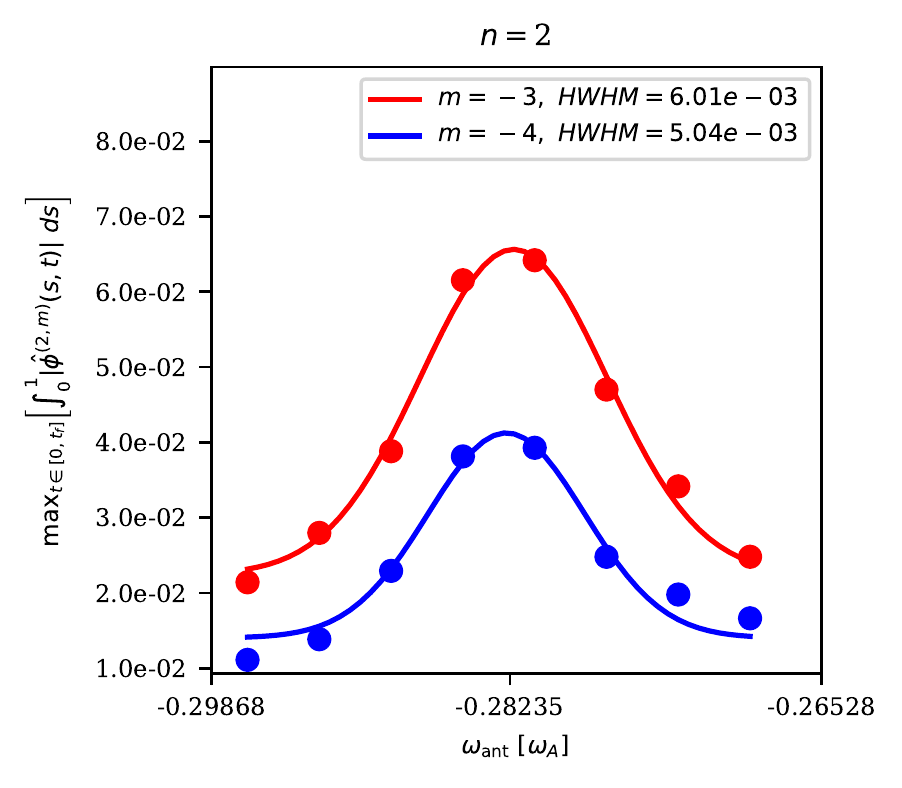}
  &
  \includegraphics[scale=0.9]{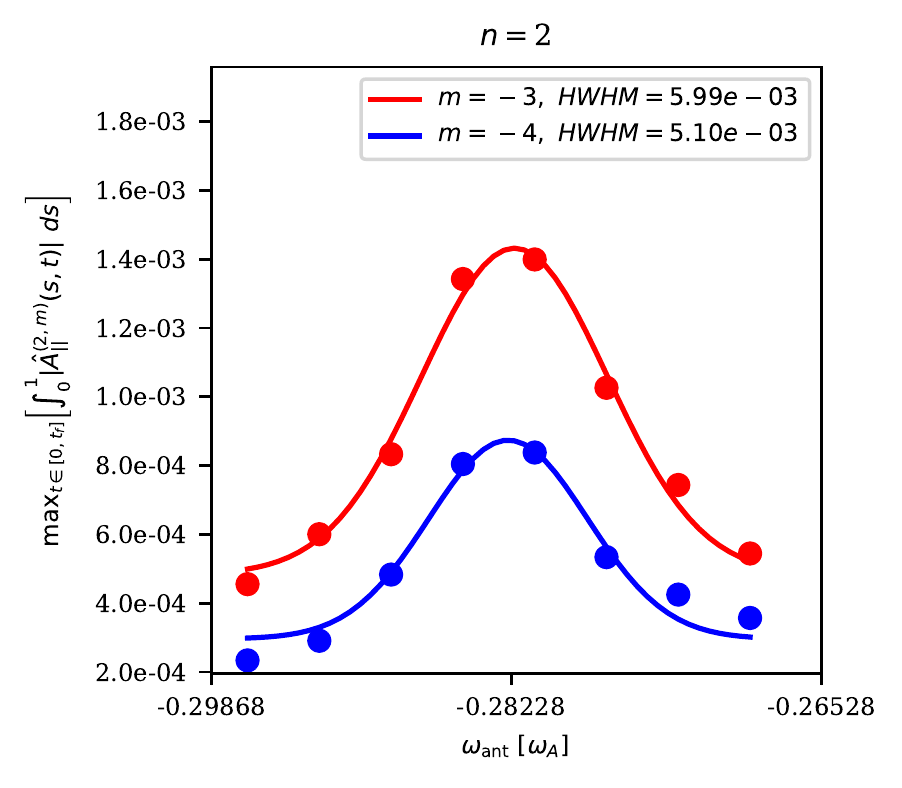}
  \\
  (c)  & (d)
  \end{tabular}
  \caption{Linear excitation of the $n=6$ (top) and the $n=2$ (bottom) TAEs  with an electromagnetic antenna at a range of frequencies near the resonance. Here, the amplitude of potential is integrated in the radial direction and its maximum over time span of $t\in[0,t_f]$ where $t_f=200000\ \omega_{ci}^{-1}\approx 1017.8\  \omega_A^{-1}$ is taken (shown as points). The solid curves indicate the profiles of a Gaussian fit to the data points.  The deployed antenna includes the mode  $(n,m)\in \{(6,-10), (6,-11)\}$ for the excitation of the $n=6$ TAE mode and  $(n,m)\in \{(2,-4), (2,-3)\}$ for the excitation of the $n=2$ TAE mode. This analysis is performed on the electrostatic $\phi$ (a)-(c) and electromagentic potentials $A_{||}$ (b)-(d).}
  \label{fig:finding_resonance}
\end{figure}

\begin{figure}
  	\centering
	\begin{tabular}{cc}
\includegraphics[scale=0.9]{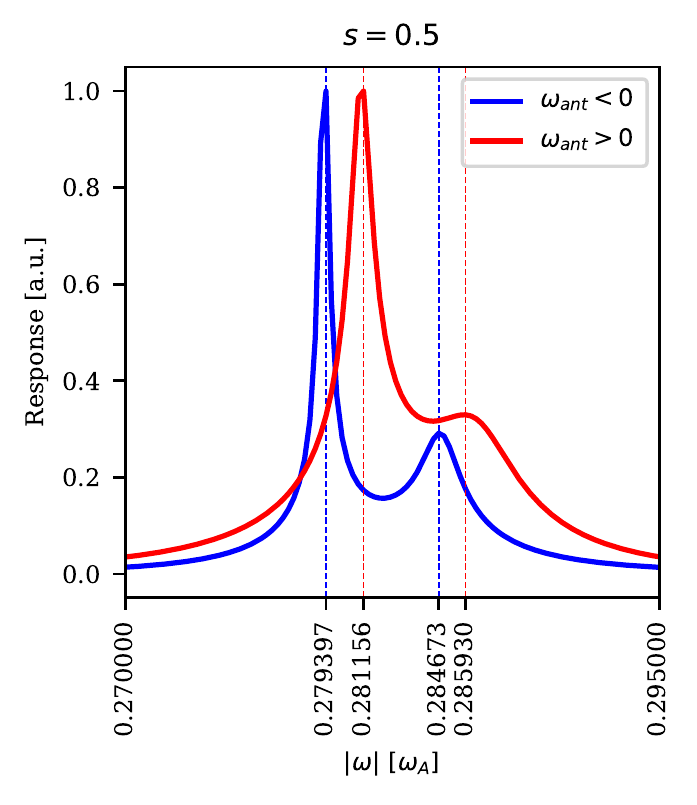}
  &
  \includegraphics[scale=0.9]{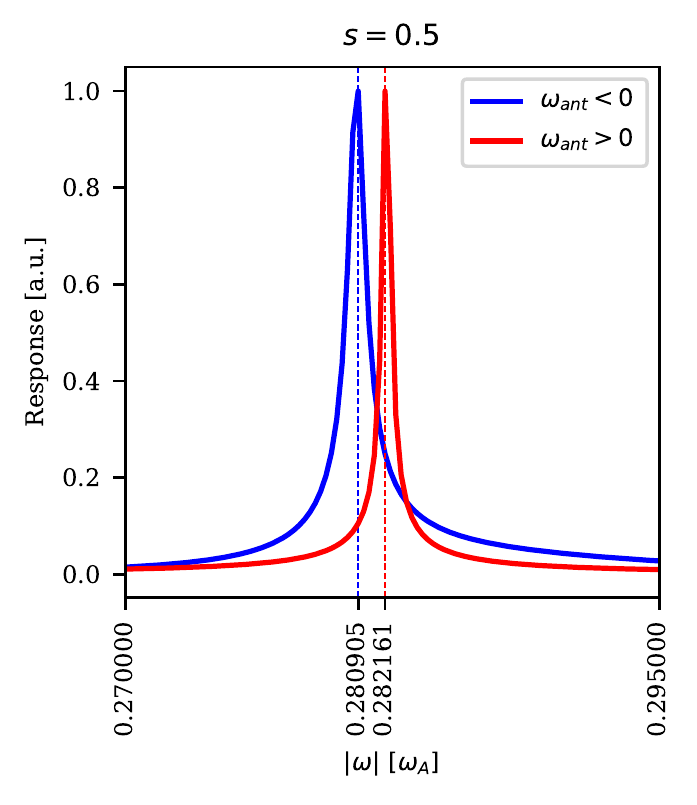}
  \\
  (a) $n=6,\ t\in[5,610] \ \omega_{A}^{-1}$& (b) $n=6,\ t>610\ \omega_{A}^{-1}$
  \\
  \includegraphics[scale=0.9]{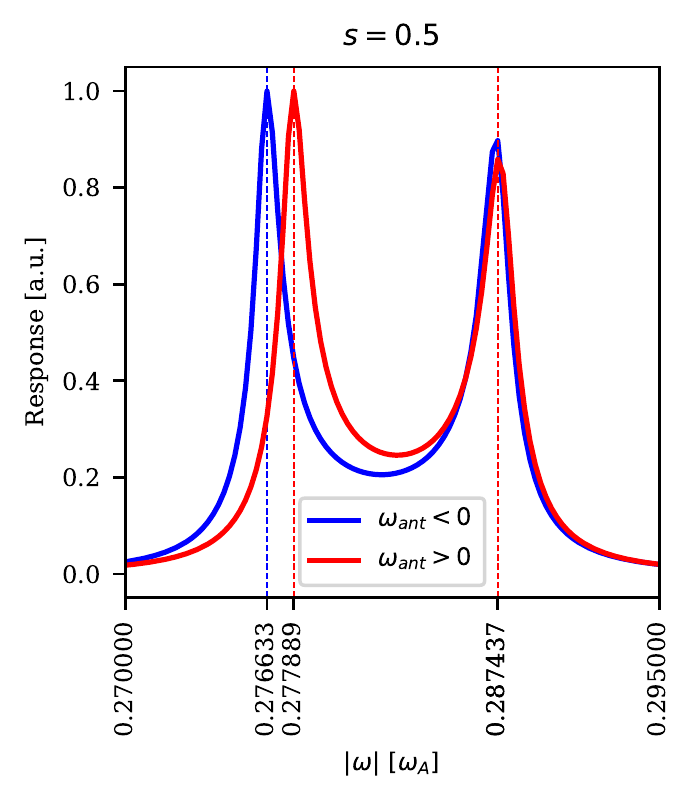}
  &
  \includegraphics[scale=0.9]{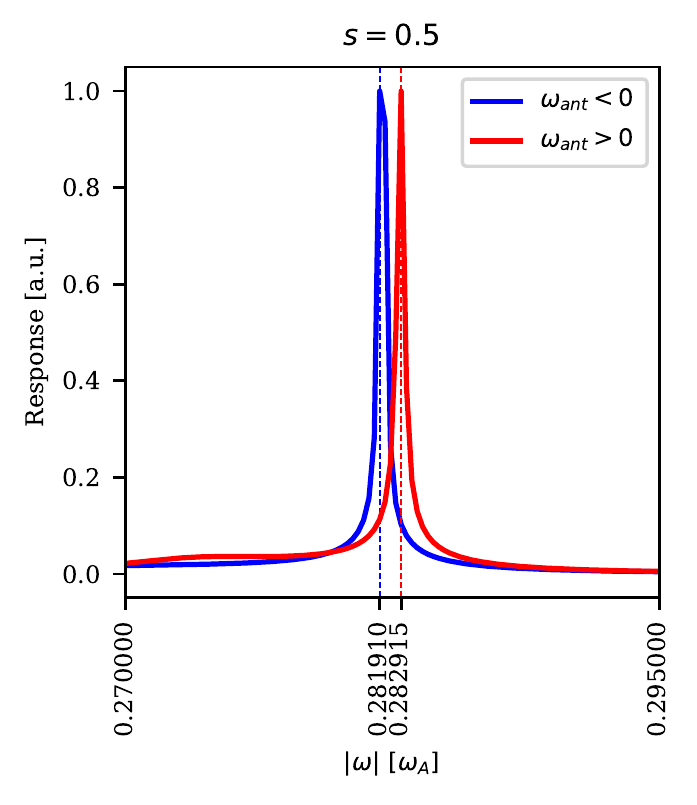}
  \\
  (a) $n=2,\ t\in[5,610] \ \omega_{A}^{-1}$& (b) $n=2,\ t>610\ \omega_{A}^{-1}$
  \end{tabular}
  \caption{The frequency scan of the electrostatic potential at $s=1/2$ for the antenna excitation of (a)-(b) the $n=6$ TAE mode and (c)-(d) the $n=2$ TAE mode. The DMUSIC was deployed to analyse the frequencies.}
  \label{fig:freq_scan_n2-6}
\end{figure}

\begin{figure}
  	\centering
	\begin{tabular}{cc}
\includegraphics[scale=0.9]{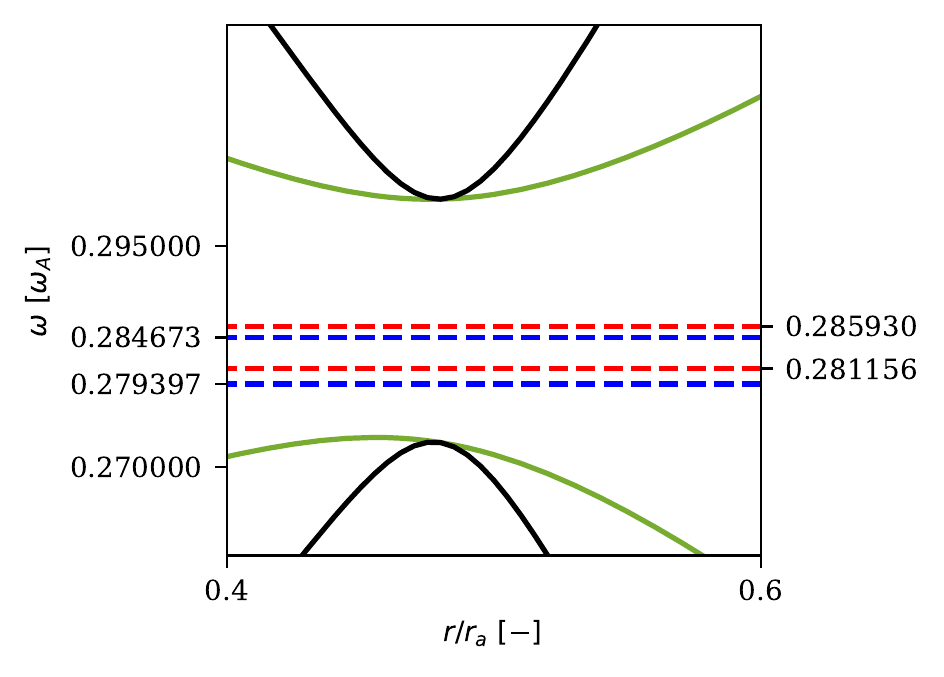}
  &
  \includegraphics[scale=0.9]{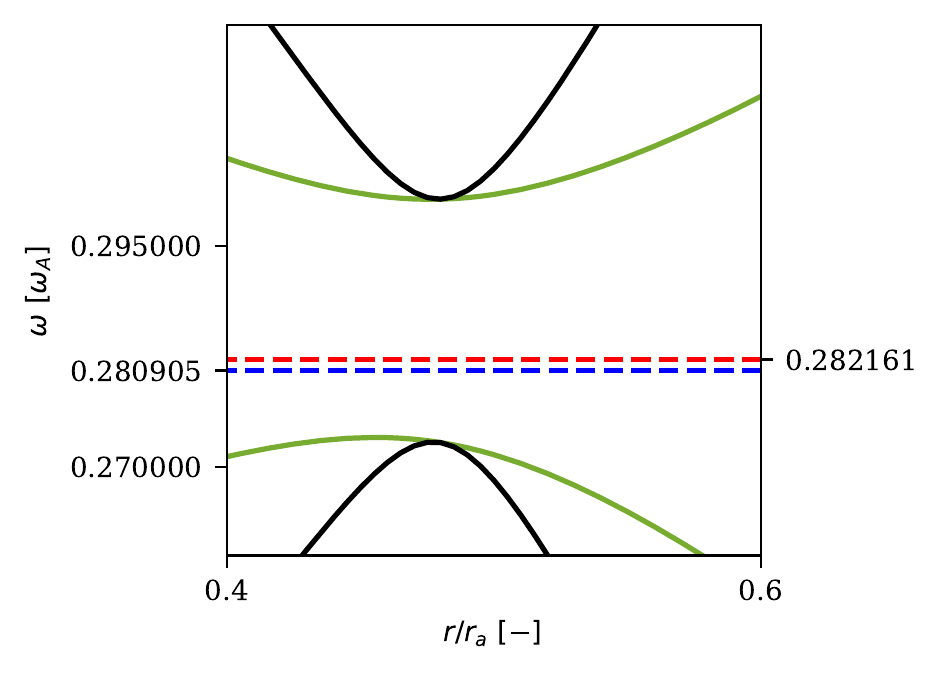}
  \\
(a) $n=6,\ t\in[5,610] \ \omega_{A}^{-1}$& (b) $n=6,\ t>610\ \omega_{A}^{-1}$
  \\
  \includegraphics[scale=0.9]{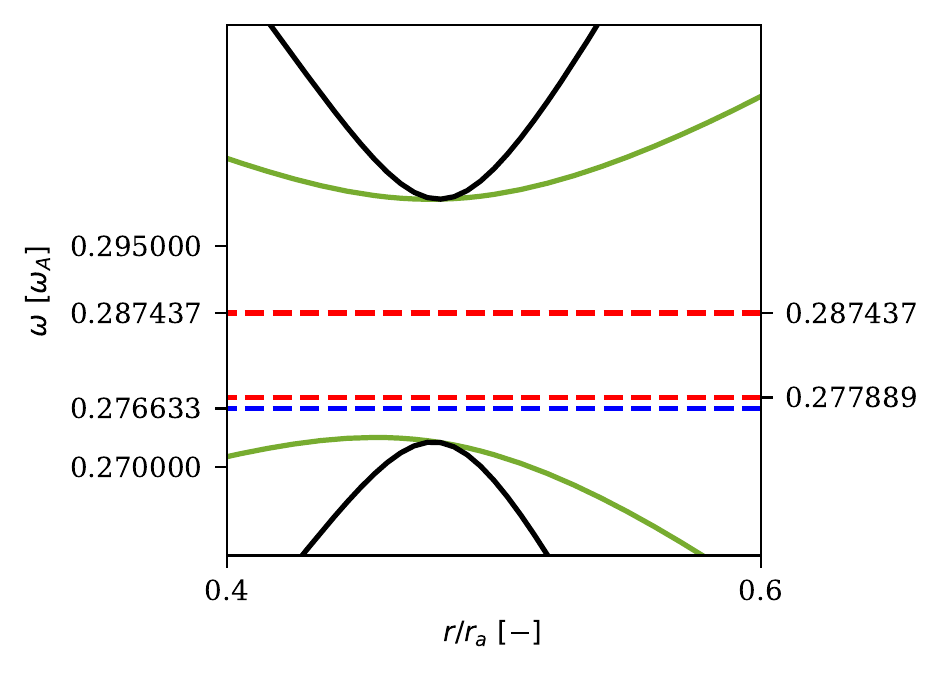}
  &
  \includegraphics[scale=0.9]{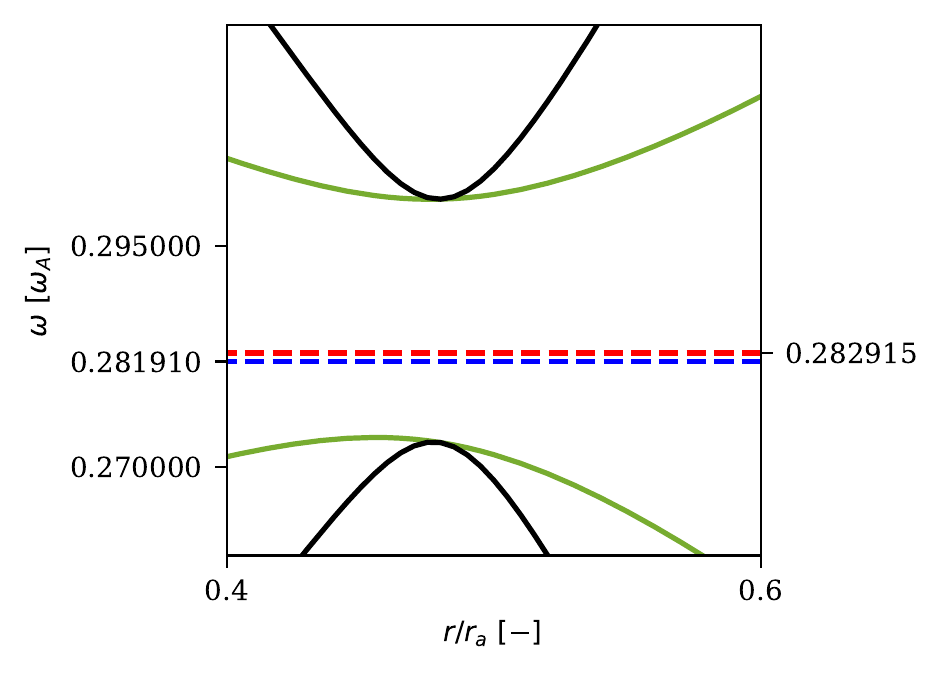}
  \\
(a) $n=2,\ t\in[5,610] \ \omega_{A}^{-1}$& (b) $n=2,\ t>610\ \omega_{A}^{-1}$
  \end{tabular}
  \caption{The peak of frequency scan for the electrostatic potential at $s=1/2$ plotted in dashed lines (blue for $\omega_\mathrm{ant}<0$ and red for $\omega_\mathrm{ant}>0$) in the frequency spectra for the antenna excitation of (a)-(b) the $n=6$ TAE mode and (c)-(d) the $n=2$ TAE mode. The black solid curve corresponds to the Alfv\'en continua of $n=6$, $m=-11,-10$ TAE mode and the green to $n=2$, $m=-4,-3$ TAE mode of the ITPA test case, see Fig.~\ref{fig:alfven_continua}.}
  \label{fig:freq_scan_n2-6_continuum_spectrum}
\end{figure}

\ \\
\noindent Furthermore, in order to find the resonance frequency numerically, simulations were repeated with the antenna at a range of frequencies, see Fig.~\ref{fig:finding_resonance}. Since the maximum point of the potentials may oscillate in the radial direction over time, we consider
\begin{flalign}
\int_0^1 |\hat{Q}|^{(n,m)}(s,t) ds
\end{flalign}
as the measure where $|\hat{Q}|^{(n,m)}$ denotes the amplitude of the mode $(n,m)$ for the potential of interest. By fitting a Gaussian function to the outcome set of points, we observe that the Half-Width-Half-Max (HWHM) of the $n=6$ TAE mode is $\mathrm{HWHM} \approx 2.2 \times 10^{-5} \omega_{ci}$ with the resonance frequency of $\omega_\mathrm{res}\approx-0.001432\ \omega_{ci} \approx -0.2813\  \omega_A$. Similar measurement for the $n=2$ TAE mode lead to $\mathrm{HWHM} \approx 2.7 \times 10^{-5}\ \omega_{ci}$ and resonance frequency is $\omega_\mathrm{res}\approx-0.001437\ \omega_{ci}\approx\ -0.2823\ \omega_A$.
\\ \ \\
Although TAE modes can be also excited similarly with the positive sign for the frequency, the excited mode with $\omega_\mathrm{ant}>0$ has a higher damping rate compared to the $\omega_\mathrm{ant}<0$, as discussed in the next section. Here, we compare the outcome frequency analysis of the electrostatic potential for the antenna excitation of the $n=6$ and $n=2$ TAE modes excited by the antenna with frequency  $\omega_\mathrm{ant}\approx \pm 0.28\ \omega_A$ where simulations for positive and negative signs are done separately. In this study, we used damped multiple signal classification (DMUSIC) \cite{kleiber2021modern} for the frequency analysis. As shown in Fig.~\ref{fig:freq_scan_n2-6}, an asymmetry in the frequency scan is observed that may be caused by the curvature drift. Furthermore, we observe that the antenna excites more than one frequency within the gap which probably corresponds to
even and odd parity TAEs,
see Fig.~\ref{fig:freq_scan_n2-6_continuum_spectrum}. Nevertheless, the closest frequency to the one of antenna becomes dominant as the system is driven with the antenna for a longer time.

\FloatBarrier
\subsubsection{Damping of TAE modes}
\label{sec:damp_TAE}
\begin{figure}
  	\centering
	\begin{tabular}{cc}
  \includegraphics[scale=1.0]{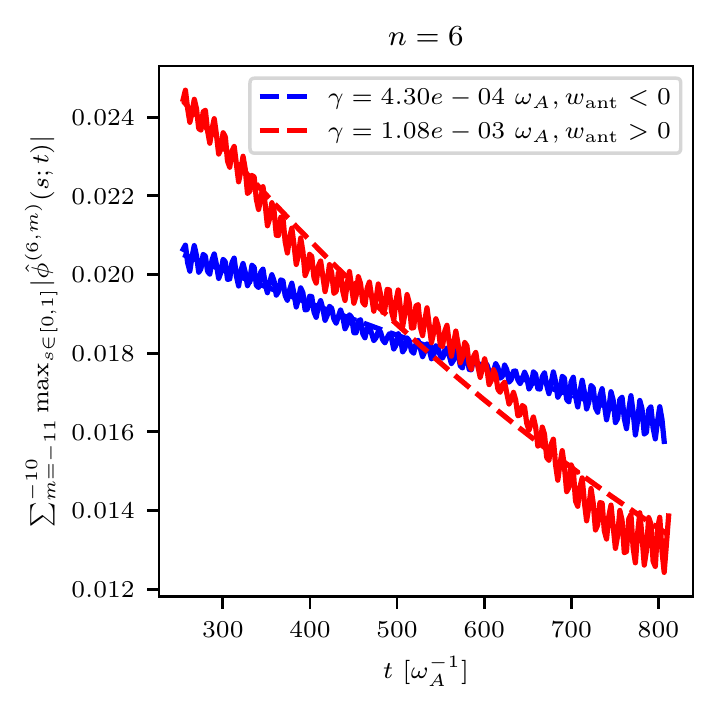}
  & \includegraphics[scale=1.0]{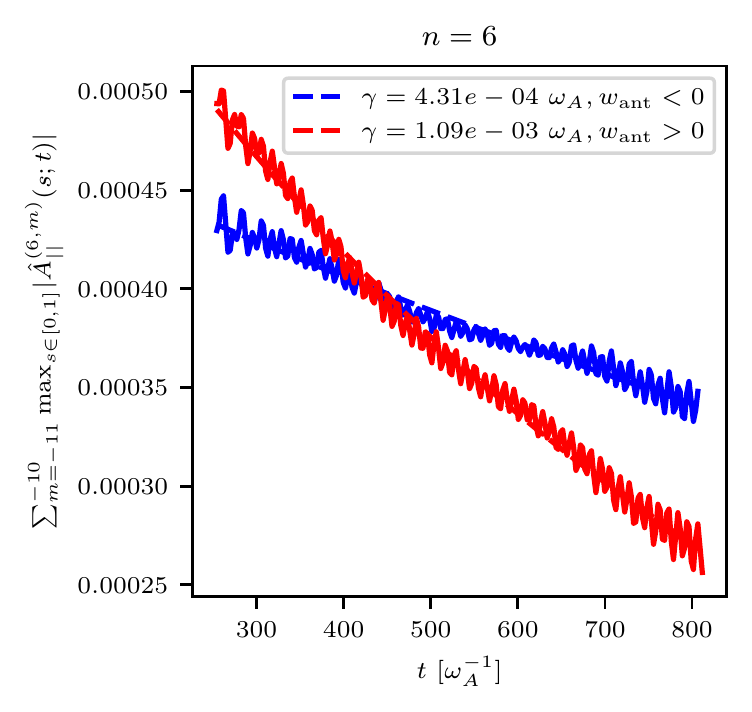}\\
  (a)  & (b)
  \end{tabular}
  \caption{Damping rates of the $n=6$ TAE mode obtained by fitting $a e^{-\gamma t}$ (dashed) to the corresponding Fourier coefficients of (a) $\phi$ and (b) $A_{||}$ integrated in the radial direction (solid curves). Here, the color blue and red denote the simulation with $\omega_\mathrm{ant}=-0.28\ \omega_A$ and $\omega_\mathrm{ant}=0.28\ \omega_A$, respectively.}
  \label{fig:damping_rate}
\end{figure}

\begin{figure}
  	\centering
	\begin{tabular}{cc}
  \includegraphics[scale=1.0]{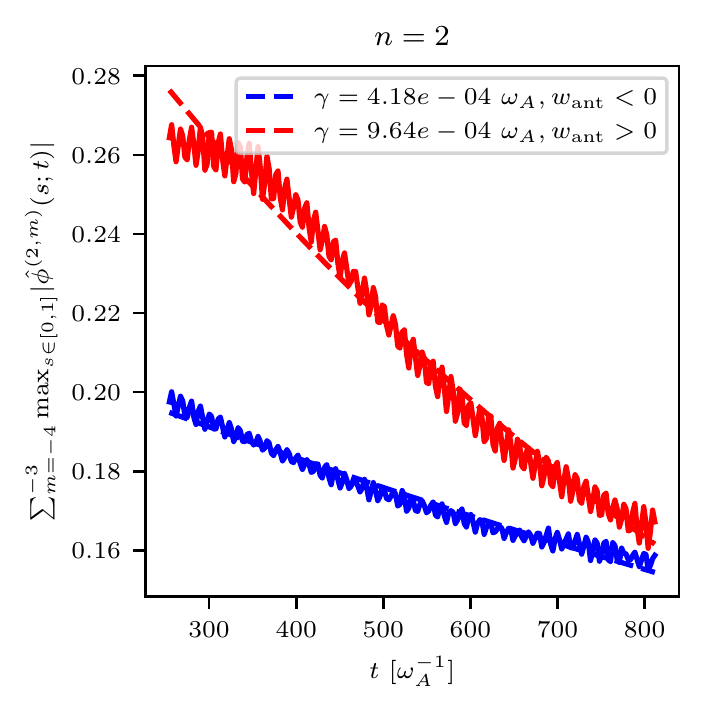}
  & \includegraphics[scale=1.0]{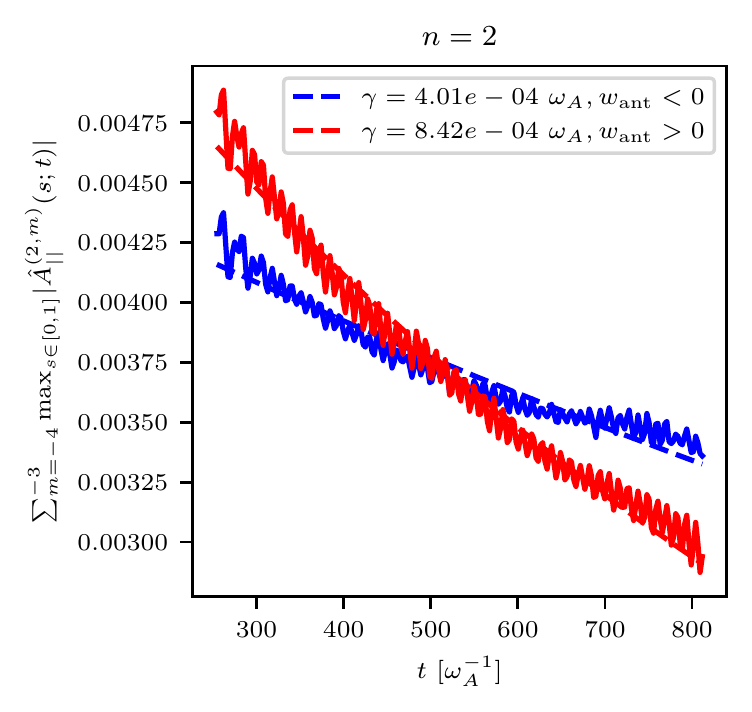}\\
  (a)  & (b)
  \end{tabular}
  \caption{Damping rates of TAE modes, i.e., $(n,m) = (2,-4)$ and $(2,-3)$, obtained by fitting $a e^{-\gamma t}$ to the corresponding Fourier coefficients of (a) $\phi$ and (b) $A_{||}$ integrated in the radial direction. Here, the color blue and red denote the simulation with $\omega_\mathrm{ant}=-0.28\ \omega_{A}$ and $\omega_\mathrm{ant}=0.28\ \omega_{A}$, respectively.}
  \label{fig:damping_rate_n2}
\end{figure}

\begin{figure}
  	\centering
  	\begin{tabular}{cc}
  \includegraphics[scale=0.9]{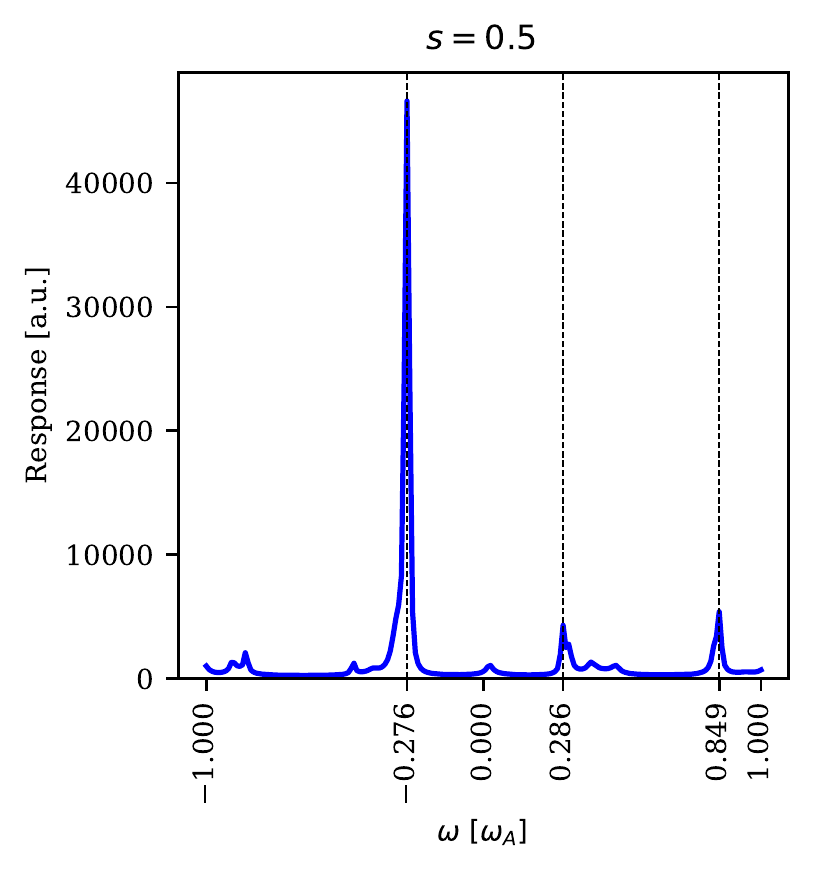} &
  \includegraphics[scale=0.9]{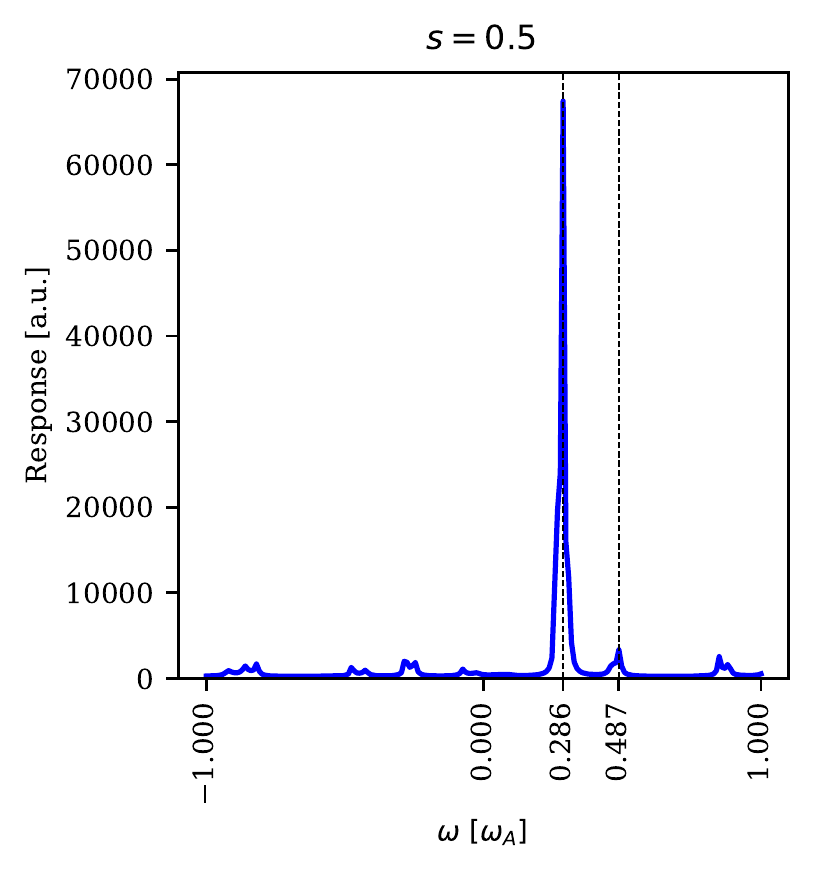}\\
  (a) $\omega_\mathrm{ant}=-0.28\ \omega_A$ & (b) $\omega_\mathrm{ant}=+0.28\ \omega_A$
  	\end{tabular}
  \caption{Frequency scan of $\phi$ at $s=1/2$ for the damping $n=6$ TAE mode for $t>254.5\ \omega_{A}^{-1}$ using DMUSIC method. The mode was initially excited by the antenna with the frequency (a) $\omega_\mathrm{ant}=-0.28\ \omega_A$ and (b) $\omega_\mathrm{ant}=0.28\ \omega_A$ for $t<254.5\ \omega_{A}^{-1}$.
}
  \label{fig:damping_frequency_n6}
\end{figure}

\begin{figure}
  	\centering
  	\begin{tabular}{cc}
  \includegraphics[scale=0.9]{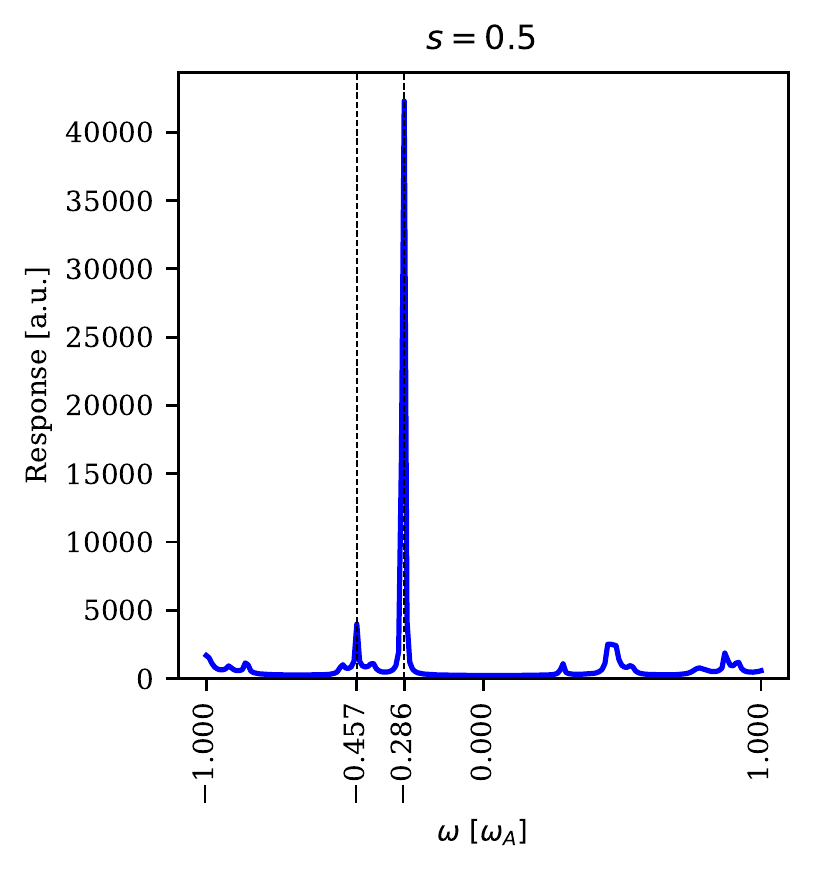} &
  \includegraphics[scale=0.9]{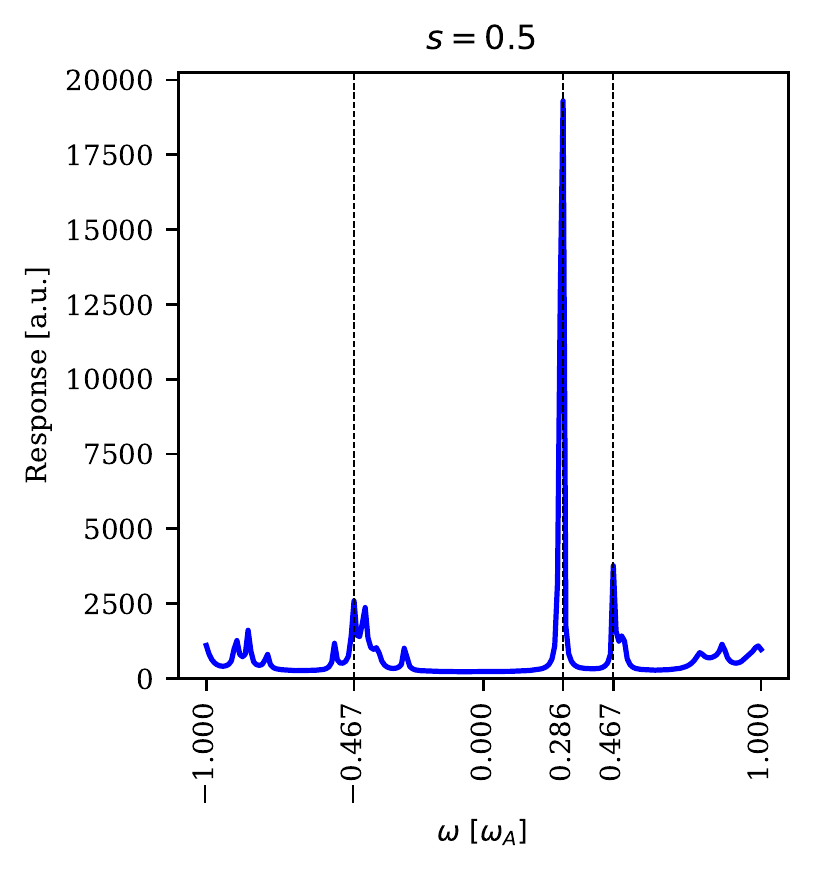}\\
  (a) $\omega_\mathrm{ant}=-0.28\ \omega_A$ & (b) $\omega_\mathrm{ant}=+0.28\ \omega_A$
  	\end{tabular}
  \caption{
  Frequency scan of $\phi$ at $s=1/2$ for the damping $n=2$ TAE mode for $t>254.5\ \omega_{A}^{-1}$ using DMUSIC method. The mode was initially excited by the antenna with the frequency (a) $\omega_\mathrm{ant}=-0.28\ \omega_A$ and (b) $\omega_\mathrm{ant}=0.28\ \omega_A$ for $t<254.5\ \omega_{A}^{-1}$.
  }
  \label{fig:damping_frequency_n2}
\end{figure}

\ \\ \ \\
After exciting the TAE mode with the antenna in  linear simulations, we turn off the antenna and measure the damping rate as well as the dominant frequency of the system. 
\ \\ \ \\
\noindent Considering the similar simulation setting as section~\ref{sec:excite_TAE_linear}, we deploy the electromagnetic antenna in the linear pullback scheme with the linear plasma response in two series of simulations; one with $\omega_\mathrm{ant}=+ 0.00143\ \omega_{ci} \approx +0.28\ \omega_A$ and the other with $\omega_\mathrm{ant}=- 0.00143\ \omega_{ci}\approx-0.28\ \omega_A$,  until $t=50000\ \omega_{ci}^{-1} \approx 254.5\ \omega_{A}^{-1}$. Then, the simulations are continued without the antenna. As shown in the Figs.~\ref{fig:damping_rate}-\ref{fig:damping_rate_n2} for the $n=6$ and the $n=2$ TAE modes, respectively, the potentials damp at a slightly faster rate with $\omega_\mathrm{ant}>0$ compared to $\omega_\mathrm{ant}<0$ which has been reported previously by \cite{kleiber2018global}. We recall that the TAE modes destabilized by  fast ions have a negative frequency, which corresponds to modes propagating in the ion diamagnetic direction. Also we note that our results are in agreement with the measured Landau damping results reported in \cite{vannini2020gyrokinetic} for electrons at temperature of $1\ \mathrm{keV}$.
\ \\ \ \\
Furthermore, we deployed the 
DMUSIC in order to numerically evaluate the frequency of the damped modes. As shown in Figs.~\ref{fig:damping_frequency_n6}-\ref{fig:damping_frequency_n2}, the dominant frequency of the system while the antenna is turned off is close to the target frequency, i.e., $\approx \pm 0.28\ \omega_A$, which confirms further that the antenna has excited the mode of interest. 

\FloatBarrier
\subsubsection{Fast particle simulations using initial condition obtained by antenna}
\label{sec:ant_fast}
\begin{figure}
  	\centering
	\begin{tabular}{cc}
  \includegraphics[scale=0.8]{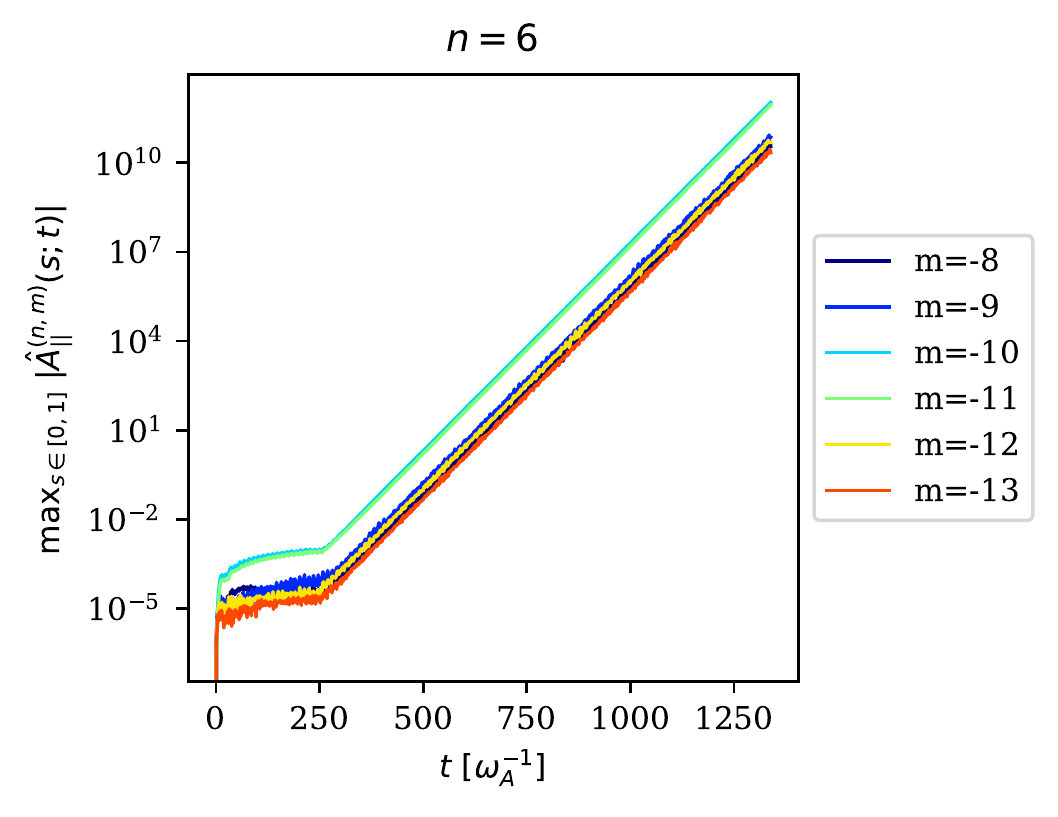}
  & \includegraphics[scale=0.8]{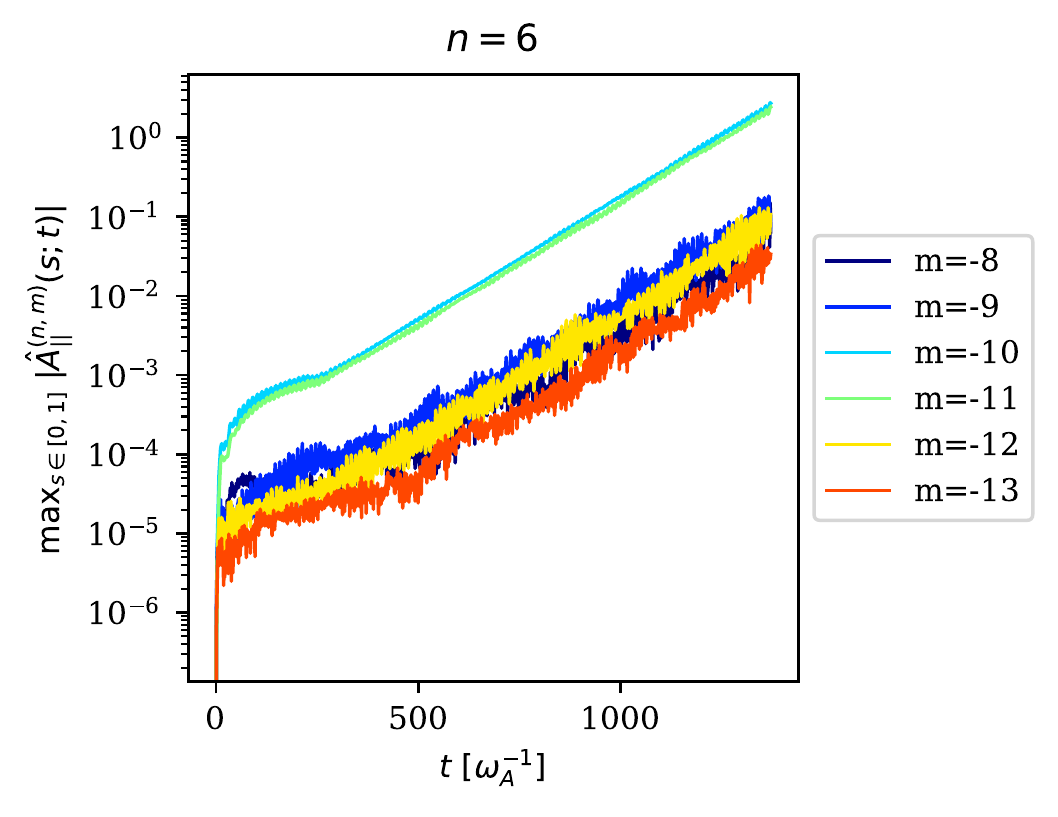}\\
  (a)  & (b)\\
   \includegraphics[scale=0.8]{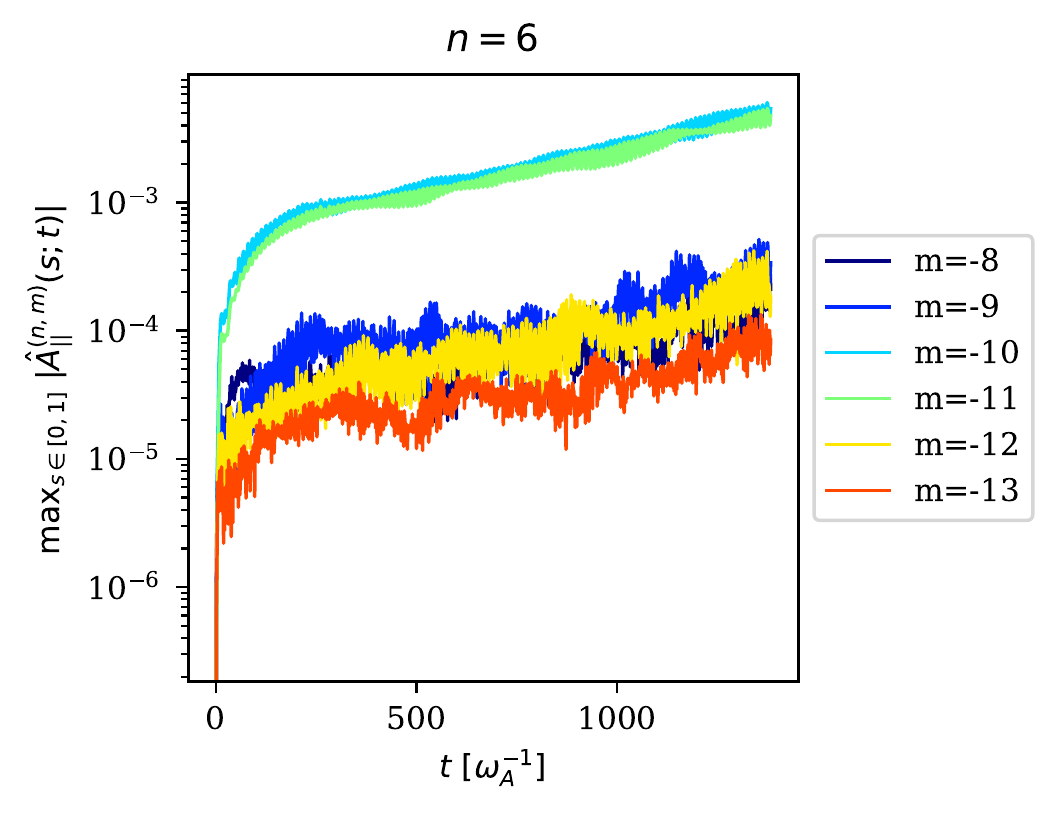}
  & \includegraphics[scale=0.8]{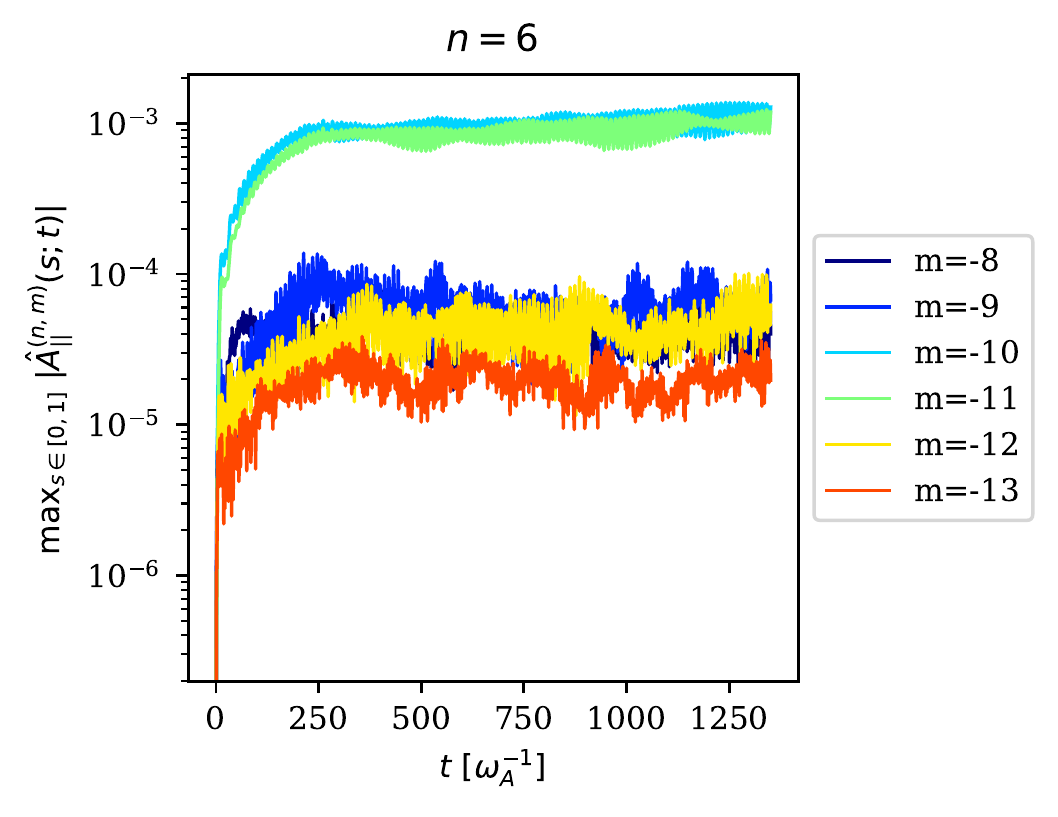}
  \\
  (c)  & (d)
  \end{tabular}
  \caption{Evolution of Fourier coefficients $|\hat{A}_{||}^{(n,m)}|$ against time. The $n=6$ TAE  mode is excited in $t\in[0,254.5]\ \omega_{A}^{-1}$ using antenna with the frequency $\omega_\mathrm{ant}=-0.00143\ \omega_{ci}\approx -0.28\ \omega_A$ while fast particles with a small density $n_\mathrm{0f}=10^{-6}$ is incorporated in the simulation. Then, considering the resulting particle positions and velocities as the initial condition, the simulation is continued  by increasing the density of fast particles to $n_\mathrm{0f}=n_0,n_0/4,n_0/16\ \text{and}\  n_0/32$ where $n_0=0.0031$ in (a), (b), (c) and (d), respectively, for $t>254.5\ \omega_{A}^{-1}$.
  }
  \label{fig:ant_freq_-TAE_fast}
\end{figure}

\begin{figure}
  	\centering
  \includegraphics[scale=1.0]{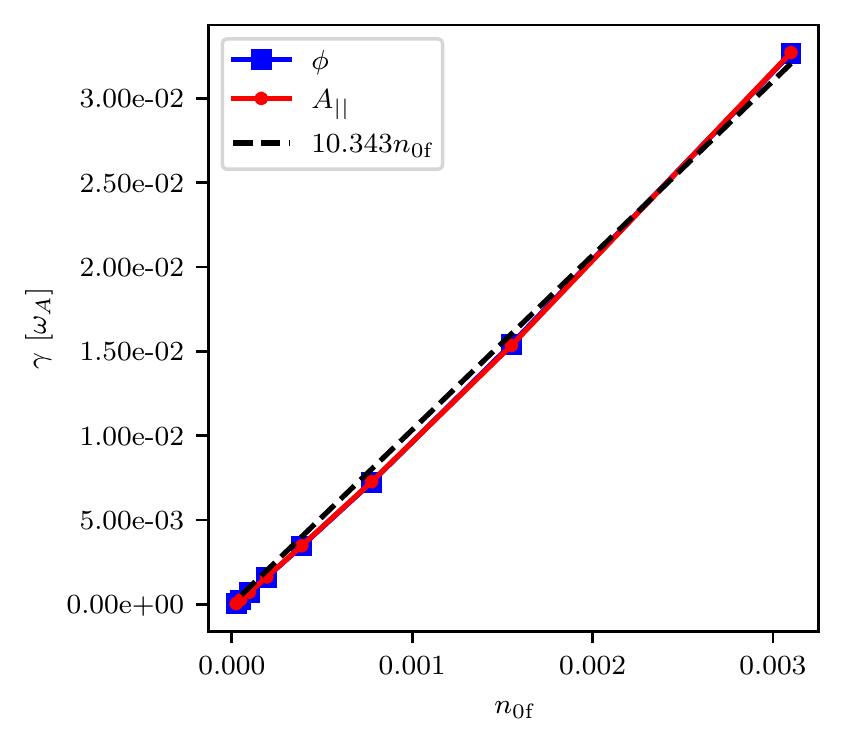}
  \caption{Growth rate $\gamma$ of the $n=6$ TAE mode against the normalized density fraction $n_{0\mathrm f}$ of fast particles is shown  where $\gamma$ is obtained by fitting (dashed lines) $a e^{\gamma t}$ to the corresponding Fourier coefficients of  $\phi$ and $A_{||}$ integrated in the radial direction for $t\in [500,1000]\ \omega_{A}^{-1}$. Here, the $n=6$ TAE mode is initially excited by the electromagnetic antenna with frequency $\omega_\mathrm{ant}=-0.28\ \omega_A$ for $t\in [0,254.5]\ \omega_{A}^{-1}$.}
  \label{fig:growth_rate}
\end{figure}

\begin{figure}
  	\centering
	\begin{tabular}{cc}
  \includegraphics[scale=0.8]{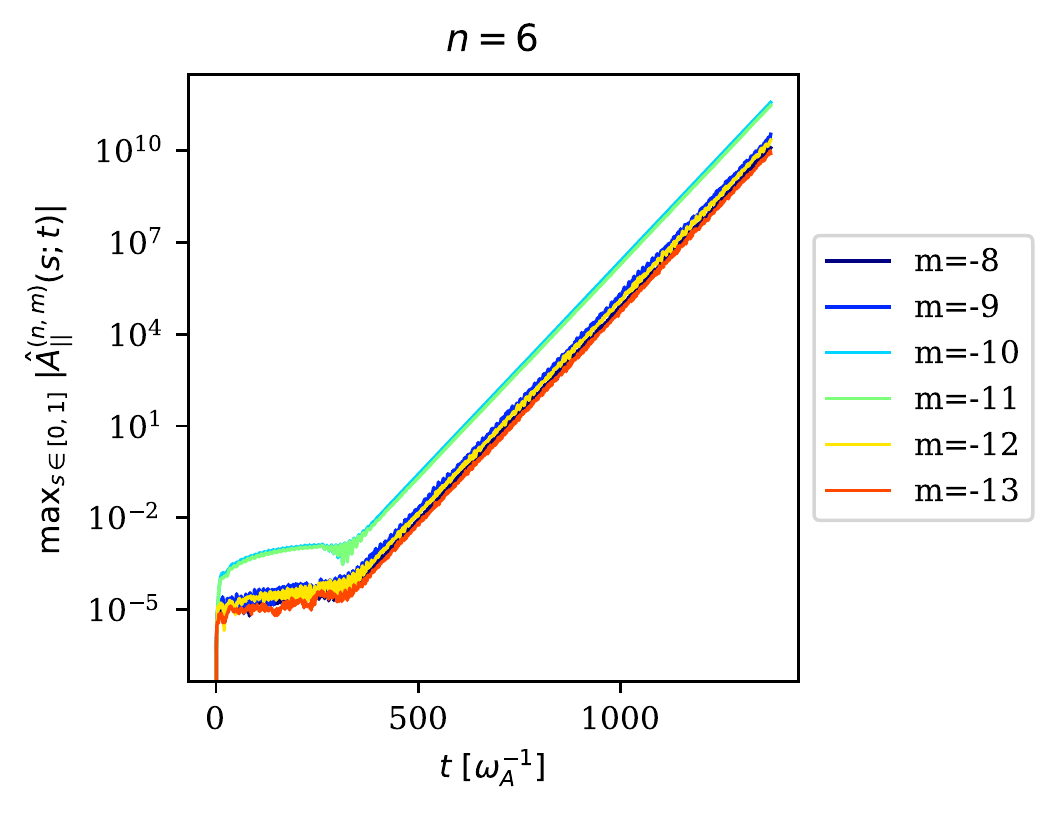}
  & \includegraphics[scale=0.8]{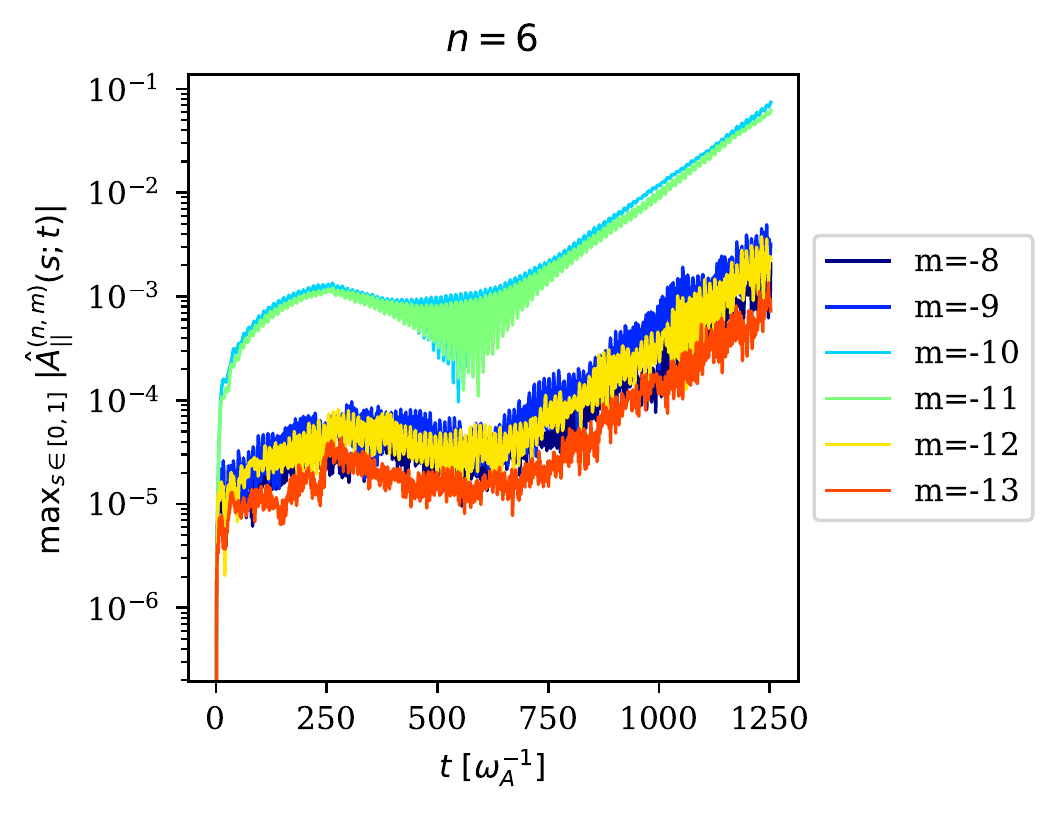}\\
  (a)  & (b)\\
   \includegraphics[scale=0.8]{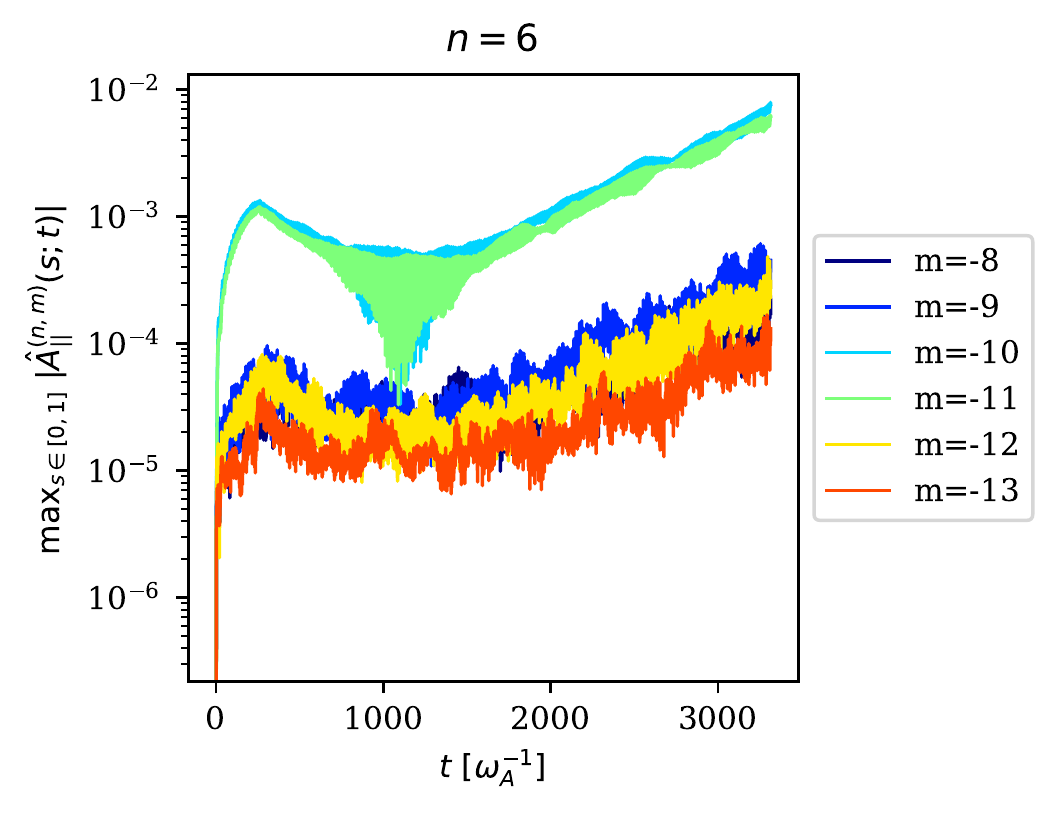}
  & \includegraphics[scale=0.8]{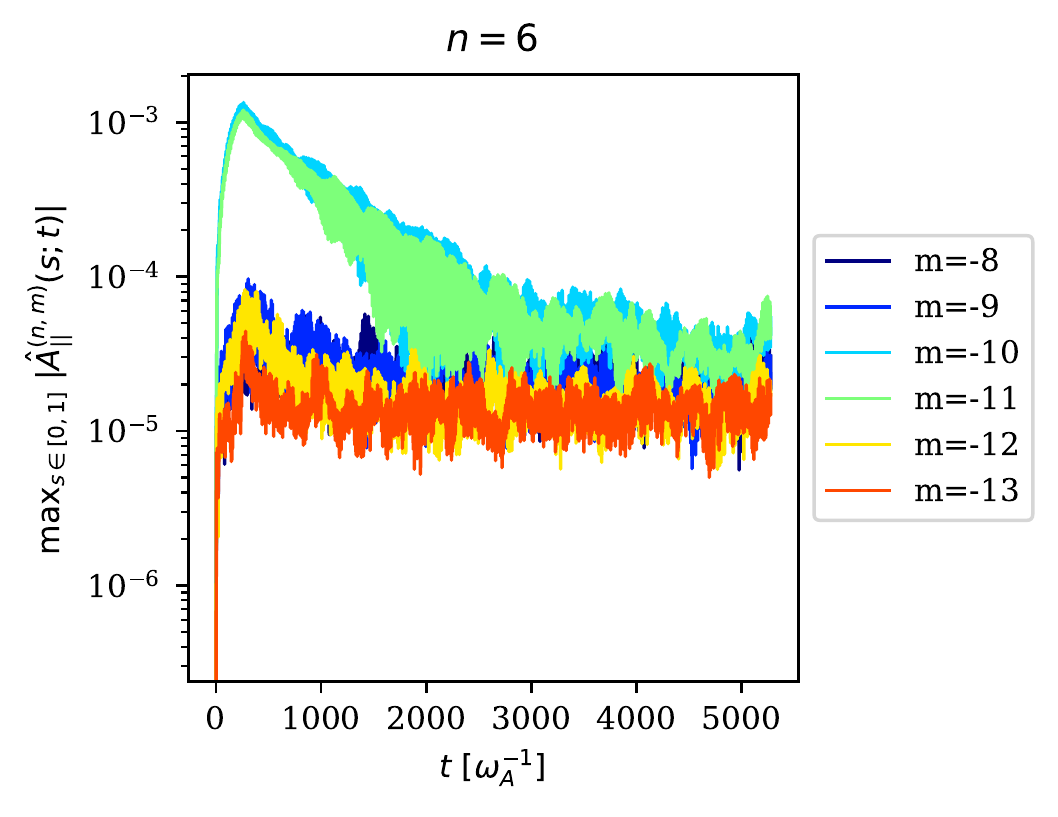}
  \\
  (c)  & (d)
  \end{tabular}
  \caption{Evolution of Fourier coefficients $|\hat{A}_{||}^{(n,m)}|$ against time. The $n=6$ TAE  mode is excited in $t < 254.5\ \omega_{A}^{-1}$ using antenna with the frequency $\omega_\mathrm{ant}=+0.00143\ \omega_{ci}\approx +0.28\ \omega_A$ while fast particles with a small density $n_\mathrm{0f}=10^{-6}$ is incorporated. Then, considering the resulting particle positions and velocities as the initial condition, the simulation is continued  by increasing the density of  fast particles to $n_\mathrm{f}=n_0,n_0/4,n_0/16\ \text{and}\  n_0/32$ where $n_0=0.0031$ in (a), (b), (c) and (d), respectively, for $t>254.5\ \omega_{A}^{-1}$. 
  }
  \label{fig:ant_freq_+TAE_fast}
\end{figure}

\begin{figure}
  	\centering
	\begin{tabular}{cc}
  \includegraphics[scale=1.0]{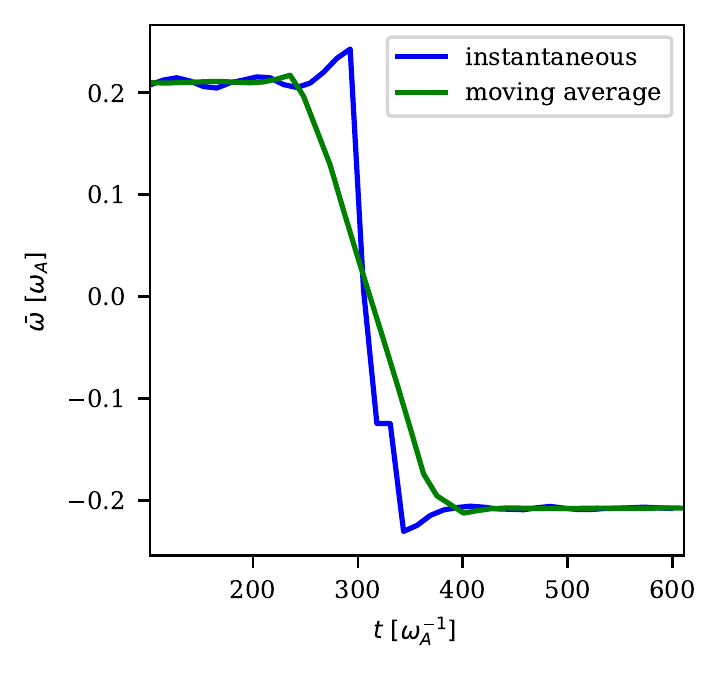}
  & \includegraphics[scale=1.0]{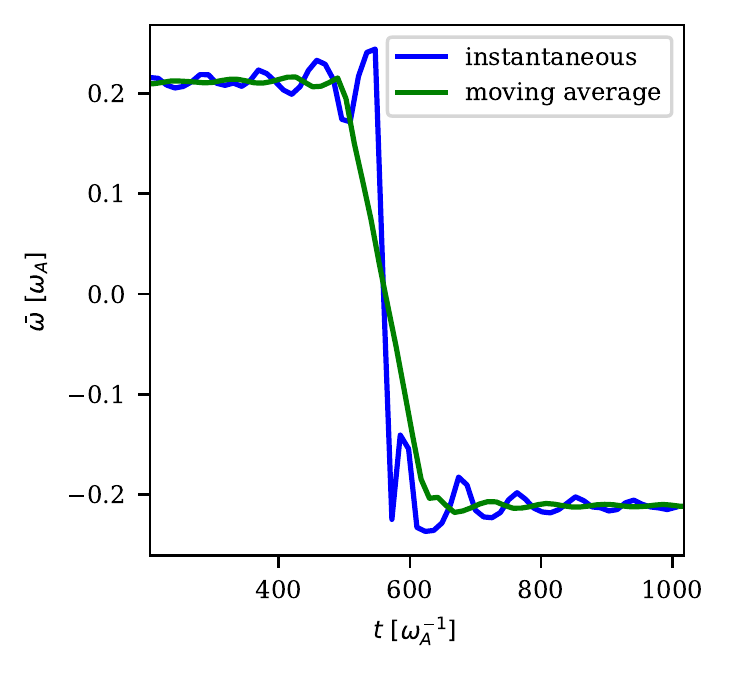}\\
  (a) $n_\mathrm{0f}=n_0$ for $t>254.5\  \omega_{A}^{-1}$ & (b) $n_\mathrm{0f}=n_0/4$ for $t>254.5\  \omega_{A}^{-1}$\\
   \includegraphics[scale=1.0]{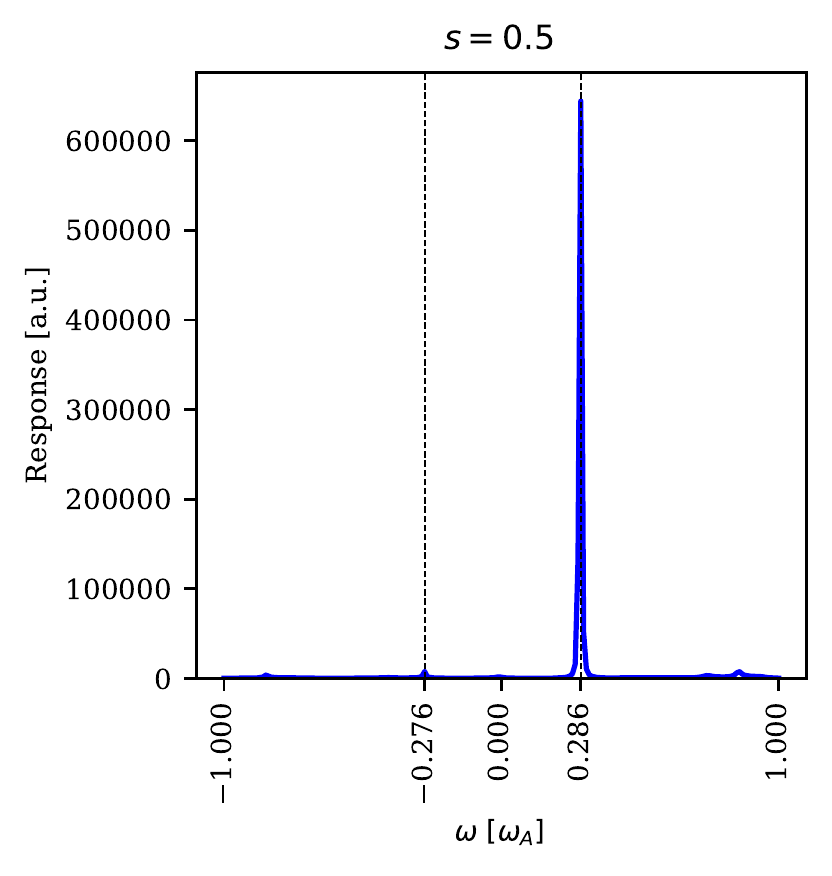}
  & \includegraphics[scale=1.0]{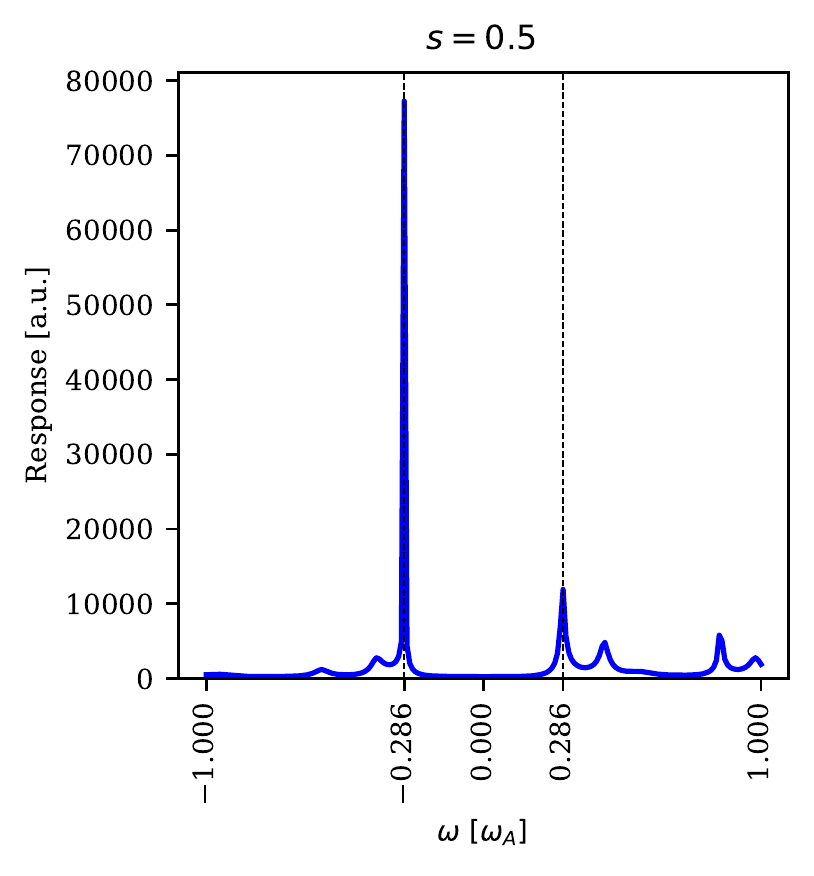}\\
  (c) $n_\mathrm{0f}=10^{-6}$ and $t<254.5\ \omega_{A}^{-1}$  & (d) $n_\mathrm{0f}=0.0031$ and $t>254.5\ \omega_{A}^{-1}$
  \end{tabular}
  \caption{
  Change of direction in the rotation of plasma where excited $n=6$ TAE mode with the frequency $\omega=+0.28\ \omega_A$ is overcomed by the dominant mode with the frequency $\omega=-0.28\ \omega_A$. 
  The frequency at which potentials rotate for the simulations of fast particles with density $n_\mathrm{0f}=n_0$ and $n_\mathrm{0f}=n_0/4$ with $n_0=0.0031$ are shown in (a) and (b), respectively. Here, the rotation frequency $\bar{\omega}$ of the field is computed instantaneously by computing the change in the phase of Fourier coefficient of $\phi$ at $s=0.5$.
  The DMUSIC frequency scan of the potential of the case (a), shown in  (c) before and (d) after injection of fast particles, shows a clear change of sign in the frequency.}
  \label{fig:ant_freq_minusTAE_fast_change_dir}
\end{figure}

\ \\ \ \\
In this section, we study the growth rate of modes where fast particles are considered in the plasma. Instead of starting from an arbitrary initial condition, here we perform a preparation step where we excite the mode of interest using the antenna where the fast particles are included in the simulation with a negligibly small density fraction $n_\mathrm{0f}=10^{-6}$. Once the mode is excited, i.e., $t= 50000 \  \omega_{ci}^{-1}\approx 254.5\ \omega_{A}^{-1}$, we turn off the antenna and continue the simulation with the fast particles at a range of density. This test case is intended to confirm that the TAE mode is excited correctly with the antenna. Here, we target the $n=6$ TAE mode.
\ \\ \ \\
\noindent As shown in Fig.~\ref{fig:ant_freq_-TAE_fast}, there is a clear exponential growth of the TAE mode for $t>254.5\ \omega_{A}^{-1}$ as antenna with $\omega_\mathrm{ant}=-0.00143\ \omega_{ci}\approx -0.28\  \omega_A<0$ is turned off and fast particles are enabled in the simulation by considering number density fraction $n_\mathrm{0f}>10^{-6}$. The measured growth rate seems to be a linear function of  fast particle number density
, see Fig.~\ref{fig:growth_rate}.
\ \\ \ \\
\noindent Furthermore, we can investigate how the system evolves as $\omega_\mathrm{ant}= 0.00143\ \omega_{ci} \approx 0.28\ \omega_A>0$ is used for excitation of the $n=6$ TAE mode for $t<254.5\ \omega_{A}^{-1}$. As smaller densities of fast particles are considered, the exponential growth of the TAE mode with the negative sign of frequency is postponed until it overcomes the damping rate of TAE mode with positive sign of frequency from the antenna phase, see Fig.~\ref{fig:ant_freq_+TAE_fast}. In fact, as the TAE mode with negative sign takes over, the direction at which potentials rotate reverses, see  Fig.~\ref{fig:ant_freq_minusTAE_fast_change_dir}.

\FloatBarrier
\subsection{Nonlinear simulations}
\label{sec:nonlin}
In this section, we deploy the antenna in order to excite TAE modes in the nonlinear setting, i.e., we solve the electromagnetic antenna adapted in the  equations of motion \eqref{eq:Rdot_antenna_nonlin}-\eqref{eq:epsilon_antenna_nonlin} which includes the zeroth and first-order plasma contribution Eqs.~\eqref{eq:R0}-\eqref{eq:v||0} and Eqs.~\eqref{eq:position_update_nonlin}-\eqref{eq:epsdot_update_nonlin} for the plasma. Here, the nonlinear pullback scheme equipped with the electromagnetic antenna where ITPA case is taken as the test case. The parameters of antenna are set similar to the linear simulations. Here, coupling among $n=0,...,18$ toroidal mode numbers with poloidal mode numbers satisfying $|m+\nint{nq}|\leq  \Delta m$  where $\Delta m=5$ was considered.
\\ \ \\
\FloatBarrier
\subsubsection{Excitation of the $n=2$ TAE mode in a nonlinear simulation}
\label{sec:nonlin_n=2}
\begin{figure}
  	\centering
	\begin{tabular}{cc}
  \includegraphics[scale=0.8]{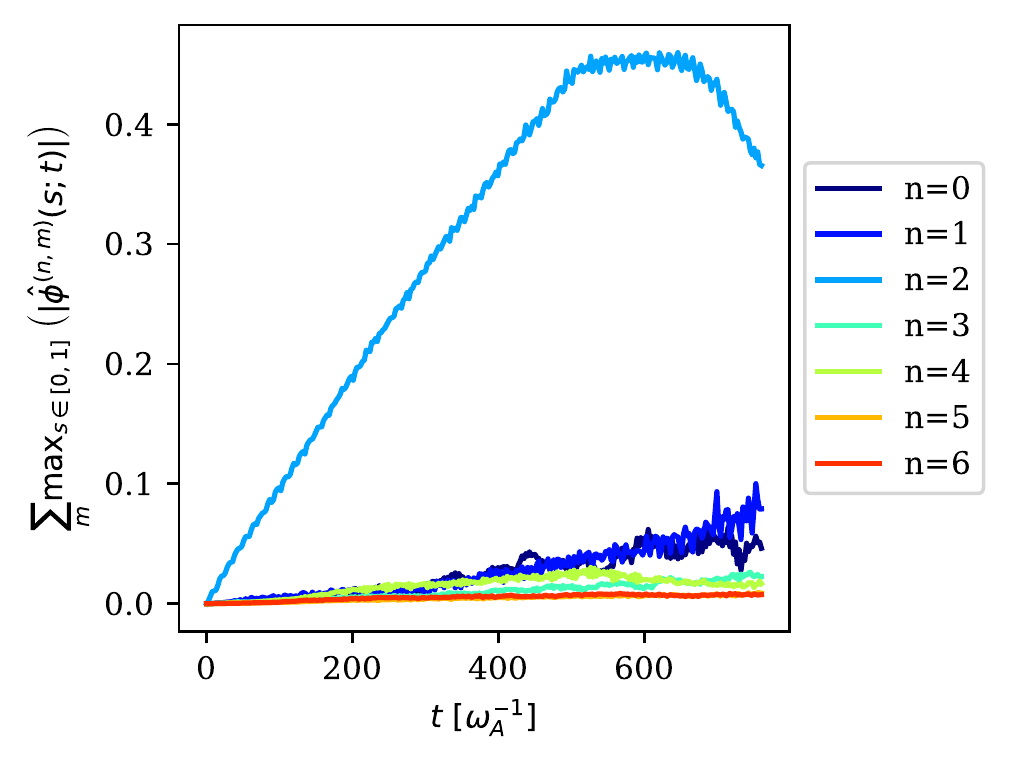}
  & \includegraphics[scale=0.8]{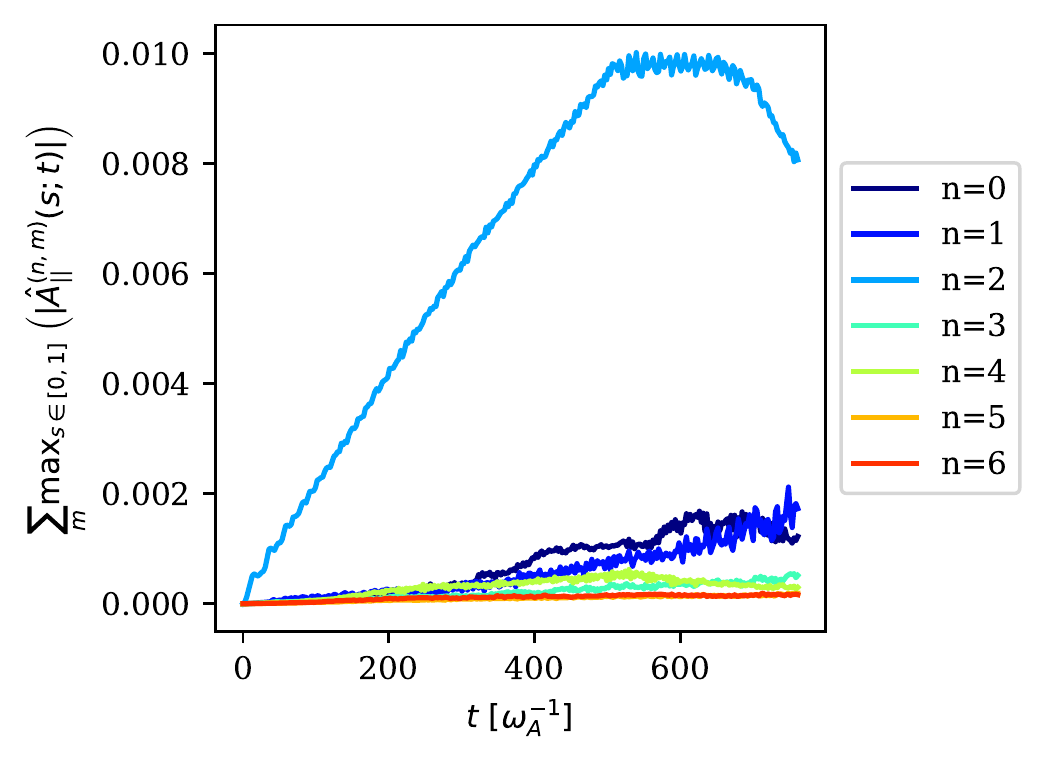}\\
  (a)  & (b)\\
   \includegraphics[scale=0.8]{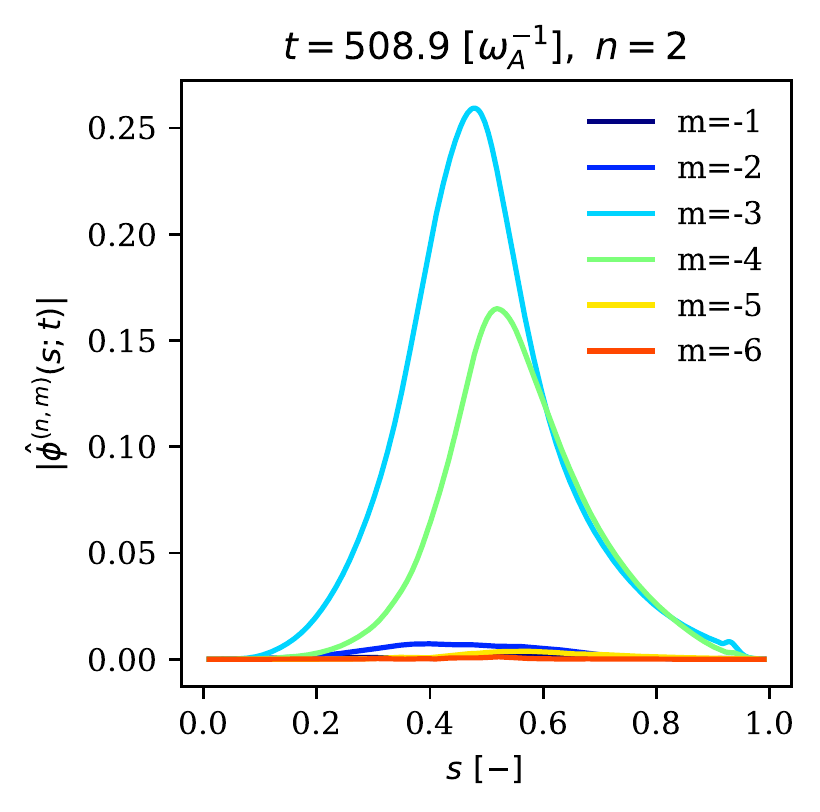}
  & \includegraphics[scale=0.8]{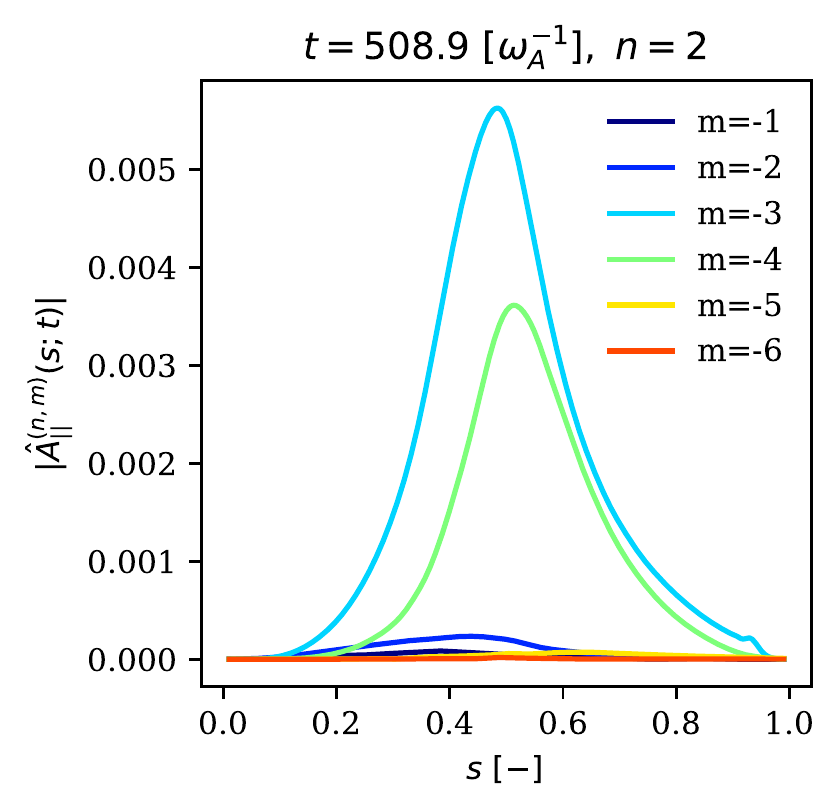}
  \\
  (c)  & (d)\\
  \includegraphics[scale=0.8]{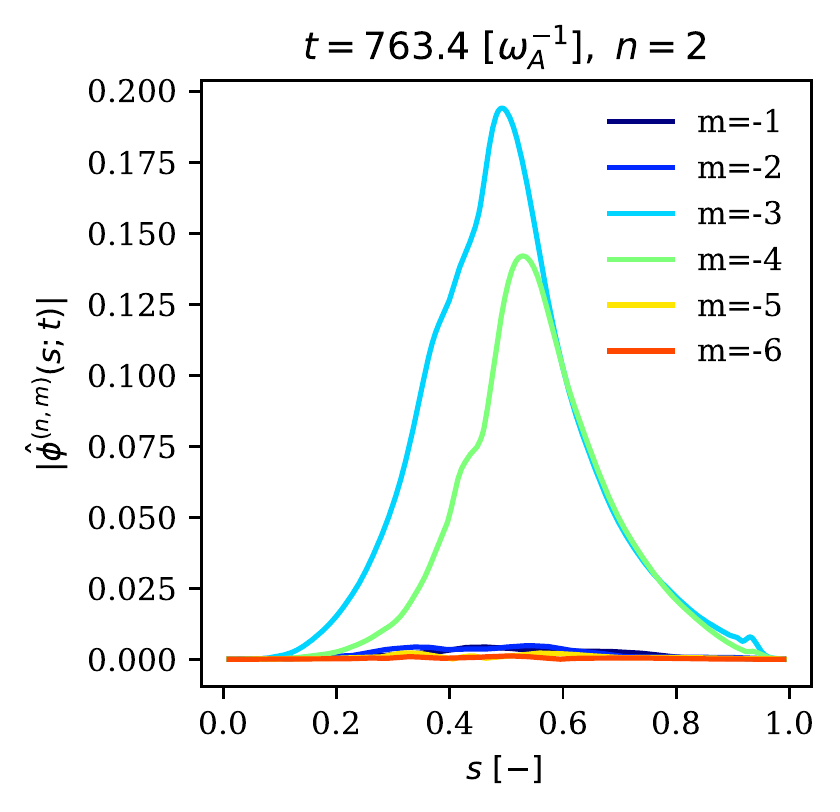}
  & \includegraphics[scale=0.8]{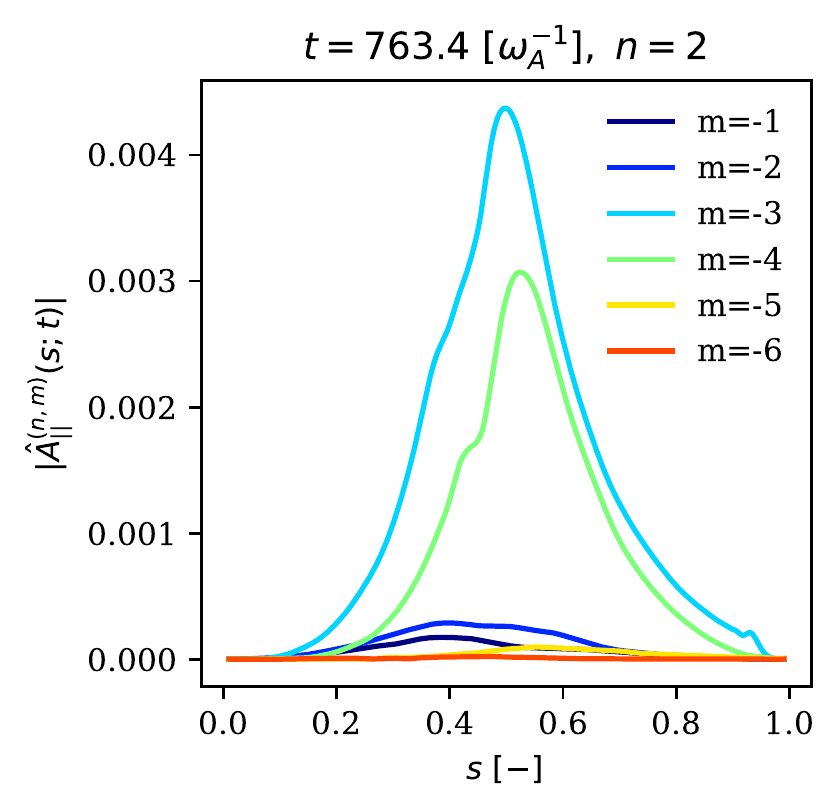}
  \\
  (e)  & (f)
  \end{tabular}
  \caption{Nonlinear excitation of TAEs using the electromagnetic antenna with $n=2,\ m=-3,-4$ and the frequency $\omega_\mathrm{ant}=-0.28\ \omega_A$. 
 Here, time traces of $\phi$ and $A_{||}$ is shown in (a) and (b), respectively.
  The radial profiles of these modes at the time of turning off the antenna $t=508.9\ \omega_{A}^{-1}$ (c)-(d) and after a period of relaxation at time $t=763.4\ \omega_{A}^{-1}$ are shown in (e)-(f).}
  \label{fig:ant_nonlinear_n2}
\end{figure}

\begin{figure}
  	\centering
	\begin{tabular}{cc}
	 \includegraphics[scale=0.9]{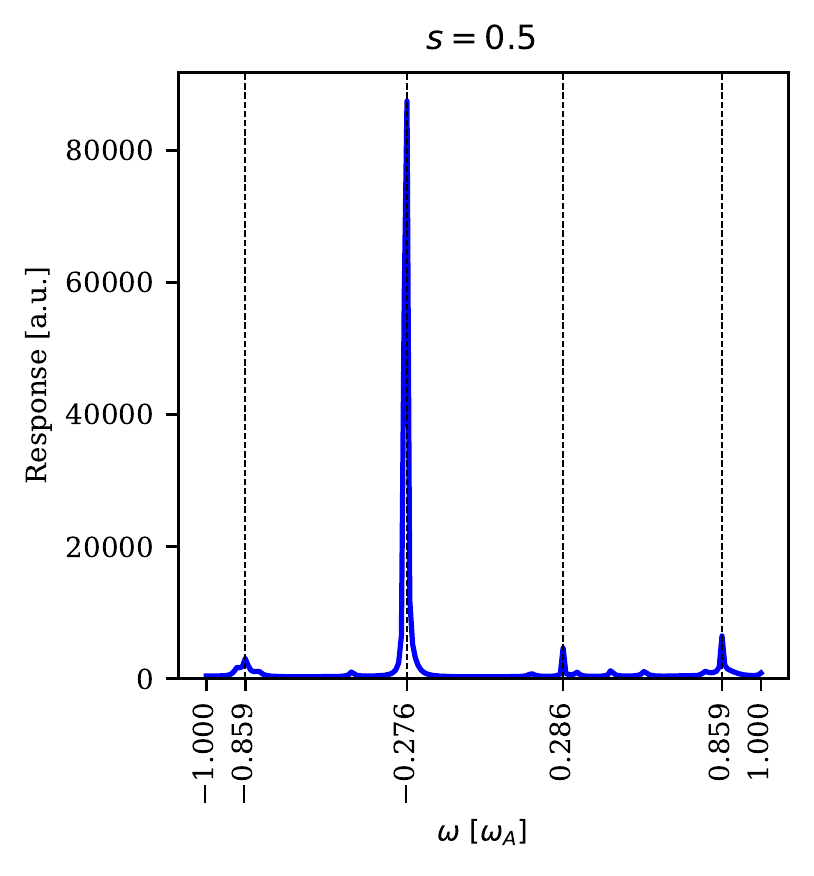}
	 &\includegraphics[scale=0.9]{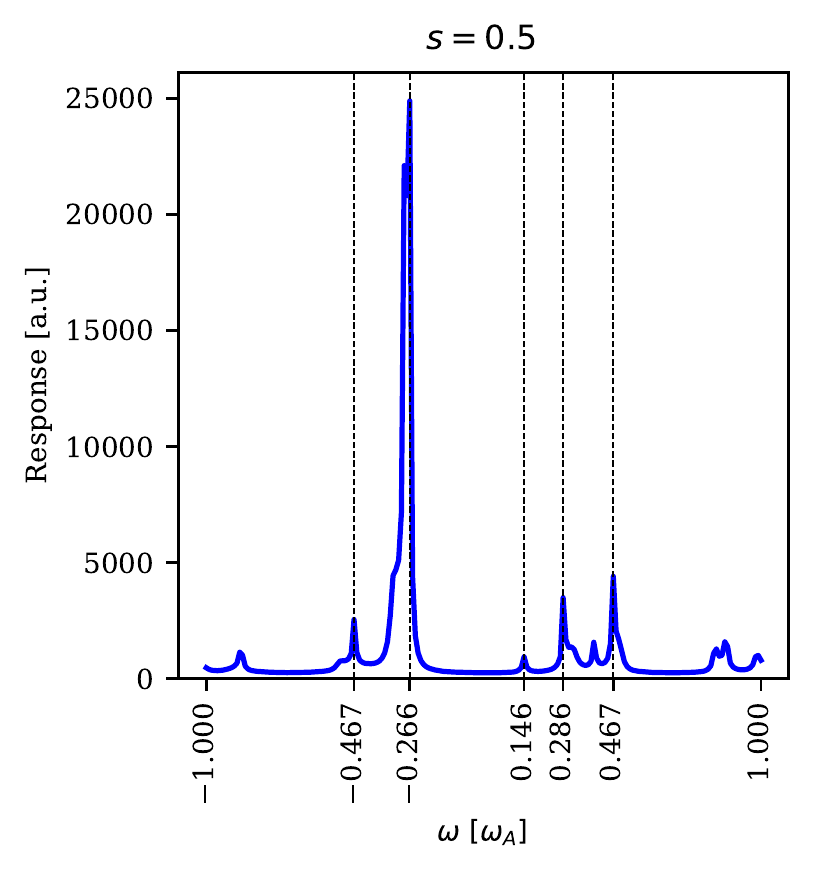}\\
  (a) $n=2$, $t<508.9\  \omega_{A}^{-1}$  & (b) $n=2$, $t>508.9\  \omega_{A}^{-1}$
  \end{tabular}
  \caption{Frequency scan of $\phi$ for $n=2$ toroidal mode number (a) during the antenna excitation $t<508.9\ \omega_{A}^{-1}$ and (b) during the damping phase with turned off antenna $t>508.9\ \omega_{A}^{-1}$ using DMUSIC for the nonlinear simulations where the $n=2$ TAE mode is the target of the  antenna.}
  \label{fig:freq_scan_nonlinear_n2_phi}
\end{figure}

\ \\ \ \\
\noindent 
In this section, we study nonlinear simulation of plasma in the ITPA test case equipped with the electromagnetic antenna to excite a TAE mode of interest. The antenna is set to target the $n=2$ TAE mode with similar parameters as the one for the linear simulation, i.e. the description given in section~\ref{sec:excite_TAE_linear}. Here, we consider $\omega_\mathrm{ant}=-0.28\ \omega_{ci}$ as the frequency of the antenna. The simulation has two phases. In the first phase, the $n=2$ TAE mode  is excited with the antenna till $t=100000\ \omega_{ci}^{-1}\approx 508.9\ \omega_A^{-1}$. Then, the antenna is switched off for the damping phase of the mode.
\\ \ \\
As depicted in Fig.~\ref{fig:ant_nonlinear_n2}, the $n=2$ TAE mode is excited clearly in the driving phase of the simulation, i.e., the $n=2$ and $m=-3, -4$ is the dominant mode numbers at $s=1/2$. However during the damping phase, the mode structure in the radial direction is slightly disturbed. A frequency analysis of the mode for before and after switch off shown in Fig.~\ref{fig:freq_scan_nonlinear_n2_phi} indicates that damping of the TAE mode allows other eigen frequencies of the system appear.

\FloatBarrier
\subsubsection{Excitation of TAE and GAE
modes in a nonlinear simulation by an $n=6$ TAE-like antenna}
\label{sec:nonlin_n=6}
\begin{figure}
  	\centering
	\begin{tabular}{cc}
  \includegraphics[scale=0.8]{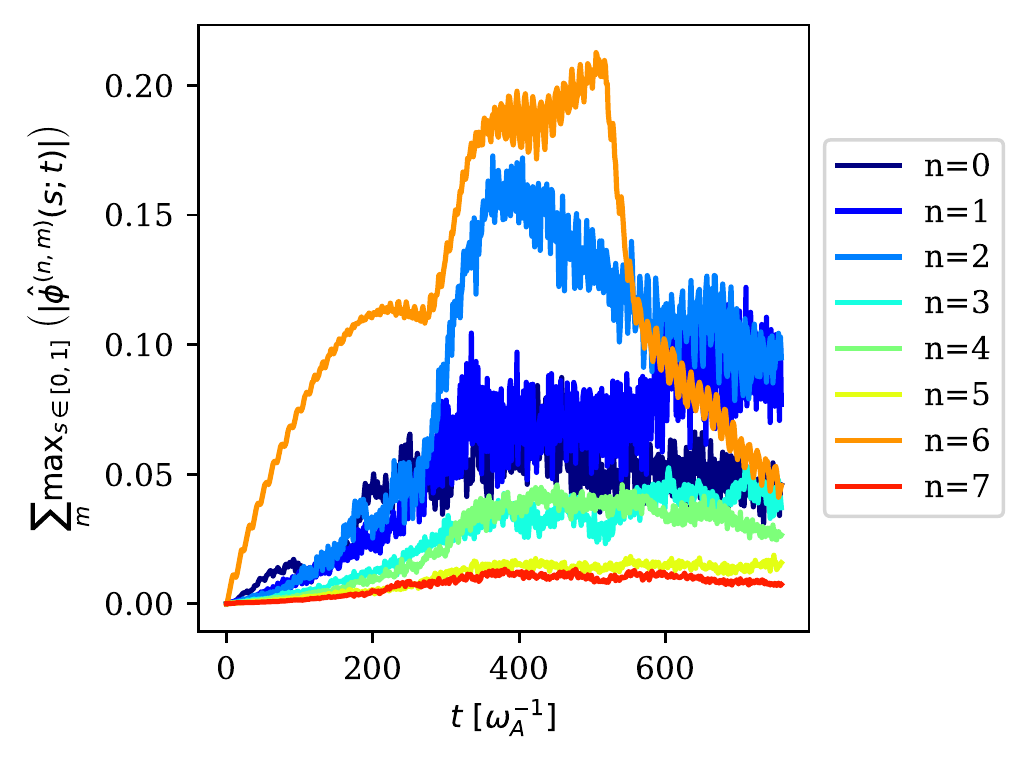}
  & \includegraphics[scale=0.8]{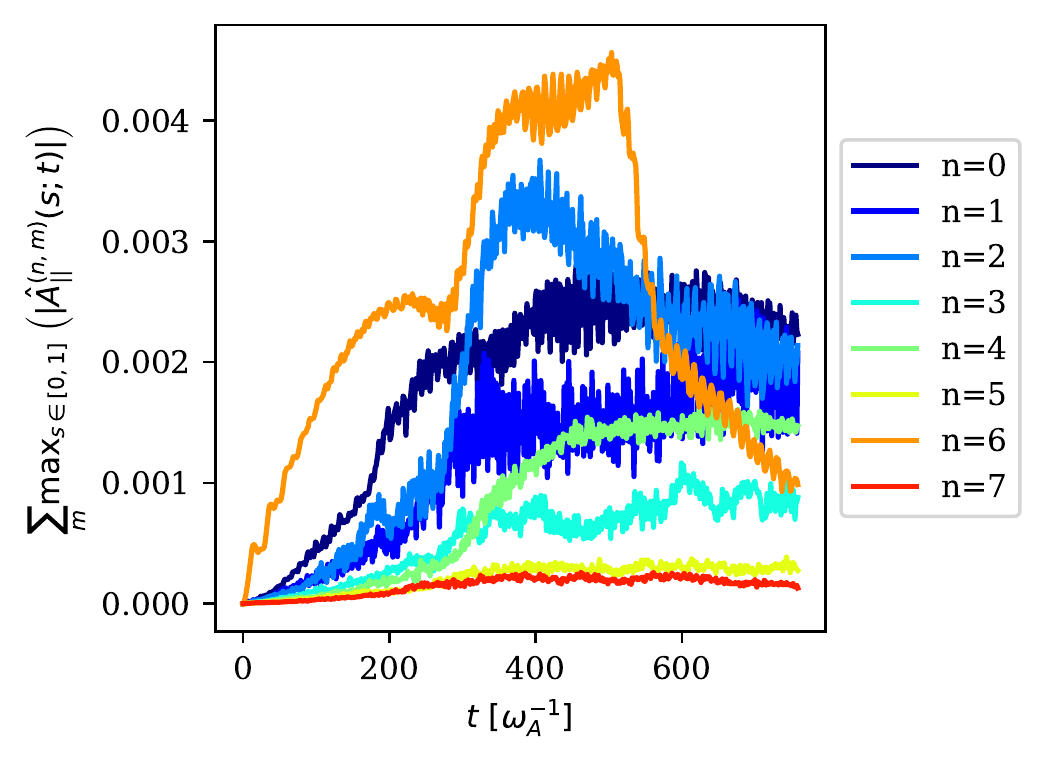}\\
  (a)  & (b)\\
  \includegraphics[scale=0.5]{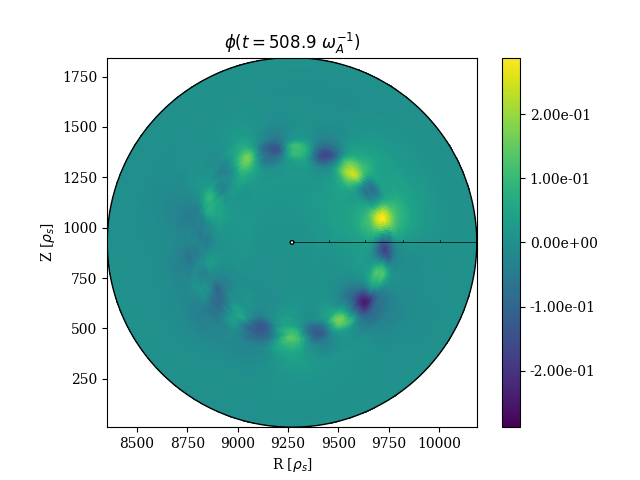}
  & \includegraphics[scale=0.5]{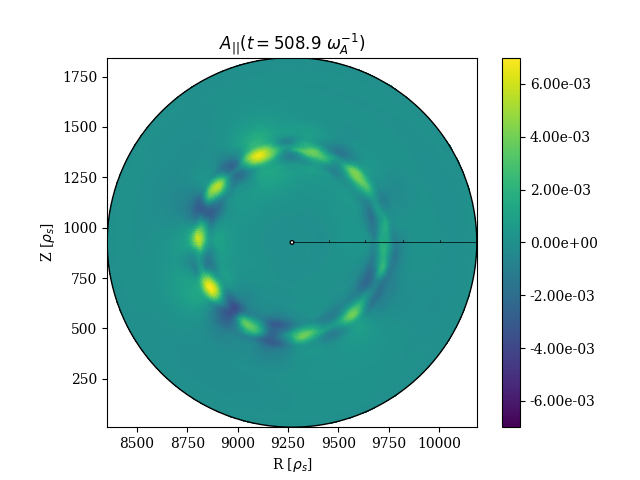}
  \\
  (c)  & (d)
  \end{tabular}
  \caption{Evolution of (a) $\phi$ and (b) $A_{||}$ for the nonlinear simulation of the ITPA case with antenna that is devised to excite the $n=6$ TAE mode. The electrostatic and magnetic potentials at $t=508.9\ \omega_{A}^{-1}$ in the poloidal plane is depicted in (c) and (d), respectively.}
  \label{fig:ant_nonlinear_n6}
\end{figure}

\begin{figure}
  	\centering
	\begin{tabular}{cc}
  \includegraphics[scale=0.8]{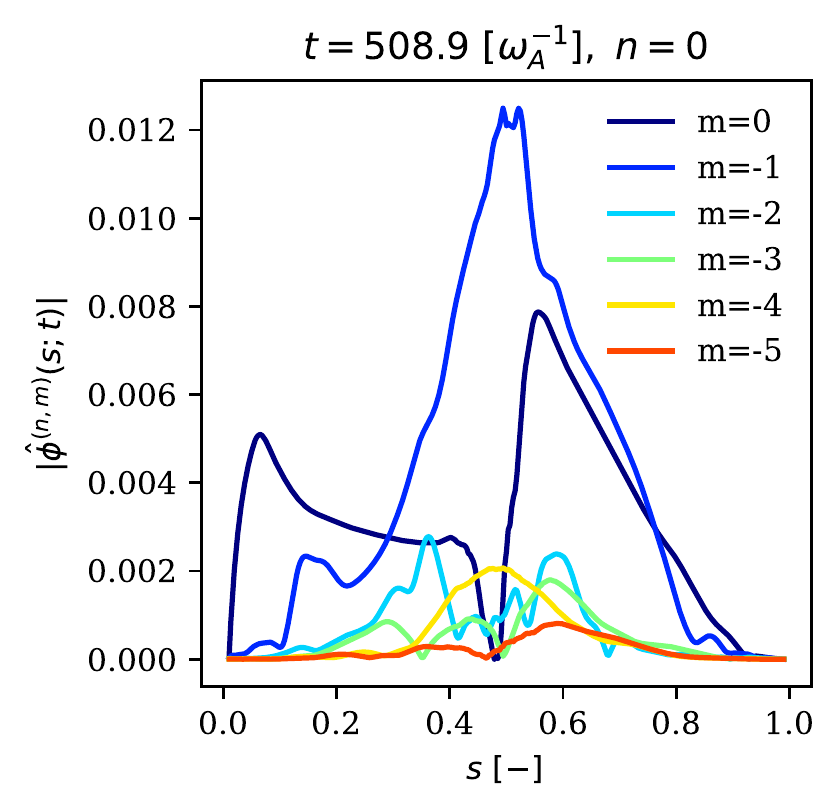}
  & \includegraphics[scale=0.8]{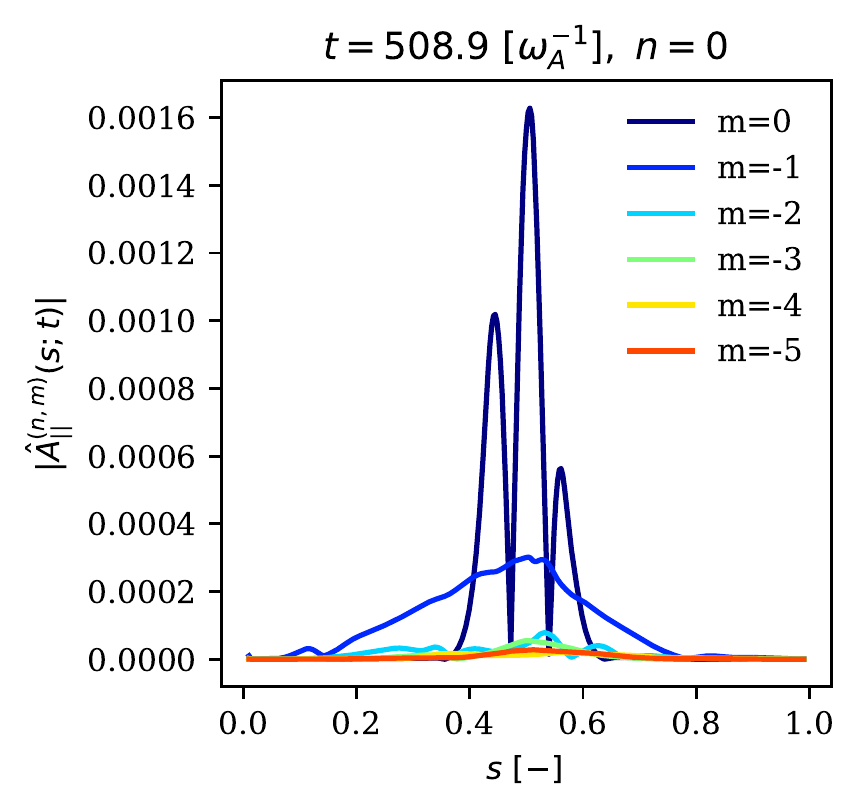}\\
  (a)  & (b)\\
  \includegraphics[scale=0.8]{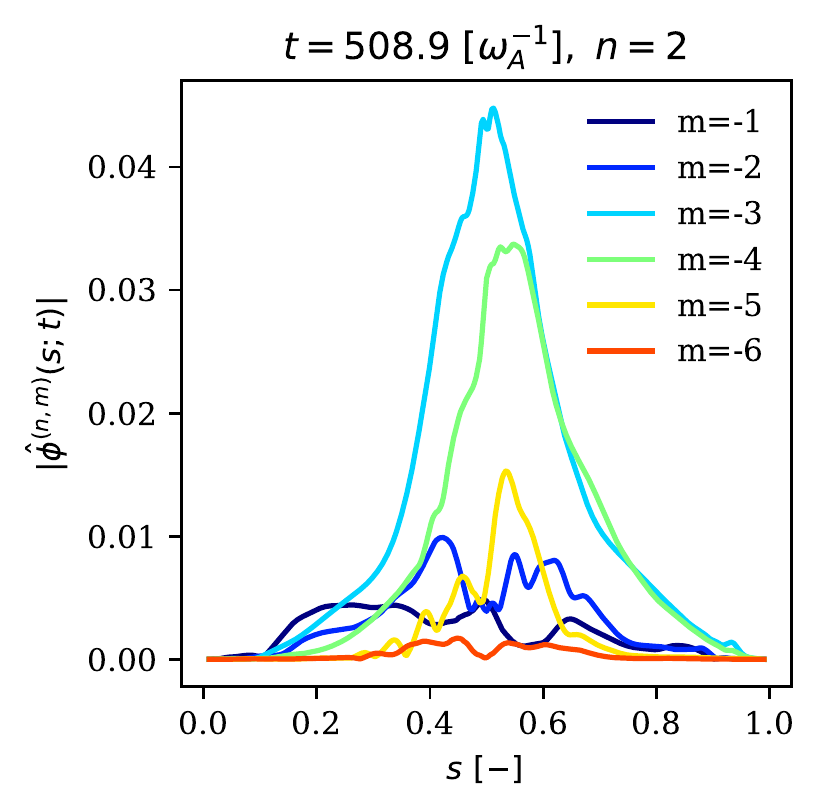}
  & \includegraphics[scale=0.8]{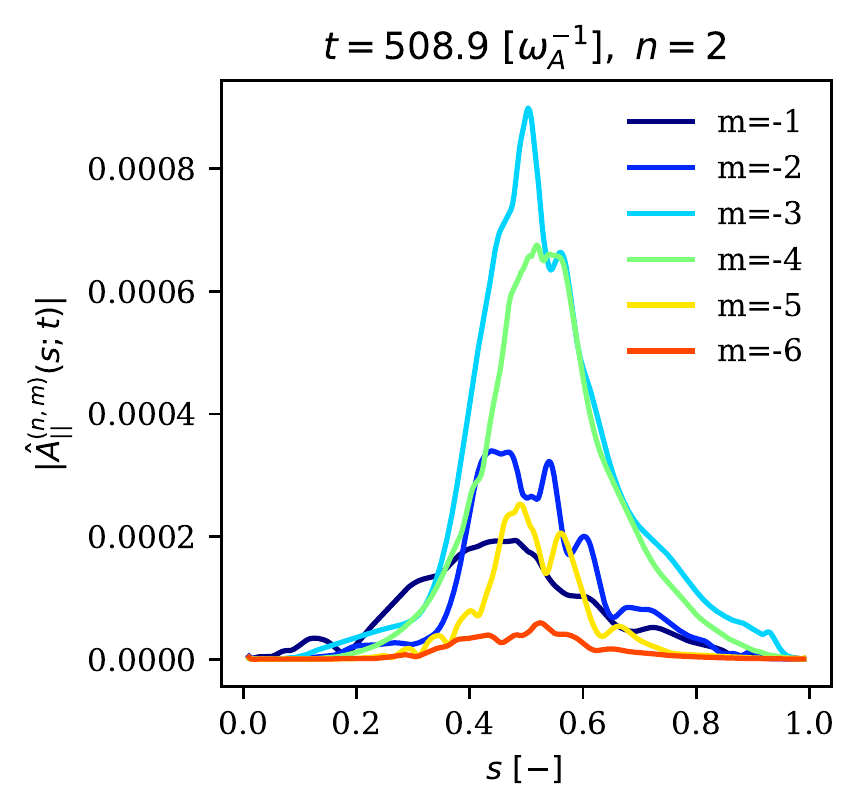}\\
  (c)  & (d)\\
  \includegraphics[scale=0.8]{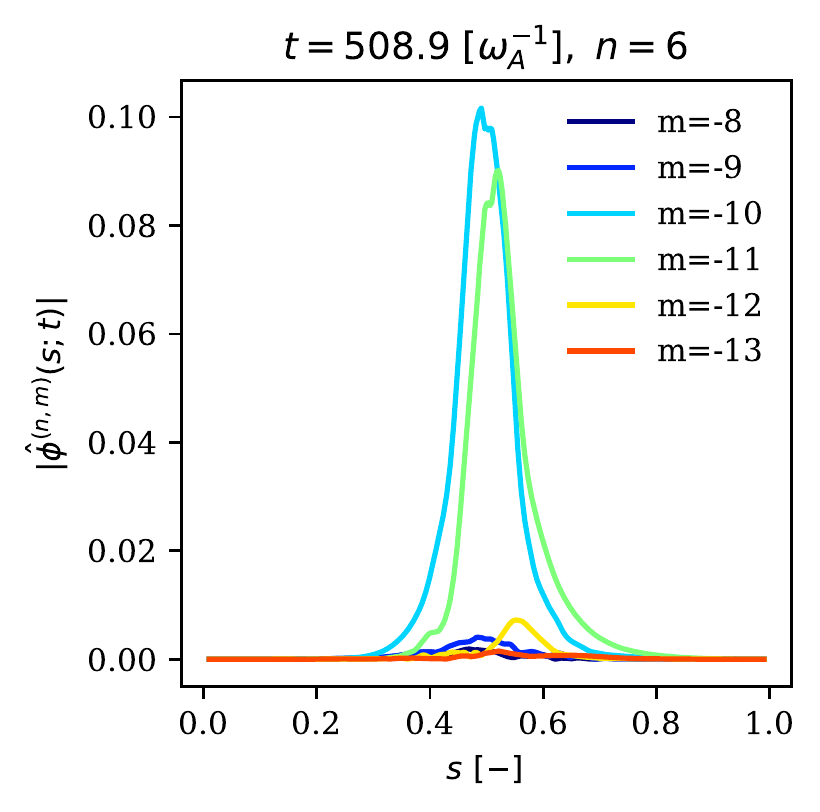}
  & \includegraphics[scale=0.8]{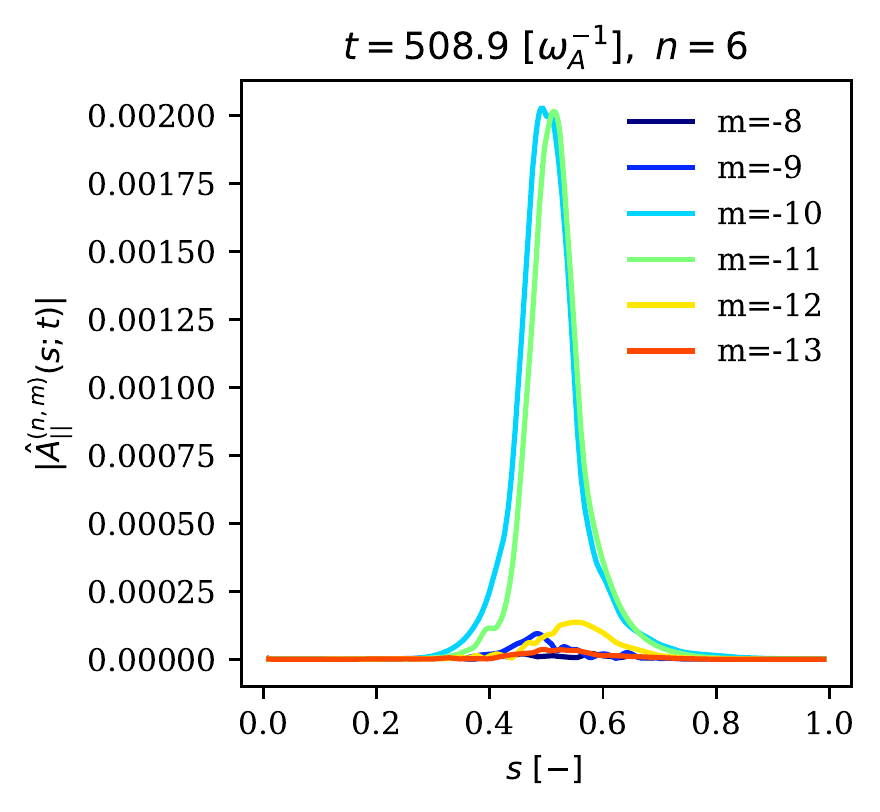}\\
  (e)  & (f)
  \end{tabular}
  \caption{Radial profile of $\phi$ and $A_{||}$ at $t=508.9\ \omega_{A}^{-1}$ for (a)-(b) $n=0$, (c)-(d) $n=2$ and (e)-(f) $n=6$, respectively, obtained from nonlinear simulation where the $n=6$ TAE mode is the target of the  antenna.}
  \label{fig:ant_nonlinear_n6_phi_apar}
\end{figure}

\begin{figure}
  	\centering
	\begin{tabular}{cc}
	 \includegraphics[scale=0.9]{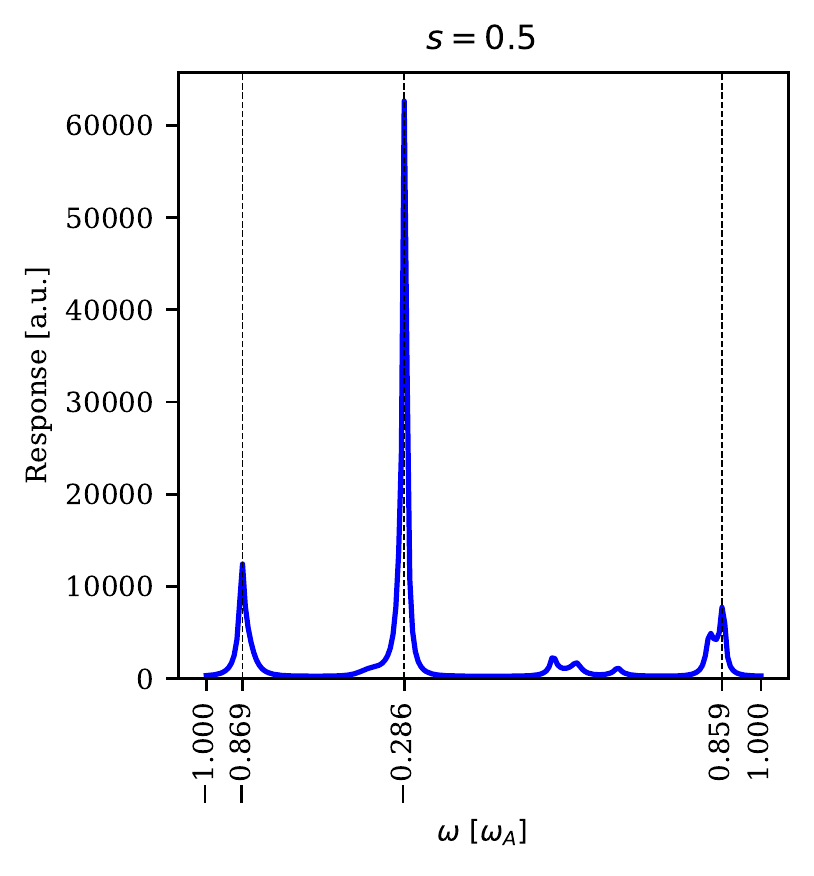}&\includegraphics[scale=0.9]{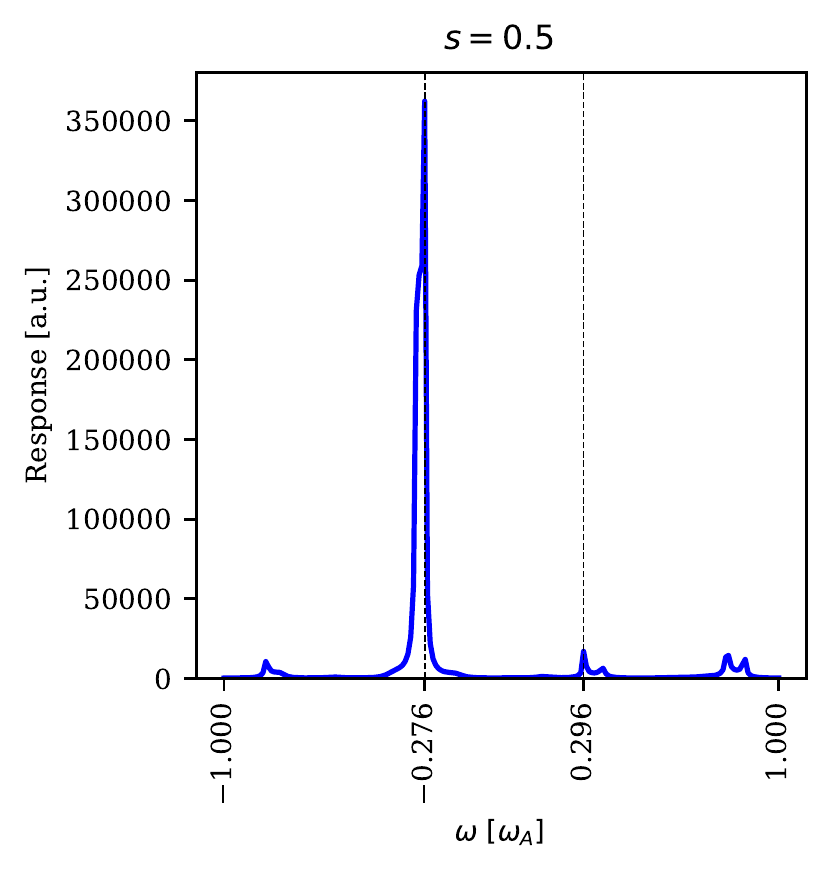}\\
  (a) $n=2$, $t>300\  \omega_{A}^{-1}$  & (b) $n=6$, $t>300\  \omega_{A}^{-1}$
  \\
  \includegraphics[scale=0.9]{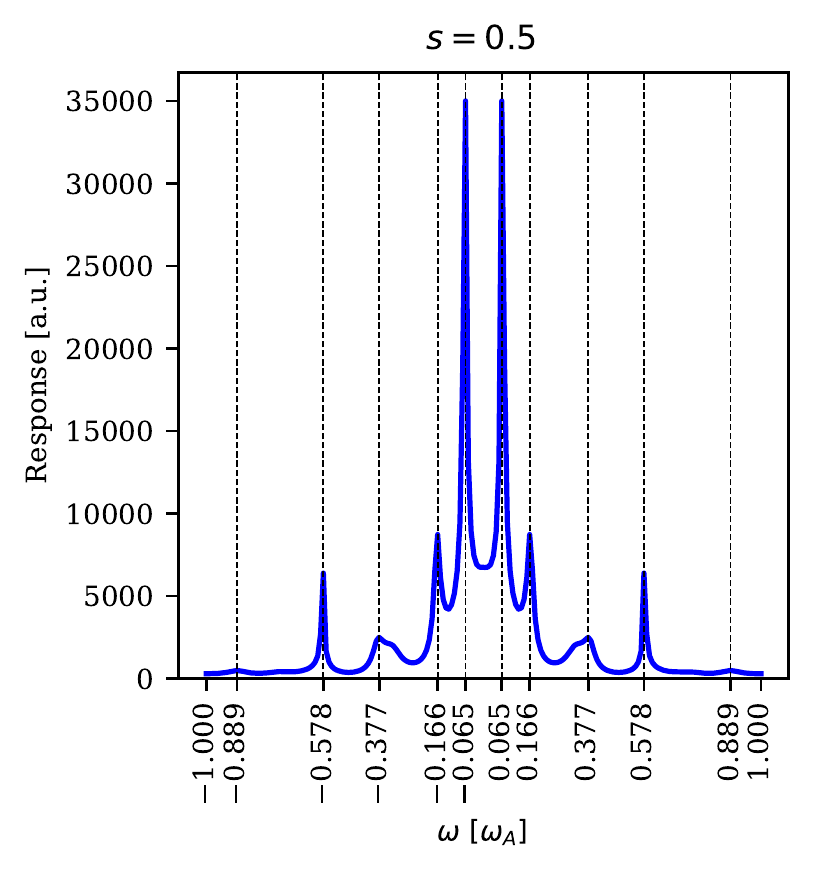}
  & \includegraphics[scale=0.9]{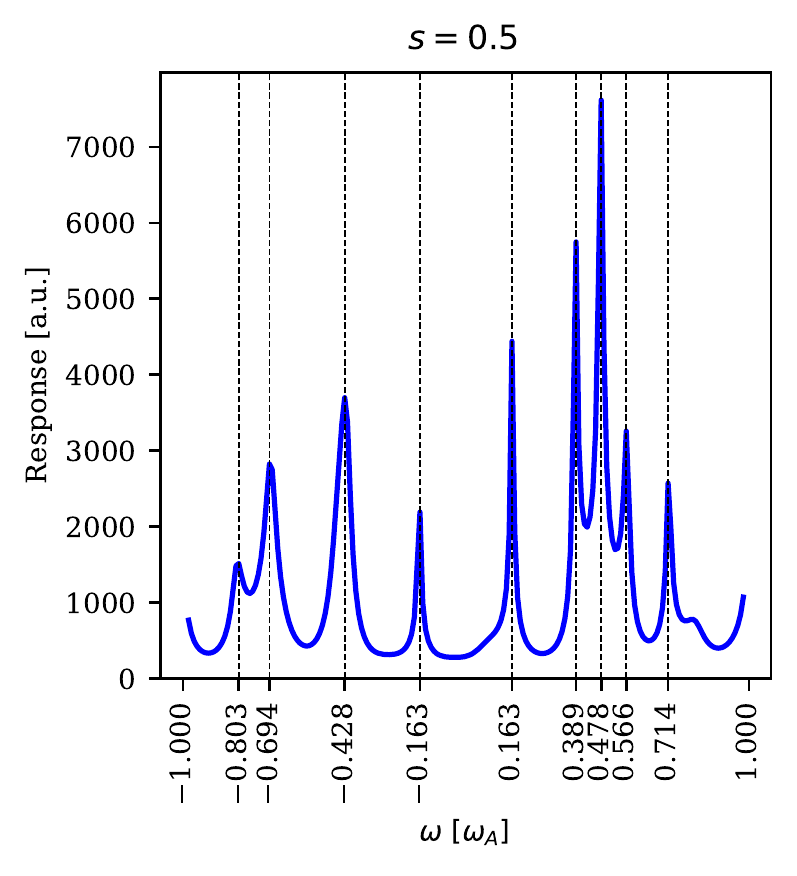}\\\
  (c) $n=0$, $t>300\  \omega_{A}^{-1}$ &  (d) $n=1$,  $t>300 \ \omega_{A}^{-1}$
  \end{tabular}
  \caption{Frequency scan of $\phi$ for (a) $n=2$, (b) $n=6$, (c) $n=0$, and (d) $n=1$ toroidal mode numbers using DMUSIC for the nonlinear simulations where the $n=6$ TAE mode is the target of the  antenna.}
  \label{fig:freq_scan_nonlinear_n6_phi}
\end{figure}

\ \\ \ \\
\noindent Although a clear excitation of the $n=2$ TAE mode was obtained with the antenna in the nonlinear setting in section \ref{sec:nonlin_n=2}, the plasma response seems different in the case of exciting the $n=6$ TAE mode. Here, we set the parameters of the antenna similar to section~\ref{sec:excite_TAE_linear} and study the plasma response in $A_{||}$ and $\phi$. The simulation has two steps. In the first part, the antenna is used to excite the desired mode and then, for $t>100000\ \omega_{ci}^{-1}\approx 508.9\ \omega_A^{-1}$ the antenna is turned off.
\\ \ \\
As shown in Figs.~\ref{fig:ant_nonlinear_n6}, \ref{fig:ant_nonlinear_n6_phi_apar}, and \ref{fig:freq_scan_nonlinear_n6_phi}, 
while the  $n=6$ TAE mode is excited as expected, several other modes appear to have been excited as well. Firstly, we observe that  the interplay of modes has lead to the excitation of the $n=2$ TAE mode along with toroidal mode numbers of  $n=0$ and $n=1$. 
The frequency analysis of $n=0$ (Fig.~\ref{fig:freq_scan_nonlinear_n6_phi} (c)) shows the nonlinear excitation of zonal structures, with frequencies very different from the antenna. The peaks at $\omega=\pm 0.578\ \omega_A$ correspond to a $m=\pm 1,\ n=0$ axisymmetric Global Alfv\'en Eigenmode (GAE) .
While those at $\omega=\pm 0.166\  \omega_A$ are of unclear origin and deserve further investigations. The frequency analysis of $n=1$, (Fig.~\ref{fig:freq_scan_nonlinear_n6_phi} (d)) also demonstrates the nonlinear excitation of frequencies different from the antenna. The frequency peak about $\omega=0.4 \omega_A$ corresponds to a $m=-1,n=1$ GAE.

\FloatBarrier
\section{Conclusion}
\label{sec:conclusion}
In this study, we provide a detailed description of an antenna's implementation in ORB5 which allows exciting an eigenmode of the system. First, the antenna is devised as an electrostatic potential, then the electromagnetic counter-part is computed using Ohm's law in order to deal with the cancellation problem. In order to obtain numerically stable simulations, the antenna's fields are integrated into the mixed-variable formulation and the  pullback scheme. We deployed the antenna in linear and nonlinear simulations that aim to excite TAE modes. We showed here that the frequency and the radial mode structure of the outcome excited mode is in good agreement with the one obtained from fast particle simulations. Furthermore, we show that measuring the damping rate of TAE mode once the antenna is switched off is possible. Although the target TAE mode can be excited in  linear simulations as expected, the antenna is shown to excite other modes in  nonlinear simulations due to the coupling between the modes. In the example shown, a $n=6$ TAE-like antenna is nonlinearly exciting a $n=2$ TAE which has almost the same frequency as the antenna, but also $n=0$ and  $n=1$ GAEs, 
at very different frequencies. Using forced external electromagnetic perturbations with the antenna setup is thus a useful tool to probe various nonlinear mode couplings.

\section*{Acknowledgment}
The authors would like to thank Ralf Kleiber and Ben McMillan for their useful comments.  
We acknowledge PRACE for awarding us access to Marconi100 at CINECA, Italy.
This work has been carried out within the framework of the EUROfusion Consortium and has received funding from the Euratom research and training programme 2014-2018 and 2019-2020 under grant agreement No 633053. The views and opinions expressed herein do not necessarily reflect those of the European Commission.
This work is also supported by a grant from the Swiss National Supercomputing Centre (CSCS) under project ID s1067.

\section*{References}

\bibliographystyle{abbrv}
\bibliography{refs}

\begin{thebibliography}{10}

\bibitem{bass2010gyrokinetic}
E.~Bass and R.~Waltz.
\newblock Gyrokinetic simulations of mesoscale energetic particle-driven
  {Alfv{\'e}nic} turbulent transport embedded in microturbulence.
\newblock {\em Physics of Plasmas}, 17(11):112319, 2010.

\bibitem{biglari1992resonant}
H.~Biglari, F.~Zonca, and L.~Chen.
\newblock On resonant destabilization of toroidal {Alfv{\'e}n} eigenmodes by
  circulating and trapped energetic ions/alpha particles in tokamaks.
\newblock {\em Physics of Fluids B: Plasma Physics}, 4(8):2385--2388, 1992.

\bibitem{briguglio1995hybrid}
S.~Briguglio, G.~Vlad, F.~Zonca, and C.~Kar.
\newblock Hybrid magnetohydrodynamic-gyrokinetic simulation of toroidal
  {Alfv{\'e}n} modes.
\newblock {\em Physics of Plasmas}, 2(10):3711--3723, 1995.

\bibitem{brunner1998global}
S.~Brunner, M.~Fivaz, T.~Tran, and J.~Vaclavik.
\newblock Global approach to the spectral problem of microinstabilities in
  tokamak plasmas using a gyrokinetic model.
\newblock {\em Physics of Plasmas}, 5(11):3929--3949, 1998.

\bibitem{cheng1986low}
C.~Cheng and M.~Chance.
\newblock Low-n shear {Alfv{\'e}n} spectra in axisymmetric toroidal plasmas.
\newblock {\em The Physics of fluids}, 29(11):3695--3701, 1986.

\bibitem{cheng1991alpha}
C.-Z. Cheng.
\newblock Alpha particle destabilization of the toroidicity-induced
  {Alfv{\'e}n} eigenmodes.
\newblock {\em Physics of Fluids B: Plasma Physics}, 3(9):2463--2471, 1991.

\bibitem{fasoli2000fast}
A.~Fasoli, D.~Borba, B.~Breizman, C.~Gormezano, R.~Heeter, A.~Juan,
  M.~Mantsinen, S.~Sharapov, and D.~Testa.
\newblock Fast particles-wave interaction in the {Alfv{\'e}n} frequency range
  on the joint european torus tokamak.
\newblock {\em Physics of Plasmas}, 7(5):1816--1824, 2000.

\bibitem{fu1989excitation}
G.~Fu and J.~Van~Dam.
\newblock Excitation of the toroidicity-induced shear {Alfv{\'e}n} eigenmode by
  fusion alpha particles in an ignited tokamak.
\newblock {\em Physics of Fluids B: Plasma Physics}, 1(10):1949--1952, 1989.

\bibitem{kleiber2018global}
R.~Kleiber, M.~Borchardt, A.~K{\"o}nies, A.~Mishchenko, J.~Riemann, C.~Slaby,
  and R.~Hatzky.
\newblock Global gyrokinetic multi-model simulations of {ITG} and
  {Alfv{\'e}nic} modes for tokamaks and the first operational phase of
  {Wendelstein 7-X}.
\newblock In {\em 27th IAEA Fusion Energy Conference, Gandhinagar}, 2018.

\bibitem{kleiber2021modern}
R.~Kleiber, M.~Borchardt, A.~K{\"o}nies, and C.~Slaby.
\newblock Modern methods of signal processing applied to gyrokinetic
  simulations.
\newblock {\em Plasma Physics and Controlled Fusion}, 63(3):035017, 2021.

\bibitem{konies2018benchmark}
A.~K{\"o}nies, S.~Briguglio, N.~Gorelenkov, T.~Feh{\'e}r, M.~Isaev, P.~Lauber,
  A.~Mishchenko, D.~Spong, Y.~Todo, W.~Cooper, et~al.
\newblock Benchmark of gyrokinetic, kinetic mhd and gyrofluid codes for the
  linear calculation of fast particle driven {TAE} dynamics.
\newblock {\em Nuclear Fusion}, 58(12):126027, 2018.

\bibitem{lanti2020orb5}
E.~Lanti, N.~Ohana, N.~Tronko, T.~Hayward-Schneider, A.~Bottino, B.~McMillan,
  A.~Mishchenko, A.~Scheinberg, A.~Biancalani, P.~Angelino, et~al.
\newblock {ORB5}: a global electromagnetic gyrokinetic code using the pic
  approach in toroidal geometry.
\newblock {\em Computer Physics Communications}, 251:107072, 2020.

\bibitem{lauber2003linear}
P.~Lauber.
\newblock {\em Linear gyrokinetic description of fast particle effects on the
  MHD stability in tokamaks}.
\newblock PhD thesis, Technische Universit{\"a}t M{\"u}nchen, 2003.

\bibitem{lauber2007ligka}
P.~Lauber, S.~G{\"u}nter, A.~K{\"o}nies, and S.~D. Pinches.
\newblock {LIGKA}: A linear gyrokinetic code for the description of background
  kinetic and fast particle effects on the {MHD} stability in tokamaks.
\newblock {\em Journal of Computational Physics}, 226(1):447--465, 2007.

\bibitem{lauber2005kinetic}
P.~Lauber, S.~G{\"u}nter, and S.~Pinches.
\newblock Kinetic properties of shear {Alfv{\'e}n} eigenmodes in tokamak
  plasmas.
\newblock {\em Physics of plasmas}, 12(12):122501, 2005.

\bibitem{MISHCHENKO2019194}
A.~Mishchenko, A.~Bottino, A.~Biancalani, R.~Hatzky, T.~Hayward-Schneider,
  N.~Ohana, E.~Lanti, S.~Brunner, L.~Villard, M.~Borchardt, R.~Kleiber, and
  A.~Könies.
\newblock Pullback scheme implementation in {ORB5}.
\newblock {\em Computer Physics Communications}, 238:194--202, 2019.

\bibitem{mishchenko2014pullback}
A.~Mishchenko, A.~K{\"o}nies, R.~Kleiber, and M.~Cole.
\newblock Pullback transformation in gyrokinetic electromagnetic simulations.
\newblock {\em Physics of Plasmas}, 21(9):092110, 2014.

\bibitem{nabais2018tae}
F.~Nabais, V.~Aslanyan, D.~Borba, R.~Coelho, R.~Dumont, J.~Ferreira,
  A.~Figueiredo, M.~Fitzgerald, E.~Lerche, J.~Mailloux, et~al.
\newblock {TAE} stability calculations compared to {TAE} antenna results in
  {JET}.
\newblock {\em Nuclear Fusion}, 58(8):082007, 2018.

\bibitem{nishimura2009excitation}
Y.~Nishimura.
\newblock Excitation of low-n toroidicity induced {Alfv{\'e}n} eigenmodes by
  energetic particles in global gyrokinetic tokamak plasmas.
\newblock {\em Physics of Plasmas}, 16(3):030702, 2009.

\bibitem{ohana2020using}
N.~T.~E. Ohana.
\newblock {\em Using an antenna as a tool for studying microturbulence and
  zonal structures in tokamaks with a global gyrokinetic {GPU}-enabled
  particle-in-cell code}.
\newblock PhD thesis, EPFL, No.10127, 2020.

\bibitem{vannini2020gyrokinetic}
F.~Vannini, A.~Biancalani, A.~Bottino, T.~Hayward-Schneider, P.~Lauber,
  A.~Mishchenko, I.~Novikau, E.~Poli, and A.~U. Team.
\newblock Gyrokinetic investigation of the damping channels of alfv{\'e}n modes
  in asdex upgrade.
\newblock {\em Physics of Plasmas}, 27(4):042501, 2020.

\bibitem{villard1995global}
L.~Villard, S.~Brunner, and J.~Vaclavik.
\newblock Global marginal stability of {TAE}s in the presence of fast ions.
\newblock {\em Nuclear Fusion}, 35(10):1173, 1995.

\bibitem{zhang2012global}
W.~Zhang, I.~Holod, Z.~Lin, and Y.~Xiao.
\newblock Global gyrokinetic particle simulation of toroidal {Alfv{\'e}n}
  eigenmodes excited by antenna and fast ions.
\newblock {\em Physics of Plasmas}, 19(2):022507, 2012.

\end{thebibliography}
\end{document}